%% file: ys-202611-019-resolvent-ideals-sigma-algebras.tex
\documentclass[11pt,a4paper]{article}

\usepackage{fontspec}
\usepackage[titletoc,title,header]{appendix}

\usepackage{amsmath}
\usepackage{amssymb}
\usepackage{amsthm}
\usepackage{mathtools}
\usepackage{xparse}    % for \NewDocumentCommand

\usepackage[margin=1in]{geometry}

\usepackage{graphicx}
\usepackage{xcolor}

\usepackage{hyperref}
\hypersetup{
  colorlinks=true,
  linkcolor=blue,
  citecolor=blue,
  urlcolor=blue,
  pdfauthor={Yoshitsugu Sekine},
  pdftitle={A Probabilistic Characterization of the Ideals of Resolvent Algebras}
}

\usepackage[numbers]{natbib}
\theoremstyle{plain}
\newtheorem{thm}{Theorem}[section]
\AtEndEnvironment{thm}{\qed}
\newtheorem{lem}[thm]{Lemma}
\AtEndEnvironment{lem}{\qed}
\newtheorem{prop}[thm]{Proposition}
\AtEndEnvironment{prop}{\qed}
\newtheorem{cor}[thm]{Corollary}
\AtEndEnvironment{cor}{\qed}

\AtEndEnvironment{conj}{\qed}

\AtEndEnvironment{fact}{\qed}

\theoremstyle{definition}
\newtheorem{defn}[thm]{Definition}
\AtEndEnvironment{defn}{\qed}

\AtEndEnvironment{ex}{\qed}

\theoremstyle{remark}
\newtheorem{rem}[thm]{Remark}
\AtEndEnvironment{rem}{\qed}

\input{mycommands.tex}

\title{A Probabilistic Characterization of the Ideals of Resolvent Algebras}

\author{%
Yoshitsugu Sekine\\{\small\texttt{4429sekine@gmail.com}}%
}

\date{\today}

\begin{document}

\maketitle

\begin{abstract}
We characterize the ideal theory of Buchholz for resolvent algebras in probabilistic terms. Regarding closed two-sided ideals as kernels of representations, we pass from representations to the Gelfand--Naimark--Segal representations of states, and describe the singular, trivially regular, and regular parts, which Buchholz gives through the one-particle space, by the behavior of the corresponding probability measures of the fields. On an isotropic subspace, the resolvent algebra is a commutative \(C^{\ast}\)-algebra whose states correspond to probability measures, and this correspondence carries the description further to sub-\(\sigma\)-algebras. Starting from Bose--Einstein condensation of the free Bose gas with its direct integral representation and from the treatment of the infrared singularity of the van Hove model, we connect the ideal-theoretic and the probabilistic arguments. Since the ideals carry physical information in these arguments, we show that the corresponding sub-\(\sigma\)-algebras and their inclusions carry this information in correspondence with the ideals.

\noindent\textbf{Keywords:} resolvent algebra, closed two-sided ideal, quotient algebra, field law, sub-\(\sigma\)-algebra, independence, conditional expectation, KMS state, \(\beta\)-Markov path space, Bose--Einstein condensation, infrared singularity
\end{abstract}

\setcounter{tocdepth}{2}
\tableofcontents

\section{Introduction}\label{sec:introduction}

Unlike the Weyl algebra, which is simple, the resolvent algebra of Buchholz and Grundling has a rich structure of closed two-sided ideals \cite[Theorem 3.8]{BuchholzGrundling002}. The ideals carry physical information: they are the kernels of the representations in which some fields take infinite values \cite[Section 9]{BuchholzGrundling002}, a unital \(\oacstar\)-algebra admitting a sufficiently diverse variety of dynamics cannot be simple, and the nesting of the primitive ideals encodes the dimension of the symplectic space \cite[Section 1]{DetlevBuchholz001}. Buchholz describes the ideals through the subspaces of the symplectic space, the one-particle space of a Bose field. A closed two-sided ideal is the kernel of a representation, and a representation is described through the one-particle space by its regular and trivially regular subspaces and the complementary singular set \cite[Lemma 3.1]{DetlevBuchholz001}. The primitive ideals of a finite-dimensional resolvent algebra correspond to the subspaces with a character of the commutative algebra of the radical of \(\opform{\sigma}\) on the subspace that does not vanish at its generating resolvents \cite[Definition 3.2 and Propositions 3.3 and 3.5]{DetlevBuchholz001}, the maximal ideals to the isotropic subspaces with such a character \cite[Theorem 4.6]{DetlevBuchholz001}, and all ideals are built from principal ideals generated by resolvents and shifted resolvents, related to sharp field values including \(\infty\) \cite[Lemma 4.1, Proposition 4.2, and Section 4]{DetlevBuchholz001}.

This paper characterizes this ideal theory in probabilistic terms. We pass from representations to the Gelfand--Naimark--Segal (GNS) representations of states, and describe these parts by the laws of the fields, the probability measures on the one-point compactification of the line by which Buchholz and Grundling describe nonregular states \cite[Section 4]{BuchholzGrundling002}: in a factor state, the singular, the trivially regular, and the other regular directions have as field laws the Dirac measure at infinity, Dirac measures at real values, and measures on the line that are not Dirac measures. On an isotropic subspace, the resolvent algebra is a commutative \(\oacstar\)-algebra, and its states correspond to probability measures on its characters. This correspondence carries the description further for the ideals generated by the resolvents outside subspaces: the information carried by these ideals is carried by sub-\(\sigma\)-algebras, as information is in probability theory, with the von Neumann algebras generated by resolvent subalgebras corresponding to sub-\(\sigma\)-algebras up to null sets, the quotients to restrictions, an inclusion of the ideals implying the reverse inclusion of the sub-\(\sigma\)-algebras, and independence characterized by conditional expectations. Klein and Landau showed that stochastically positive Kubo--Martin--Schwinger (KMS) systems are equivalent to periodic stochastic processes satisfying a positivity of Osterwalder--Schrader type \cite[Theorem 6.1 and Sections 7 and 8]{KleinLandau001}, and Dereziński and Gérard construct KMS states from \(\sminvtemperature\)-Markov path spaces \cite[Section 21.4, Theorems 21.43 and 21.47]{DerezinskiGerard001}; the sub-\(\sigma\)-algebras then become those of these processes. Probabilistic techniques often solve problems that are hard for operator theory and operator algebras alone, as in constructive quantum field theory \cite{GlimmJaffe001,BarrySimon004,LorincziHiroshimaBetz002,LorincziHiroshimaBetz003} and in the uses of path and function space methods recalled in \cite[Introduction to Chapter 21]{DerezinskiGerard001} and \cite[Introduction]{KleinLandau001}, and the probabilistic description gives such techniques access to the ideals.

The concrete motivation is Bose--Einstein condensation (BEC) of quasiparticles whose number is not conserved. According to Yukalov, such a condensation occurs only in nonequilibrium systems with external pumping, and self-consistently defined collective excitations such as phonons do not condense \cite[Abstract and Conclusion]{VIYukalov001}. A series of preprints examines this statement mathematically for the free Bose gas \cite{YoshitsuguSekine004} and for Bose fields coupled to a source or to particles, the van Hove model \cite{YoshitsuguSekine006,YoshitsuguSekine007}, the spin--boson model \cite{YoshitsuguSekine008}, and the Nelson and Pauli--Fierz models \cite{YoshitsuguSekine009}, both on the resolvent algebra and on the singular Gaussian \(\sminvtemperature\)-Markov path spaces \cite{YoshitsuguSekine018}, which extend those of \cite[Section 21.4]{DerezinskiGerard001} to one-particle Hamiltonians without a spectral gap and realize the Araki--Woods representations \cite[Theorem 5.4]{YoshitsuguSekine018}. Two mechanisms, from the free Bose gas and from the van Hove model, give ideals that enter in opposite ways. Above the critical density, the BEC state of the free Bose gas is a mixture of states whose fields are shifted by the condensate, and its GNS representation is a direct integral over the order coordinate of the condensate. The condensate is carried by the test functions on which the zero-mode form does not vanish, and the resolvents of these condensate directions generate an ideal. This ideal is not visible on the level of the \(\oacstar\)-algebra, since the representation of the BEC state is faithful; in the weak closure, strong limits of resolvents of zero-mode probe test functions, which are condensate directions, are central and give the order coordinate. In the van Hove model, the linear coupling shifts the fields by a mean functional defined only on a subspace \(Y\) of the test functions, and the infrared-singular directions outside \(Y\) generate an ideal of the other kind. The treatment of the infrared singularity reduces the algebra of observables to a subalgebra: the quotient by this ideal is the resolvent algebra of \(Y\), so that the observables of the infrared-singular directions are discarded and those of \(Y\) are kept. The removal of the infrared cutoff makes the fields of the directions with divergent cutoff means escape to infinity, and the limit state annihilates the ideal of these directions, which is contained in the ideal of the infrared-singular directions; when all fields outside a subspace escape and the laws of the fields on it converge, the GNS representation of the limit state maps the whole resolvent algebra onto the image of the resolvent algebra of this subspace. The condensate is removed when the first ideal is contained in the second. On the path space, the fields of the condensate directions are correlated with the order coordinate, and the fields of the escaping directions take the value \(\infty\) in the limit. In the Pauli--Fierz model without spin, the minimal coupling enters the functional integral through a phase, the fields are not shifted, and no real-valued direction escapes; in both Pauli--Fierz models, the condensate directions are those of the free field. The revisions of \cite{YoshitsuguSekine004,YoshitsuguSekine006} and the other models thus need a general theory of these ideals and of their probabilistic counterparts on an abstract symplectic space. The results of Buchholz do not yet cover the ideal generated by the resolvents outside an infinite-dimensional, possibly degenerate subspace, such as the domain of a mean functional or the real-valued test functions, and its quotient. For a finite-dimensional symplectic space, the kernels of irreducible representations of Buchholz are generated by such resolvents and shifted resolvents of the trivially regular part, and the quotient is the resolvent algebra of a nondegenerate subspace. The description of the ideals through field laws, sub-\(\sigma\)-algebras with their inclusion, independence, and conditional expectations, and the processes of Klein and Landau is not covered either.

The results are as follows. For real linear subspaces \(Y\subset W\), the ideal \(\oaideal{J}_{W\setminus Y}\) generated by the resolvents of \(W\setminus Y\) is the kernel of the representation obtained from a faithful representation of the resolvent algebra of \(Y\) by letting the resolvents of the other directions vanish, and the quotient is the resolvent algebra of \(Y\) (Theorem \ref{thm:subspace-complement-quotient}). After the classification of the characters (Proposition \ref{prop:character-classification}), the ideals of Buchholz are described by the laws of single fields. Principal ideals are described by sharp or infinite values (Proposition \ref{prop:principal-ideal-sharp-value}), and the singular, the trivially regular, and the other regular directions of a factor state by the Dirac measure at infinity, Dirac measures at real values, and the other laws (Proposition \ref{prop:factor-field-laws}). Primitive ideals of a finite-dimensional resolvent algebra are described by the directions with infinite fields, the directions with sharp fields, and the sharp values (Proposition \ref{prop:primitive-ideal-field-laws}), and characters by Dirac laws and the commutator ideal by mixtures of characters (Proposition \ref{prop:partial-configurations}). Limits of states with infinite fields in some directions annihilate the ideal of these directions (Proposition \ref{prop:infinite-field-reduction} and Corollary \ref{cor:shifted-vacuum-reduction}). For commuting random variables on an isotropic subspace, the GNS representation is the multiplication representation on \(\lp^{2}\) of the \(\sigma\)-algebra of the fields, the von Neumann algebras generated by resolvent subalgebras correspond to sub-\(\sigma\)-algebras up to null sets, and the quotient by \(\oaideal{J}_{W\setminus Y}\) gives the discarded fields the value \(\infty\) (Theorems \ref{thm:random-variable-ideals} and \ref{thm:random-variable-gns} and Corollary \ref{cor:random-variable-subalgebras}); an inclusion of the ideals \(\oaideal{J}_{W\setminus Y}\) implies the reverse inclusion of the sub-\(\sigma\)-algebras, and independence is characterized by conditional expectations (Theorem \ref{thm:ideal-inclusion-sigma-independence} and Remark \ref{rem:information}). For fields that are sums of a family independent of a sub-\(\sigma\)-algebra and of shifts measurable with respect to it, the correlated directions generate an ideal, removed by a subspace exactly when its fields are independent of the sub-\(\sigma\)-algebra (Theorem \ref{thm:shifted-fields-correlated-ideal}). For the corresponding direct integral of shifted representations, the representation is faithful, so that this ideal is not visible on the level of the \(\oacstar\)-algebra, and, along directions in which the unshifted fields vanish in the limit and the shifts converge to a function that is not almost everywhere constant, strong limits of the images of its generators are nonscalar elements of the center of the weak closure (Propositions \ref{prop:direct-integral-shifted-quotient} and \ref{prop:direct-integral-faithful-center}). The escape of all fields of \(W\setminus Y\) to infinity, the abstract infrared limit, gives with the convergence of the laws of the fields on \(Y\) as the limit the state that extends the state of the limit fields on \(Y\) and vanishes on the ideal of the escaping directions; this ideal lies in the kernel of its GNS representation, which maps the resolvent algebra of \(W\) onto the image of that of \(Y\) (Theorem \ref{thm:escape-to-infinity}). Conversely, fields whose laws are symmetric and whose characteristic functions dominate those of a fixed family do not escape, and no limit state vanishes on such an ideal (Proposition \ref{prop:field-law-domination}). For KMS states (Definitions \ref{def:kms-condition} and \ref{def:stochastic-positivity}), the time translates of the resolvent algebra of an isotropic subspace generate in general a proper subalgebra (Proposition \ref{prop:time-translates-generation}), on whose quotient by the kernel of the representation the state is faithful (Proposition \ref{prop:kms-reduction}); a resolvent of the subspace lies in the kernel exactly when its field is \(\infty\) almost surely, and under stochastic positivity the process of Klein and Landau has the measure representing the state on the characters as its one-time law and the reduced correlation functions as its expectations (Corollary \ref{cor:kms-reduction-process}). This applies to the sharp-time fields of the Gaussian \(\sminvtemperature\)-Markov path spaces (Proposition \ref{prop:path-space-random-variables}). A Gaussian field coupled to a source along paths is shifted by random variables of the paths for a real coupling constant with a finite normalization; for an imaginary coupling constant, fields with the resulting characteristic functions, when they exist, have symmetric laws whose characteristic functions dominate those of the uncoupled field. A source orthogonal to the condensate form leaves the condensate directions unchanged, provided that the fields of the reduced covariance form exist (Proposition \ref{prop:gaussian-source}). Section \ref{sec:applications} indicates how the results enter the models.

\section{Setting}\label{sec:setting}

Throughout, \(\pairbk{X,\opform{\sigma}}\) is a nondegenerate real symplectic space of arbitrary dimension, \(\semigrposint\) is the set of positive integers, \(\fndef{B}\) is the indicator function of a set \(B\), and \(\opspbddlin{\sphilbfrak{H}}\) is the algebra of bounded operators on a Hilbert space \(\sphilbfrak{H}\). Its resolvent algebra \(\fun{\oaresolventalgebra}{X,\opform{\sigma}}\) is generated by the resolvents \(\fun{\oaresolvent}{\lambda,f}\), \(\lambda\in\fldreal\setminus\setone{0}\), \(f\in X\), with the relations of \cite[Definition 3.1, (1)--(6)]{BuchholzGrundling002}: for \(\lambda,\mu,\nu\in\fldreal\setminus\setone{0}\) and \(f,g\in X\), with \(\lambda+\mu\neq0\) in \eqref{eq:resolvent-relation-sum}, \begin{align}
\fun{\oaresolvent}{\lambda,0}
&=
-\frac{\imunit}{\lambda},
\label{eq:resolvent-relation-zero}
\\ %%%%%%%%%%%%%%%%
\faadj{\fun{\oaresolvent}{\lambda,f}}
&=
\fun{\oaresolvent}{-\lambda,f},
\label{eq:resolvent-relation-adjoint}
\\ %%%%%%%%%%%%%%%%
\nu\fun{\oaresolvent}{\nu\lambda,\nu f}
&=
\fun{\oaresolvent}{\lambda,f},
\label{eq:resolvent-relation-scaling}
\\ %%%%%%%%%%%%%%%%
\fun{\oaresolvent}{\lambda,f}-\fun{\oaresolvent}{\mu,f}
&=
\imunit\rbk{\mu-\lambda}\fun{\oaresolvent}{\lambda,f}\fun{\oaresolvent}{\mu,f},
\label{eq:resolvent-relation-identity}
\\ %%%%%%%%%%%%%%%%
\commutator{\fun{\oaresolvent}{\lambda,f}}{\fun{\oaresolvent}{\mu,g}}
&=
\imunit\fun{\opform{\sigma}}{f,g}\fun{\oaresolvent}{\lambda,f}\fun{\oaresolvent}{\mu,g}^{2}\fun{\oaresolvent}{\lambda,f},
\label{eq:resolvent-relation-commutator}
\\ %%%%%%%%%%%%%%%%
\fun{\oaresolvent}{\lambda,f}\fun{\oaresolvent}{\mu,g}
&=
\fun{\oaresolvent}{\lambda+\mu,f+g}\sqbk{\fun{\oaresolvent}{\lambda,f}+\fun{\oaresolvent}{\mu,g}+\imunit\fun{\opform{\sigma}}{f,g}\fun{\oaresolvent}{\lambda,f}^{2}\fun{\oaresolvent}{\mu,g}};
\label{eq:resolvent-relation-sum}
\end{align} its norm is that of \cite[Definition 3.4]{BuchholzGrundling002}. By the construction, every assignment of elements of a unital \(\oacstar\)-algebra to the generators that satisfies these relations extends to a unital \(\ast\)-homomorphism of \(\fun{\oaresolventalgebra}{X,\opform{\sigma}}\), and \(\norm{\fun{\oaresolvent}{\lambda,f}}=\abs{\lambda}^{-1}\) \cite[Theorem 3.6(iii)]{BuchholzGrundling002}. The resolvents \(\fun{\oaresolvent}{z,f}\) with complex parameters \(z\in\fldcmp\setminus\imunit\fldreal\) are those of the paragraph after \cite[Theorem 3.6]{BuchholzGrundling002}, and for a linear map \(h\colon X\to\fldreal\), the automorphism \(\Theta_{h}\) with \(\fun{\Theta_{h}}{\fun{\oaresolvent}{z,f}}=\fun{\oaresolvent}{z+\imunit\fun{h}{f},f}\) is that of \cite[Proposition 3.7]{BuchholzGrundling002}. A symplectic linear bijection \(T\) of \(X\) gives the automorphism \(\alpha_T\) with \(\fun{\alpha_T}{\fun{\oaresolvent}{\lambda,f}}=\fun{\oaresolvent}{\lambda,Tf}\) \cite[Theorem 3.6(v)]{BuchholzGrundling002}. In a representation \(\oarepn\) with \(\Ker\fun{\oarepn}{\fun{\oaresolvent}{1,f}}=\setone{0}\), the operator \(\fun{\oarepn}{\fun{\oaresolvent}{\lambda,f}}\) is \(\rbk{\imunit\lambda-\fun{\phi_{\oarepn}}{f}}^{-1}\) for a self-adjoint field \(\fun{\phi_{\oarepn}}{f}\) \cite[Theorem 4.2]{BuchholzGrundling002}. For a state \(\oastate[\psi]\), the Gelfand--Naimark--Segal triple is written \(\pairbk{\sphilbfrak{H}_{\psi},\oarepn_{\psi},\oagnsvector[\Psi]}\), with \begin{equation}\label{eq:gns-vector-state}
\begin{aligned}
\fun{\oastate[\psi]}{A}
&=
\bkt{\oagnsvector[\Psi]}{\fun{\oarepn_{\psi}}{A}\oagnsvector[\Psi]}.
\end{aligned}
\end{equation} The GNS triple of \(\oastate[\psi_{\xi}]\) is \(\pairbk{\sphilbfrak{H}_{\xi},\oarepn_{\xi},\oagnsvector[\Psi_{\xi}]}\), and similarly for other subscripts.

The functions \begin{equation}\label{eq:resolvent-value-function}
\begin{aligned}
\fun{r_{\lambda}}{u}
=
\begin{dcases}
\frac{1}{\imunit\lambda-u} & \rbk{u\in\fldreal},
\\ %%%%%%%%%%%%%%%%
0 & \rbk{u=\infty}
\end{dcases}
\end{aligned}
\end{equation} on the one-point compactification \(\gtoneptcpt{\fldreal}=\fldreal\cup\setone{\infty}\) are continuous and satisfy \(\cmpconj{r_{\lambda}}=r_{-\lambda}\). Each \(r_{\lambda}\) is a bijection of \(\gtoneptcpt{\fldreal}\) onto the circle \begin{equation}\label{eq:resolvent-value-circle}
\begin{aligned}
C_{\lambda}
&=
\set{z\in\fldcmp}{\abs{z+\frac{\imunit}{2\lambda}}=\frac{1}{2\abs{\lambda}}},
\end{aligned}
\end{equation} since \(\fun{r_{\lambda}}{u}\neq0\) and \(u=\imunit\lambda-\fun{r_{\lambda}}{u}^{-1}\) for \(u\in\fldreal\), and \(z\in C_{\lambda}\setminus\setone{0}\) exactly when \(\opimag z^{-1}=\lambda\).

For a real linear subspace \(S\subset X\), let \(\oaresolventalgebra(S)\) be the \(\oacstar\)-subalgebra of \(\fun{\oaresolventalgebra}{X,\opform{\sigma}}\) generated by the resolvents \(\fun{\oaresolvent}{\lambda,f}\) with \(\lambda\in\fldreal\setminus\setone{0}\) and \(f\in S\); it is the norm closure of the linear span of these resolvents alone \cite[Theorem and Section 3]{BuchholzVanNuland001}, since a product of two of them is a norm limit of linear combinations of the resolvents of the linear combinations of the two vectors, which lie in \(S\), and \(\idone=\imunit\lambda\fun{\oaresolvent}{\lambda,0}\). A real linear subspace \(W\subset X\) is isotropic if \begin{equation}\label{eq:isotropic-subspace}
\begin{aligned}
\fun{\opform{\sigma}}{f,g}
&=
0
&&\rbk{f,g\in W}.
\end{aligned}
\end{equation} For real linear subspaces \(Y\subset W\subset X\), let \(\oaideal{J}_{W\setminus Y}\) be the closed two-sided ideal of \(\oaresolventalgebra(W)\) generated by the resolvents \(\fun{\oaresolvent}{\lambda,f}\) with \(\lambda\in\fldreal\setminus\setone{0}\) and \(f\in W\setminus Y\). For \(f\in X\), \(\lambda\in\fldreal\setminus\setone{0}\), and \(\rho\in\fldcmp\), the closed two-sided ideal of \(\fun{\oaresolventalgebra}{X,\opform{\sigma}}\) generated by \(\fun{\oaresolvent}{\lambda,f}-\rho\) is written \(\fun{I}{\lambda,f;\rho}\), as in \cite[Proposition 4.2]{DetlevBuchholz001}. For a representation \(\oarepn\), the regular, trivially regular, and singular parts of \(X\) are \begin{equation}\label{eq:regular-trivial-singular-parts}
\begin{aligned}
X_{\txtregular,\oarepn}
&=
\set{f\in X}{\Ker\fun{\oarepn}{\fun{\oaresolvent}{1,f}}=\setone{0}},
\\ %%%%%%%%%%%%%%%%
X_{\mathrm{t},\oarepn}
&=
\set{f\in X_{\txtregular,\oarepn}}{\fun{\oarepn}{\fun{\oaresolvent}{1,f}}^{-1}\in\opspbddlin{\sphilbfrak{H}_{\oarepn}}},
\\ %%%%%%%%%%%%%%%%
X_{\txtsingular,\oarepn}
&=
X\setminus X_{\txtregular,\oarepn},
\end{aligned}
\end{equation} the sets \(X_R\), \(X_T\), and \(X_S\) of \cite[Proposition 4.7]{BuchholzGrundling002} and \cite[Lemma 3.1]{DetlevBuchholz001}. The subscripts \(\txtregular\), \(\mathrm{t}\), and \(\txtsingular\) stand for regular, trivially regular, and singular: on \(X_{\mathrm{t},\oarepn}\) the field \(\fun{\phi_{\oarepn}}{f}\) is a bounded operator, and on \(X_{\txtsingular,\oarepn}\) the resolvent \(\fun{\oarepn}{\fun{\oaresolvent}{1,f}}\) has a nonzero kernel.

\section{Ideals of subspaces}\label{sec:ideals-of-subspaces}

In this section, the ideal generated by the resolvents of the vectors outside a subspace is identified as the kernel of a representation in which these resolvents vanish, the quotient by it is identified with the resolvent algebra of the subspace, and the distance of such a resolvent from the resolvent algebra of the subspace is bounded below.

\begin{lem}[Resolvent families vanishing outside a subspace]\label{lem:zero-extension}
Let $Y\subset X$ be a real linear subspace.
\begin{enumerate}
\item Let $\oa{B}$ be a unital $\oacstar$-algebra, and let
$\fun{d}{\lambda,f}\in\oa{B}$, $\lambda\in\fldreal\setminus\setone{0}$,
$f\in Y$, satisfy
\eqref{eq:resolvent-relation-zero}--\eqref{eq:resolvent-relation-sum} with $d$
in place of $\oaresolvent$ for $f,g\in Y$.
Then there is exactly one unital $\ast$-homomorphism of
$\fun{\oaresolventalgebra}{X,\opform{\sigma}}$ into $\oa{B}$ that maps
$\fun{\oaresolvent}{\lambda,f}$ to $\fun{d}{\lambda,f}$ for $f\in Y$ and to $0$
for $f\in X\setminus Y$.
\item For $\lambda,\mu,\nu\in\fldreal\setminus\setone{0}$ and $a,b\in\fldreal$,
\begin{equation}\label{eq:value-function-relations}
\begin{aligned}
\fun{r_{\lambda}}{0}
&=
-\frac{\imunit}{\lambda},
\quad
\cmpconj{\fun{r_{\lambda}}{a}}
=
\fun{r_{-\lambda}}{a},
\quad
\nu\fun{r_{\nu\lambda}}{\nu a}
=
\fun{r_{\lambda}}{a},
\\ %%%%%%%%%%%%%%%%
\fun{r_{\lambda}}{a}-\fun{r_{\mu}}{a}
&=
\imunit\rbk{\mu-\lambda}\fun{r_{\lambda}}{a}\fun{r_{\mu}}{a},
\end{aligned}
\end{equation}
and, for $\lambda+\mu\neq0$,
\begin{equation}\label{eq:value-function-sum}
\begin{aligned}
\fun{r_{\lambda}}{a}\fun{r_{\mu}}{b}
&=
\fun{r_{\lambda+\mu}}{a+b}\rbk{\fun{r_{\lambda}}{a}+\fun{r_{\mu}}{b}}.
\end{aligned}
\end{equation}
If $Y$ is isotropic and $\ell\colon Y\to\fldreal$ is linear, the scalars
$\fun{r_{\lambda}}{\fun{\ell}{f}}$, $f\in Y$, satisfy the hypothesis of (1)
with $\oa{B}=\fldcmp$.
\end{enumerate}
\end{lem}

\begin{proof}
(1): Extend $d$ by $0$ to $X\setminus Y$.
A relation that involves a vector outside $Y$ holds: in
\eqref{eq:resolvent-relation-adjoint}--\eqref{eq:resolvent-relation-commutator},
every term contains $\fun{d}{\lambda,f}$ or $\fun{d}{\mu,g}$ with a vector
outside $Y$, since $\nu f\notin Y$ for $f\notin Y$; in
\eqref{eq:resolvent-relation-sum} with $f\notin Y$ or $g\notin Y$, the left-hand
side is $0$, and so is the right-hand side, by its first factor if
$f+g\notin Y$ and by the bracket if $f+g\in Y$, in which case $f,g\notin Y$.
The universal property of Section \ref{sec:setting} gives the homomorphism, which
is unique since it is determined on the generators.

(2): The identities \eqref{eq:value-function-relations} follow from
\eqref{eq:resolvent-value-function}, and \eqref{eq:value-function-sum} from
$$\begin{aligned}
\fun{r_{\lambda}}{a}+\fun{r_{\mu}}{b}
&=
\rbk{\imunit\rbk{\lambda+\mu}-a-b}\fun{r_{\lambda}}{a}\fun{r_{\mu}}{b}.
\end{aligned}$$
For $f,g\in Y$, the linearity of $\ell$ and $\fun{\opform{\sigma}}{f,g}=0$
reduce the relations to these identities.
\end{proof}

\begin{thm}[Ideal generated by the resolvents outside a subspace]\label{thm:subspace-complement-quotient}
Let $Y\subset W\subset X$ be real linear subspaces.
\begin{enumerate}
\item For every unital representation $\oarepn$ of $\oaresolventalgebra(Y)$ on a
Hilbert space $\sphilbfrak{H}$, there is exactly one representation $\oarepn_{\infty}$
of $\fun{\oaresolventalgebra}{X,\opform{\sigma}}$ on $\sphilbfrak{H}$ with
\begin{equation}\label{eq:subspace-infinite-representation}
\begin{aligned}
\fun{\oarepn_{\infty}}{\fun{\oaresolvent}{\lambda,f}}
&=
\begin{cases}
\fun{\oarepn}{\fun{\oaresolvent}{\lambda,f}},&f\in Y,\\
0,&f\in X\setminus Y,
\end{cases}
\end{aligned}
\end{equation}
for $\lambda\in\fldreal\setminus\setone{0}$.
Its restriction to $\oaresolventalgebra(Y)$ is $\oarepn$, and
$\fun{\oarepn_{\infty}}{\oaresolventalgebra(W)}=\fun{\oarepn}{\oaresolventalgebra(Y)}$.
\item Every $A\in\oaresolventalgebra(W)$ is $A=B+C$ with exactly one
$B\in\oaresolventalgebra(Y)$ and one $C\in\oaideal{J}_{W\setminus Y}$, and
$B\mapsto B+\oaideal{J}_{W\setminus Y}$ is a $\ast$-isomorphism from
$\oaresolventalgebra(Y)$ onto
$\setquot{\oaresolventalgebra(W)}{\oaideal{J}_{W\setminus Y}}$.
For $\oarepn$ as in (1), the kernel of the restriction of $\oarepn_{\infty}$ to
$\oaresolventalgebra(W)$ is $\Ker\oarepn+\oaideal{J}_{W\setminus Y}$; it is
$\oaideal{J}_{W\setminus Y}$ if and only if $\oarepn$ is faithful, and then
$A+\oaideal{J}_{W\setminus Y}\mapsto\fun{\oarepn_{\infty}}{A}$ is a faithful
representation of the quotient.
\item For every state $\oastate[\psi]$ of $\oaresolventalgebra(Y)$, the
functional
\begin{equation}\label{eq:subspace-extension-state}
\begin{aligned}
A
&\mapsto
\bkt{\oagnsvector[\Psi]}{\fun{\oarepn_{\psi,\infty}}{A}\oagnsvector[\Psi]}
&&\rbk{A\in\oaresolventalgebra(W)},
\end{aligned}
\end{equation}
with $\oarepn_{\psi,\infty}$ the representation of (1) for the GNS
representation $\oarepn_{\psi}$, is the only state of
$\oaresolventalgebra(W)$ that extends $\oastate[\psi]$ and vanishes on
$\oaideal{J}_{W\setminus Y}$, and the restriction of $\oarepn_{\psi,\infty}$
to $\oaresolventalgebra(W)$, with $\sphilbfrak{H}_{\psi}$ and
$\oagnsvector[\Psi]$, is a GNS triple of it.
A state of $\oaresolventalgebra(W)$ vanishes on $\oaideal{J}_{W\setminus Y}$ if
and only if it is the state \eqref{eq:subspace-extension-state} for its
restriction to $\oaresolventalgebra(Y)$.
\item For $f\in W$ and $\lambda\in\fldreal\setminus\setone{0}$,
$\fun{\oaresolvent}{\lambda,f}\in\oaideal{J}_{W\setminus Y}$ if and only if
$f\notin Y$.
In particular, for real linear subspaces $Y_1,Y_2\subset W$,
$\oaideal{J}_{W\setminus Y_2}\subset\oaideal{J}_{W\setminus Y_1}$ if and only
if $Y_1\subset Y_2$.
In this case,
$\oaresolventalgebra(Y_2)\cap\oaideal{J}_{W\setminus Y_1}=\oaideal{J}_{Y_2\setminus Y_1}$,
and $A+\oaideal{J}_{W\setminus Y_2}\mapsto A+\oaideal{J}_{W\setminus Y_1}$ is a
surjective $\ast$-homomorphism whose kernel is the image of
$\oaideal{J}_{Y_2\setminus Y_1}$ under the isomorphism of (2) for $Y_2$.
\end{enumerate}
\end{thm}

No dimension condition and no nondegeneracy of \(\opform{\sigma}\) on \(Y\) or \(W\) is assumed. For a finite-dimensional \(X\), a nondegenerate \(Y\), and \(W=X\), the isomorphism of item (2) is contained in the proof of \cite[Proposition 3.5]{DetlevBuchholz001}. In the representation \(\oarepn_{\infty}\), the resolvents of the directions of \(X\setminus Y\) vanish, so that the fields of these directions are infinite in the sense of \cite[Section 9]{BuchholzGrundling002}, while those of \(Y\) are represented as in \(\oarepn\); the passage to the quotient by \(\oaideal{J}_{W\setminus Y}\) is the passage to such a representation. The extension \eqref{eq:subspace-extension-state} of the restriction of a state of \(\oaresolventalgebra(W)\) is in general a new state: for \(Y\neq W\), the directions of \(W\setminus Y\) belong to the singular part \eqref{eq:regular-trivial-singular-parts} of its GNS representation, and it agrees with the given state on \(\oaresolventalgebra(Y)\), and on all of \(\oaresolventalgebra(W)\) exactly when the given state vanishes on \(\oaideal{J}_{W\setminus Y}\).

\begin{proof}[Proof of Theorem \ref{thm:subspace-complement-quotient}]
(1): The operators $\fun{\oarepn}{\fun{\oaresolvent}{\lambda,f}}$, $f\in Y$,
satisfy the relations for $f,g\in Y$, and
Lemma \ref{lem:zero-extension}(1) with $\oa{B}=\opspbddlin{\sphilbfrak{H}}$
gives $\oarepn_{\infty}$ and its uniqueness.
It agrees with $\oarepn$ on the generators of $\oaresolventalgebra(Y)$, hence on
$\oaresolventalgebra(Y)$, and maps the resolvents of the vectors of $W$ into the
norm-closed algebra $\fun{\oarepn}{\oaresolventalgebra(Y)}$, which gives
$\fun{\oarepn_{\infty}}{\oaresolventalgebra(W)}=\fun{\oarepn}{\oaresolventalgebra(Y)}$.

(2): Let $q$ be the quotient map of $\oaresolventalgebra(W)$ onto
$\setquot{\oaresolventalgebra(W)}{\oaideal{J}_{W\setminus Y}}$, and $q_Y$ its
restriction to $\oaresolventalgebra(Y)$.
The image of $q_Y$ is closed and contains the classes of the resolvents of the
vectors of $W$, which are $0$ for $W\setminus Y$; hence $q_Y$ is surjective.
For $\oarepn$ as in (1), $\oarepn_{\infty}$ vanishes on the generators of
$\oaideal{J}_{W\setminus Y}$, so that the homomorphism theorem gives a
representation $\widetilde{\oarepn}$ of the quotient with
$\oarepn_{\infty}=\widetilde{\oarepn}\circ q$ on $\oaresolventalgebra(W)$ and
$\widetilde{\oarepn}\circ q_Y=\oarepn$.
For a faithful $\oarepn$, $q_Y$ is therefore injective, hence a
$\ast$-isomorphism, and
$\oaresolventalgebra(Y)\cap\oaideal{J}_{W\setminus Y}=\Ker q_Y=\setone{0}$.
Consequently, $A=B+C$ with $B=\fun{\inv{q_Y}}{\fun{q}{A}}$ and
$C=A-B\in\oaideal{J}_{W\setminus Y}$, and the decomposition is unique because
$\oaresolventalgebra(Y)\cap\oaideal{J}_{W\setminus Y}=\setone{0}$.
For an arbitrary $\oarepn$, $\widetilde{\oarepn}=\oarepn\circ\inv{q_Y}$, so that
the kernel of $\oarepn_{\infty}$ on $\oaresolventalgebra(W)$ is
$\fun{\inv{q}}{\fun{q_Y}{\Ker\oarepn}}=\Ker\oarepn+\oaideal{J}_{W\setminus Y}$;
it is $\oaideal{J}_{W\setminus Y}$ if and only if $\Ker\oarepn=\setone{0}$, and
then $\widetilde{\oarepn}$ is faithful.

(3): The functional \eqref{eq:subspace-extension-state} is a state, extends
$\oastate[\psi]$ by (1), and vanishes on $\oaideal{J}_{W\setminus Y}$.
Its triple is a GNS triple, since
$\fun{\oarepn_{\psi,\infty}}{\oaresolventalgebra(W)}=\fun{\oarepn_{\psi}}{\oaresolventalgebra(Y)}$
by (1) and $\oagnsvector[\Psi]$ is cyclic for the latter.
A state $\oastate[\psi_1]$ of $\oaresolventalgebra(W)$ that vanishes on
$\oaideal{J}_{W\setminus Y}$ satisfies $\fun{\oastate[\psi_1]}{B+C}=\fun{\oastate[\psi_1]}{B}$
for the decomposition of (2), so it is determined by its restriction to
$\oaresolventalgebra(Y)$; this gives the uniqueness and the last statement.

(4): For $f\in W\setminus Y$, the resolvent is a generator.
For $f\in Y$, the resolvent is a nonzero element of $\oaresolventalgebra(Y)$, of
norm $\abs{\lambda}^{-1}$, and is not in $\oaideal{J}_{W\setminus Y}$ by (2).
If $Y_1\subset Y_2$, the generators of $\oaideal{J}_{W\setminus Y_2}$ are
among those of $\oaideal{J}_{W\setminus Y_1}$.
If $\oaideal{J}_{W\setminus Y_2}\subset\oaideal{J}_{W\setminus Y_1}$ and
$f\in W\setminus Y_2$, then
$\fun{\oaresolvent}{\lambda,f}\in\oaideal{J}_{W\setminus Y_1}$, and the first
statement gives $f\notin Y_1$; hence $Y_1\subset Y_2$.
In this case, the generators of $\oaideal{J}_{Y_2\setminus Y_1}$ are among those
of $\oaideal{J}_{W\setminus Y_1}$.
Conversely, (2) with $Y_1\subset Y_2$ in place of $Y\subset W$ writes
$B\in\oaresolventalgebra(Y_2)\cap\oaideal{J}_{W\setminus Y_1}$ as $B=B_1+C_1$
with $B_1\in\oaresolventalgebra(Y_1)$ and
$C_1\in\oaideal{J}_{Y_2\setminus Y_1}\subset\oaideal{J}_{W\setminus Y_1}$, and
$B_1=B-C_1\in\oaresolventalgebra(Y_1)\cap\oaideal{J}_{W\setminus Y_1}=\setone{0}$
by (2) for $Y_1\subset W$; hence $B\in\oaideal{J}_{Y_2\setminus Y_1}$.
The map of the quotients is well defined and surjective by the inclusion of the
ideals, and its kernel is the image of $\oaideal{J}_{W\setminus Y_1}$; by the
decomposition of (2) for $Y_2$, this image consists of the classes of the
elements of $\oaresolventalgebra(Y_2)\cap\oaideal{J}_{W\setminus Y_1}=\oaideal{J}_{Y_2\setminus Y_1}$.
\end{proof}

\begin{lem}[Resolvents outside a subspace]\label{lem:resolvent-outside-subspace}
Let $S\subset X$ be a real linear subspace, let $f\in X\setminus S$, and let
$\lambda\in\fldreal\setminus\setone{0}$.
Then
\begin{equation}\label{eq:resolvent-outside-subspace-distance}
\begin{aligned}
\norm{\fun{\oaresolvent}{\lambda,f}-A}
&\geq
\frac{1}{2\abs{\lambda}}
&&\rbk{A\in\oaresolventalgebra(S)};
\end{aligned}
\end{equation}
in particular, $\fun{\oaresolvent}{\lambda,f}\notin\oaresolventalgebra(S)$.
\end{lem}

\begin{proof}
By Theorem \ref{thm:subspace-complement-quotient}(2) with $Y=S$ and $W=X$, the
quotient map by $\oaideal{J}_{X\setminus S}$ is nonexpansive, maps
$\fun{\oaresolvent}{\lambda,f}$ to $0$, and is isometric on
$\oaresolventalgebra(S)$.
Hence
$\norm{\fun{\oaresolvent}{\lambda,f}-A}\geq\norm{A+\oaideal{J}_{X\setminus S}}=\norm{A}$,
and
$\norm{\fun{\oaresolvent}{\lambda,f}-A}\geq\abs{\lambda}^{-1}-\norm{A}$.
The larger of the two lower bounds is at least $\frac{1}{2\abs{\lambda}}$.
\end{proof}

\section{Characters and commutative subalgebras}\label{sec:characters}

In this section, the characters of \(\fun{\oaresolventalgebra}{X,\opform{\sigma}}\) are constructed from isotropic subspaces with linear functionals and are classified, the algebra generated by the resolvents of one vector is identified with the continuous functions on \(\gtoneptcpt{\fldreal}\), and the characters of the commutative algebras of isotropic subspaces are described.

For a real linear subspace \(S\subset X\), let \(\Delta_S\) be the set of characters of \(\oaresolventalgebra(S)\) with the weak-\(\ast\) topology.

\begin{lem}[Characters of an isotropic subspace and a linear functional]\label{lem:character-existence}
Let $Z\subset X$ be an isotropic subspace, and let $\ell\colon Z\to\fldreal$ be
linear.
There is exactly one character $\chi_{Z,\ell}\in\Delta_X$ with
\begin{equation}\label{eq:character-values}
\begin{aligned}
\fun{\chi_{Z,\ell}}{\fun{\oaresolvent}{\lambda,f}}
&=
\begin{cases}
\fun{r_{\lambda}}{\fun{\ell}{f}},&f\in Z,\\
\fun{r_{\lambda}}{\infty},&f\in X\setminus Z,
\end{cases}
\end{aligned}
\end{equation}
for $\lambda\in\fldreal\setminus\setone{0}$.
\end{lem}

\begin{proof}
Lemma \ref{lem:zero-extension}(2) with $Y=Z$ and
Lemma \ref{lem:zero-extension}(1) with $\oa{B}=\fldcmp$ give the character,
since $\fun{r_{\lambda}}{\infty}=0$.
\end{proof}

For \(f\in X\setminus\setone{0}\), let \(\oa{C}_f\) be the unital \(\oacstar\)-subalgebra generated by the resolvents \(\fun{\oaresolvent}{\lambda,f}\), \(\lambda\in\fldreal\setminus\setone{0}\). For \(u\in\fldreal\), let \(\chi_{f,u}\in\Delta_X\) be the character \(\chi_{\fldreal f,\ell}\) with \(\fun{\ell}{cf}=cu\), and let \(\chi_{f,\infty}=\chi_{\setone{0},0}\in\Delta_X\).

\begin{lem}[Continuous functions of one field]\label{lem:single-field-algebra}
Let $f\in X\setminus\setone{0}$.
\begin{enumerate}
\item The algebra $\oa{C}_f$ is commutative, and it is the unital
$\oacstar$-algebra generated by the normal element $\fun{\oaresolvent}{1,f}$.
\item There is exactly one $\ast$-isomorphism
$\gamma_f\colon\conti(\gtoneptcpt{\fldreal})\to\oa{C}_f$ with
$\fun{\gamma_f}{r_{\lambda}}=\fun{\oaresolvent}{\lambda,f}$ for every
$\lambda\in\fldreal\setminus\setone{0}$, and
$\fun{\chi_{f,u}}{\fun{\gamma_f}{h}}=\fun{h}{u}$ for
$h\in\conti(\gtoneptcpt{\fldreal})$ and $u\in\gtoneptcpt{\fldreal}$.
\item The spectrum of $\fun{\oaresolvent}{\lambda,f}$ in
$\fun{\oaresolventalgebra}{X,\opform{\sigma}}$ is
$r_{\lambda}(\gtoneptcpt{\fldreal})=C_{\lambda}$.
\end{enumerate}
\end{lem}

\begin{proof}
(1): The resolvents of $f$ commute \cite[Theorem 3.6(i)]{BuchholzGrundling002}.
Relation \eqref{eq:resolvent-relation-identity} with $\mu=1$ gives
$\fun{\oaresolvent}{\lambda,f}\rbk{\idone-\imunit\rbk{1-\lambda}\fun{\oaresolvent}{1,f}}=\fun{\oaresolvent}{1,f}$,
and the function $z\mapsto1-\imunit\rbk{1-\lambda}z$ does not vanish on the
spectrum of $\fun{\oaresolvent}{1,f}$ by (3) for $\lambda=1$, whose proof uses
only $\fun{\oaresolvent}{1,f}$; hence $\fun{\oaresolvent}{\lambda,f}$ is a
continuous function of $\fun{\oaresolvent}{1,f}$.

(3): For $\lambda=1$, relations \eqref{eq:resolvent-relation-adjoint} and
\eqref{eq:resolvent-relation-identity} with $\mu=-1$ give
\begin{equation}\label{eq:single-field-imaginary-part}
\begin{aligned}
\fun{\oaresolvent}{1,f}-\faadj{\fun{\oaresolvent}{1,f}}
&=
-2\imunit\fun{\oaresolvent}{1,f}\faadj{\fun{\oaresolvent}{1,f}},
\end{aligned}
\end{equation}
so every point $z$ of the spectrum of the normal element
$\fun{\oaresolvent}{1,f}$ satisfies $z-\cmpconj{z}=-2\imunit\abs{z}^{2}$, that is,
$z\in C_1$.
Every point $\fun{r_1}{u}$ of $C_1$ is the value of the character $\chi_{f,u}$ at
$\fun{\oaresolvent}{1,f}$ and hence belongs to the spectrum.
The same argument applies to $\fun{\oaresolvent}{\lambda,f}$, with
$\fun{\oaresolvent}{\lambda,f}-\faadj{\fun{\oaresolvent}{\lambda,f}}=-2\imunit\lambda\fun{\oaresolvent}{\lambda,f}\faadj{\fun{\oaresolvent}{\lambda,f}}$;
the spectrum in the $\oacstar$-subalgebra $\oa{C}_f$ is the same.

(2): By (1) and (3), the functional calculus of $\fun{\oaresolvent}{1,f}$ is an
isometric $\ast$-isomorphism of $\fun{\conti}{C_1}$ onto $\oa{C}_f$, and
composition with the homeomorphism $r_1$ gives $\gamma_f$ with
$\fun{\gamma_f}{r_1}=\fun{\oaresolvent}{1,f}$.
The function $r_{\lambda}$ is $r_1\rbk{1-\imunit\rbk{1-\lambda}r_1}^{-1}$, so that
$\fun{\gamma_f}{r_{\lambda}}=\fun{\oaresolvent}{\lambda,f}$ by (1).
For a character $\chi_{f,u}$, the map $h\mapsto\fun{\chi_{f,u}}{\fun{\gamma_f}{h}}$
is a character of $\conti(\gtoneptcpt{\fldreal})$ with the value
$\fun{r_1}{u}$ at $r_1$, which is the evaluation at $u$ because $r_1$ separates
the points.
Uniqueness holds because the functions $r_{\lambda}$ generate
$\conti(\gtoneptcpt{\fldreal})$ by the Stone--Weierstrass theorem.
\end{proof}

In a representation \(\oarepn\) with \(f\in X_{\txtregular,\oarepn}\), \begin{equation}\label{eq:single-field-functional-calculus}
\begin{aligned}
\fun{\oarepn}{\fun{\gamma_f}{h}}
&=
\fun{h}{\fun{\phi_{\oarepn}}{f}}
&&\rbk{h\in\conti(\gtoneptcpt{\fldreal})},
\end{aligned}
\end{equation} where the right-hand side is the functional calculus of the self-adjoint field \(\fun{\phi_{\oarepn}}{f}\) of Section \ref{sec:setting} applied to the restriction of \(h\) to \(\fldreal\); both sides are unital \(\ast\)-homomorphisms that agree at the functions \(r_{\lambda}\). Thus \(\fun{\oarepn}{\oa{C}_f}\) consists of the bounded continuous functions of the field with one common limit at \(+\infty\) and \(-\infty\), and \(\fun{h}{\infty}\) is this limit, the value of \(h\) at large field values. The field itself takes the value \(\infty\) only in representations with \(f\notin X_{\txtregular,\oarepn}\).

\begin{prop}[Classification of the characters]\label{prop:character-classification}
Every character $\chi\in\Delta_X$ is
$\chi_{Z,\ell}$ for exactly one isotropic subspace $Z$ and exactly one linear
$\ell\colon Z\to\fldreal$, namely
\begin{equation}\label{eq:character-data}
\begin{aligned}
Z
&=
\set{f\in X}{\fun{\chi}{\fun{\oaresolvent}{1,f}}\neq0},
\\ %%%%%%%%%%%%%%%%
\fun{\chi}{\fun{\oaresolvent}{\lambda,f}}
&=
\fun{r_{\lambda}}{\fun{\ell}{f}}
\quad\rbk{f\in Z,\ \lambda\in\fldreal\setminus\setone{0}}.
\end{aligned}
\end{equation}
The same holds for the characters of the algebra $\oaresolventalgebra(W)$ of an
isotropic subspace $W$, with $Z\subset W$: every $\chi\in\Delta_W$ is the
restriction of $\chi_{Z,\ell}$ for exactly one subspace $Z\subset W$ and one
linear $\ell\colon Z\to\fldreal$.
\end{prop}

\begin{proof}
For $f\neq0$, the restriction of $\chi$ to $\oa{C}_f$ composed with $\gamma_f$ is a
character of $\conti(\gtoneptcpt{\fldreal})$, the evaluation at one point
$u_f\in\gtoneptcpt{\fldreal}$; hence
$\fun{\chi}{\fun{\oaresolvent}{\lambda,f}}=\fun{r_{\lambda}}{u_f}$ for every
$\lambda$, and $u_f\in\fldreal$ exactly when $f\in Z$.
Put $u_0=0$.
For $c\neq0$, relation \eqref{eq:resolvent-relation-scaling} with $\nu=c$
gives
$c\fun{r_{c\lambda}}{u_{cf}}=\fun{r_{\lambda}}{u_f}=c\fun{r_{c\lambda}}{cu_f}$
under $\chi$
with $c \cdot \infty=\infty$, and hence $u_{cf}=cu_f$, since $r_{c\lambda}$ is
injective.
For $f,g\in Z$, relation \eqref{eq:resolvent-relation-commutator} gives
$0=\imunit\fun{\opform{\sigma}}{f,g}\fun{r_{\lambda}}{u_f}^{2}\fun{r_{\mu}}{u_g}^{2}$
under $\chi$,
so $\fun{\opform{\sigma}}{f,g}=0$.
Relation \eqref{eq:resolvent-relation-sum} under $\chi$ then reads
\begin{equation}\label{eq:character-sum-relation}
\begin{aligned}
\fun{r_{\lambda}}{u_f}\fun{r_{\mu}}{u_g}
&=
\fun{r_{\lambda+\mu}}{u_{f+g}}\rbk{\fun{r_{\lambda}}{u_f}+\fun{r_{\mu}}{u_g}}
\quad\rbk{\lambda+\mu\neq0},
\end{aligned}
\end{equation}
and \eqref{eq:value-function-sum} gives
$\fun{r_{\lambda+\mu}}{u_{f+g}}=\fun{r_{\lambda+\mu}}{u_f+u_g}$, since
$\fun{r_{\lambda}}{u_f}+\fun{r_{\mu}}{u_g}\neq0$; hence $f+g\in Z$ and
$u_{f+g}=u_f+u_g$.
Thus $Z$ is an isotropic subspace, $\fun{\ell}{f}=u_f$ is linear, and $\chi$
and $\chi_{Z,\ell}$ agree on the generators.
The data are determined by $\chi$ through \eqref{eq:character-data}.
For $\chi\in\Delta_W$, the same argument applies to the
vectors of $W$, and $\chi_{Z,\ell}$ of
Lemma \ref{lem:character-existence} restricts to it.
\end{proof}

For an isotropic subspace \(W\), the algebra \(\oaresolventalgebra(W)\) is commutative, since the right-hand side of \eqref{eq:resolvent-relation-commutator} vanishes for \(f,g\in W\) by \eqref{eq:isotropic-subspace}. By Proposition \ref{prop:character-classification}, its spectrum \(\Delta_W\) is in bijection with the set of pairs \(\pairbk{Z,\ell}\) of a subspace \(Z\subset W\) and a linear \(\ell\colon Z\to\fldreal\): a character \(\chi\in\Delta_W\) assigns a real value to the fields of the directions of \(Z\) and the value \(\infty\) to the fields of the other directions.

\section{Probabilistic description of ideals}\label{sec:resolvent-ideals-probabilistic}

In this section, the ideals of the resolvent algebra, which Buchholz describes through subspaces of the symplectic space, are expressed through the laws of single fields. Throughout this section, \(X\) carries a complex structure and a complex inner product with \(\fun{\opform{\sigma}}{f,g}=\opimag\bkt{f}{g}\), the setting of \cite{DetlevBuchholz001}, whose results are used.

\subsection{Field laws and principal ideals}\label{sec:field-laws}

\begin{defn}[Field laws of a state]\label{def:field-law}
Let $\oastate[\psi]$ be a state of $\fun{\oaresolventalgebra}{X,\opform{\sigma}}$
and $f\in X\setminus\setone{0}$.
The field law $\msrprb_{\psi,f}$ of $f$ in $\oastate[\psi]$ is the regular Borel
probability measure on $\gtoneptcpt{\fldreal}$ with
\begin{equation}\label{eq:field-law}
\begin{aligned}
\fun{\oastate[\psi]}{\fun{\gamma_f}{h}}
&=
\sqfun{\prbexp_{\msrprb_{\psi,f}}}{h}
&&\rbk{h\in\conti(\gtoneptcpt{\fldreal})},
\end{aligned}
\end{equation}
which exists and is unique by Lemma \ref{lem:single-field-algebra}(2) and the
Riesz representation theorem.
The number $\fun{\msrprb_{\psi,f}}{\setone{\infty}}$ is the probability that the
field of $f$ is infinite.
\end{defn}

\begin{prop}[Principal ideals and sharp field values]\label{prop:principal-ideal-sharp-value}
Let $\oastate[\psi]$ be a state, $f\in X\setminus\setone{0}$,
$\lambda\in\fldreal\setminus\setone{0}$, and $u\in\gtoneptcpt{\fldreal}$.
\begin{enumerate}
\item The following are equivalent:
$\fun{I}{\lambda,f;\fun{r_{\lambda}}{u}}\subset\Ker\oarepn_{\psi}$;
$\fun{\oastate[\psi]}{\fun{\oaresolvent}{\lambda,f}}=\fun{r_{\lambda}}{u}$;
$\msrprb_{\psi,f}=\msrdirac_{u}$.
\item The points of the spectrum of $\fun{\oaresolvent}{\lambda,f}$ are the
numbers $\fun{r_{\lambda}}{u}$, and every ideal
$\fun{I}{\lambda,f;\fun{r_{\lambda}}{u}}$ is proper.
\item If $\msrprb_{\psi,f}=\msrdirac_{u}$ with $u\in\fldreal$, then
$\fun{\oarepn_{\psi}}{\fun{\oaresolvent}{\mu,g}}=0$ for all
$\mu\in\fldreal\setminus\setone{0}$ and all $g\in X$ with
$\fun{\opform{\sigma}}{f,g}\neq0$.
\end{enumerate}
\end{prop}

\begin{proof}
(1): If
$\fun{\oastate[\psi]}{\fun{\oaresolvent}{\lambda,f}}=\fun{r_{\lambda}}{u}$,
then the expectation of $r_{\lambda}$ under $\msrprb_{\psi,f}$ is a point of the
circle $C_{\lambda}$, which bounds the closed disc containing all values of
$r_{\lambda}$; since every point of a circle is an extreme point of the disc,
$\msrprb_{\psi,f}$ is concentrated on $\fun{\inv{r_{\lambda}}}{\setone{\fun{r_{\lambda}}{u}}}=\setone{u}$.
If $\msrprb_{\psi,f}=\msrdirac_{u}$, then, for the vector
$\oagnsvector[\Psi]$ of
\eqref{eq:gns-vector-state},
$\norm{\fun{\oarepn_{\psi}}{\fun{\gamma_f}{h}-\fun{h}{u}}\oagnsvector[\Psi]}^{2}=\sqfun{\prbexp_{\msrprb_{\psi,f}}}{\abs{h-\fun{h}{u}}^{2}}=0$
for every $h$, so that $\oagnsvector[\Psi]$ is an eigenvector of
$\fun{\oarepn_{\psi}}{\fun{\oaresolvent}{\lambda,f}}$ with the eigenvalue
$\fun{r_{\lambda}}{u}$, and \cite[Lemma 4.1(ii)]{DetlevBuchholz001} gives
$\fun{\oarepn_{\psi}}{\fun{\oaresolvent}{\lambda,f}-\fun{r_{\lambda}}{u}}=0$;
the kernel is a closed two-sided ideal and contains
$\fun{I}{\lambda,f;\fun{r_{\lambda}}{u}}$.
Conversely, this inclusion gives
$\fun{\oastate[\psi]}{\fun{\oaresolvent}{\lambda,f}-\fun{r_{\lambda}}{u}}=0$.

(2): The first statement is Lemma \ref{lem:single-field-algebra}(3), and the
second is \cite[Proposition 4.2(i)]{DetlevBuchholz001}.

(3): By the proof of (1),
$\fun{\oarepn_{\psi}}{\fun{\oaresolvent}{\lambda,f}}$ is the nonzero
scalar $\fun{r_{\lambda}}{u}$, so that $\oarepn_{\psi}$ maps the left-hand side of
\eqref{eq:resolvent-relation-commutator} to $0$, and the right-hand side gives
\begin{equation}\label{eq:principal-ideal-commutator}
\begin{aligned}
\fun{\opform{\sigma}}{f,g}\fun{r_{\lambda}}{u}^{2}\fun{\oarepn_{\psi}}{\fun{\oaresolvent}{\mu,g}}^{2}
&=
0;
\end{aligned}
\end{equation}
the normal operator $\fun{\oarepn_{\psi}}{\fun{\oaresolvent}{\mu,g}}$ with
vanishing square is $0$.
\end{proof}

By Proposition \ref{prop:principal-ideal-sharp-value}, the principal ideal \(\fun{I}{\lambda,f;0}\), which is the ideal generated by \(\fun{\oaresolvent}{\lambda,f}\), is annihilated exactly by the states in which the field of \(f\) is infinite with probability one, and the principal ideals \(\fun{I}{\lambda,f;\fun{r_{\lambda}}{u}}\) with \(u\in\fldreal\) exactly by the states in which the field of \(f\) has the sharp value \(u\).

\subsection{Factor states and primitive ideals}\label{sec:factor-primitive}

\begin{prop}[Field laws in factor states]\label{prop:factor-field-laws}
Let $\oastate[\psi]$ be a factor state and $f\in X\setminus\setone{0}$.
Exactly one of the following holds:
$f\in X_{\txtsingular,\oarepn_{\psi}}$ and $\msrprb_{\psi,f}=\msrdirac_{\infty}$;
$f\in X_{\mathrm{t},\oarepn_{\psi}}$ and $\msrprb_{\psi,f}=\msrdirac_{u}$ for some
$u\in\fldreal$;
$f\in X_{\txtregular,\oarepn_{\psi}}\setminus X_{\mathrm{t},\oarepn_{\psi}}$,
$\fun{\msrprb_{\psi,f}}{\setone{\infty}}=0$, and $\msrprb_{\psi,f}$ is not a Dirac
measure.
In particular, $\fun{\msrprb_{\psi,f}}{\setone{\infty}}\in\setone{0,1}$.
\end{prop}

\begin{proof}
For the factor representation $\oarepn_{\psi}$,
\cite[Proposition 4.7(iii)]{BuchholzGrundling002} gives
$\fun{\oarepn_{\psi}}{\fun{\oaresolvent}{\lambda,f}}=0$ for
$f\in X_{\txtsingular,\oarepn_{\psi}}$ and
$\fun{\oarepn_{\psi}}{\fun{\oaresolvent}{\lambda,f}}\in\fldcmp\setminus\setone{0}$
for $f\in X_{\mathrm{t},\oarepn_{\psi}}$.
In the first case, $\fun{\oastate[\psi]}{\fun{\oaresolvent}{1,f}}=0=\fun{r_1}{\infty}$,
and Proposition \ref{prop:principal-ideal-sharp-value}(1) gives
$\msrdirac_{\infty}$.
In the second case, the scalar $z=\fun{\oarepn_{\psi}}{\fun{\oaresolvent}{1,f}}$
lies in the spectrum of $\fun{\oaresolvent}{1,f}$, since
$\fun{\oarepn_{\psi}}{\fun{\oaresolvent}{1,f}-z}=0$ and the image of an
invertible element under the unital $\ast$-homomorphism $\oarepn_{\psi}$ is
invertible; hence
$z=\fun{r_1}{u}$ with $u\in\fldreal$ by Lemma \ref{lem:single-field-algebra}(3),
since $z\neq0$, and $\fun{\oastate[\psi]}{\fun{\oaresolvent}{1,f}}=z$ gives
$\msrdirac_{u}$.
In the third case, $\imunit k\fun{\oarepn_{\psi}}{\fun{\oaresolvent}{k,f}}$ tends
strongly to the identity as $k\to\infty$
\cite[Theorem 4.2(ii)]{BuchholzGrundling002}, and the functions
$\imunit kr_{k}$, bounded by $1$, converge pointwise to $\fndef{\fldreal}$ on
$\gtoneptcpt{\fldreal}$; dominated convergence gives
$\fun{\msrprb_{\psi,f}}{\fldreal}=1$.
A Dirac measure $\msrdirac_{u}$ with $u\in\fldreal$ is excluded in this case:
by the proof of Proposition \ref{prop:principal-ideal-sharp-value}(1), it gives
$\fun{\oarepn_{\psi}}{\fun{\oaresolvent}{1,f}}=\fun{r_1}{u}\neq0$, a scalar with
a bounded inverse, and hence $f\in X_{\mathrm{t},\oarepn_{\psi}}$.
\end{proof}

For a state \(\oastate[\psi]\), let \begin{equation}\label{eq:infinite-sharp-directions}
\begin{aligned}
X_{\infty,\psi}
&=
\set{f\in X\setminus\setone{0}}{\msrprb_{\psi,f}=\msrdirac_{\infty}},
\\ %%%%%%%%%%%%%%%%
X_{\mathrm{sh},\psi}
&=
\set{f\in X\setminus\setone{0}}{\msrprb_{\psi,f}=\msrdirac_{u_f}\ \text{for some}\ u_f\in\fldreal}.
\end{aligned}
\end{equation} The subscript \(\infty\) marks the directions whose field is infinite with probability one, and \(\mathrm{sh}\) the directions whose field has the sharp value \(u_f\), that is, a deterministic real value, in the sense of Proposition \ref{prop:principal-ideal-sharp-value}(1). The number \(u_f\) is determined by \(f\), since the field law determines it.

\begin{prop}[Primitive ideals and degenerate field laws]\label{prop:primitive-ideal-field-laws}
Let $X$ be finite dimensional, and let $\oastate[\psi]$ and $\oastate[\psi']$ be
pure states, with the numbers $u_f$ and $u'_f$ of
\eqref{eq:infinite-sharp-directions}.
\begin{enumerate}
\item The primitive ideal $\Ker\oarepn_{\psi}$ is the closed two-sided ideal
generated by the resolvents $\fun{\oaresolvent}{\lambda,f}$ with
$f\in X_{\infty,\psi}$ and by the elements
$\fun{\oaresolvent}{\mu,g}-\fun{r_{\mu}}{u_g}$ with $g\in X_{\mathrm{sh},\psi}$,
where $\lambda,\mu\in\fldreal\setminus\setone{0}$.
\item $\Ker\oarepn_{\psi}=\Ker\oarepn_{\psi'}$ if and only if
$X_{\infty,\psi}=X_{\infty,\psi'}$, $X_{\mathrm{sh},\psi}=X_{\mathrm{sh},\psi'}$, and
$u_g=u'_g$ for $g\in X_{\mathrm{sh},\psi}$.
\end{enumerate}
\end{prop}

\begin{proof}
By Proposition \ref{prop:factor-field-laws} for the irreducible representation
$\oarepn_{\psi}$, $X_{\infty,\psi}$ is the singular part, the complement of the
regular part, and $X_{\mathrm{sh},\psi}\cup\setone{0}$ is the
trivially regular part, on whose resolvents $\oarepn_{\psi}$ takes the values
$\fun{r_{\mu}}{u_g}$.

(1): This is \cite[Proposition 3.5(i)]{DetlevBuchholz001}, whose generating set
consists of the resolvents of the singular part and of the shifted resolvents of
the trivially regular part with the values of the character of
\cite[Definition 3.2]{DetlevBuchholz001}.

(2): Equal data give equal generating sets in (1).
If $f\in X_{\infty,\psi}\setminus X_{\infty,\psi'}$, then
$\fun{\oaresolvent}{\lambda,f}$ lies in $\Ker\oarepn_{\psi}$ and, by
Proposition \ref{prop:principal-ideal-sharp-value}(1), not in
$\Ker\oarepn_{\psi'}$; if $g\in X_{\mathrm{sh},\psi}$ and
$\msrprb_{\psi',g}\neq\msrdirac_{u_g}$, then
$\fun{\oaresolvent}{\mu,g}-\fun{r_{\mu}}{u_g}$ lies in
$\Ker\oarepn_{\psi}$ and not in $\Ker\oarepn_{\psi'}$; the same holds with
the two states exchanged.
\end{proof}

By Proposition \ref{prop:primitive-ideal-field-laws}, the data that determine a primitive ideal of a finite-dimensional resolvent algebra are the directions in which the fields are infinite, the directions in which they are sharp, and the sharp values; the directions with nondegenerate laws do not enter.

\subsection{Reduction of the algebra by infinite field values}\label{sec:infinite-field-reduction}

In the van Hove model with an infrared cutoff, the ground states converge weakly to zero in the Fock space as the cutoff is removed if and only if the coupling function is infrared singular \cite[Section 13.9 and Theorem 13.10]{AsaoArai026}. On the resolvent algebra, this divergence is described by the states: the fields of some directions become infinite, and every weak-\(\ast\) accumulation state annihilates the ideal of these directions.

\begin{prop}[Reduction by infinite field values]\label{prop:infinite-field-reduction}
Let $\net{\oastate[\psi_{\iota}]}{\iota\in I}$ be a net of states, let
$\oastate[\psi]$ be a weak-$\ast$ accumulation state of the net, let
\begin{equation}\label{eq:infinite-field-directions}
\begin{aligned}
X_{\infty}
&=
\set{f\in X\setminus\setone{0}}{\lim_{\iota}\sqfun{\prbexp_{\msrprb_{\psi_{\iota},f}}}{h}=\fun{h}{\infty}\ \text{for every}\ h\in\conti(\gtoneptcpt{\fldreal})},
\end{aligned}
\end{equation}
and let $\oaideal{J}_{\infty}$ be the closed two-sided ideal generated by the
resolvents $\fun{\oaresolvent}{\lambda,f}$ with
$\lambda\in\fldreal\setminus\setone{0}$ and $f\in X_{\infty}$.
\begin{enumerate}
\item $\msrprb_{\psi,f}=\msrdirac_{\infty}$ for $f\in X_{\infty}$, and
$\oaideal{J}_{\infty}\subset\Ker\oarepn_{\psi}$; hence $\oarepn_{\psi}$ and
$\oastate[\psi]$ factor through the quotient by $\oaideal{J}_{\infty}$.
\item For $g\in X\setminus\setone{0}$,
$\fun{\oaresolvent}{\lambda,g}\in\Ker\oarepn_{\psi}$ if and only if
$\msrprb_{\psi,g}=\msrdirac_{\infty}$; if $\fun{\msrprb_{\psi,g}}{\fldreal}>0$, then
$\fun{\oaresolvent}{\lambda,g}\notin\oaideal{J}_{\infty}$; and
$\idone\notin\oaideal{J}_{\infty}$.
\end{enumerate}
\end{prop}

\begin{proof}
(1): For $f\in X_{\infty}$ and $h\in\conti(\gtoneptcpt{\fldreal})$,
$\fun{\oastate[\psi]}{\fun{\gamma_f}{h}}$ is an accumulation point of the
convergent net $\sqfun{\prbexp_{\msrprb_{\psi_{\iota},f}}}{h}$, hence equal to
$\fun{h}{\infty}$, and $\msrprb_{\psi,f}=\msrdirac_{\infty}$.
Proposition \ref{prop:principal-ideal-sharp-value}(1) with $u=\infty$ gives
$\fun{\oaresolvent}{\lambda,f}\in\Ker\oarepn_{\psi}$, and the kernel, a closed
two-sided ideal, contains the ideal $\oaideal{J}_{\infty}$ generated by these
elements.

(2): The first statement is
Proposition \ref{prop:principal-ideal-sharp-value}(1) with $u=\infty$.
If $\fun{\msrprb_{\psi,g}}{\fldreal}>0$, then
$\msrprb_{\psi,g}\neq\msrdirac_{\infty}$, so that
$\fun{\oaresolvent}{\lambda,g}\notin\Ker\oarepn_{\psi}$ and, by (1),
$\fun{\oaresolvent}{\lambda,g}\notin\oaideal{J}_{\infty}$.
The state $\oastate[\psi]$ vanishes on $\Ker\oarepn_{\psi}$ by
\eqref{eq:gns-vector-state} and takes the value $1$ at $\idone$, so that
$\idone\notin\Ker\oarepn_{\psi}\supset\oaideal{J}_{\infty}$.
\end{proof}

For the Fock vacuum state of the regular Fock representation with shifted fields, the directions with diverging shifts generate the ideal of Proposition \ref{prop:infinite-field-reduction}. Let \(\sphilbfrak{h}\) be a complex Hilbert space, regarded as a real symplectic space with \(\fun{\opform{\sigma}}{f,g}=\opimag\bkt{f}{g}\), let \(\oarepn_{\txtfock}\) be the representation with \(\fun{\oarepn_{\txtfock}}{\fun{\oaresolvent}{\lambda,f}}=\rbk{\imunit\lambda-\fun{\opfocksegal_{\txtfock}}{f}}^{-1}\) for the Segal field \(\opfocksegal_{\txtfock}\) on the bosonic Fock space, and let \(\oastate[\psi_{\txtfock,\txtvac}]\) be the Fock vacuum state, the vector state of the Fock vacuum.

\begin{cor}[Fock vacuum states with diverging shifts]\label{cor:shifted-vacuum-reduction}
Let $\net{\ell_{\iota}}{\iota\in I}$ be a net of real-linear maps
$\ell_{\iota}\colon\sphilbfrak{h}\to\fldreal$, let
$\oastate[\psi_{\iota}]=\oastate[\psi_{\txtfock,\txtvac}]\circ\oarepn_{\txtfock}\circ\Theta_{\ell_{\iota}}$,
and let $\oastate[\psi]$ be a weak-$\ast$ accumulation state of the net.
Let $\sphilbfrak{t}$ be the set of $f\in\sphilbfrak{h}$ with
$\lim_{\iota}\abs{\fun{\ell_{\iota}}{f}}=\infty$, let $\sphilbfrak{d}$ be the
set of $f$ for
which $\fun{\ell_{\iota}}{f}$ converges in $\fldreal$, and let
$\fun{\ell_{\sphilbfrak{d}}}{f}=\lim_{\iota}\fun{\ell_{\iota}}{f}$ for
$f\in\sphilbfrak{d}$.
\begin{enumerate}
\item $\sphilbfrak{d}$ is a real linear subspace, $\ell_{\sphilbfrak{d}}$ is real linear,
$\sphilbfrak{t}\subset X_{\infty}$, and the ideal generated by the resolvents of
the vectors
of $\sphilbfrak{t}$ is contained in $\Ker\oarepn_{\psi}$.
\item For $f\in\sphilbfrak{d}\setminus\setone{0}$, $\msrprb_{\psi,f}$ is the Gaussian law
with mean $\fun{\ell_{\sphilbfrak{d}}}{f}$ and variance
$\frac{1}{2}\norm{f}^{2}$; it has no
mass at $\infty$ and is not a Dirac measure.
\item For a real linear subspace $\sphilbfrak{d}_{0}\subset\sphilbfrak{d}$ and a real-linear
$\ell\colon\sphilbfrak{h}\to\fldreal$ with $\ell=\ell_{\sphilbfrak{d}}$ on
$\sphilbfrak{d}_{0}$, the states
$\oastate[\psi]$ and
$\oastate[\psi_{\txtfock,\txtvac}]\circ\oarepn_{\txtfock}\circ\Theta_{\ell}$
coincide on $\oaresolventalgebra(\sphilbfrak{d}_{0})$.
\item If $\sphilbfrak{d}=\sphilbfrak{h}$, the net converges in the weak-$\ast$ topology to
$\oastate[\psi_{\txtfock,\txtvac}]\circ\oarepn_{\txtfock}\circ\Theta_{\ell_{\sphilbfrak{d}}}$.
\end{enumerate}
\end{cor}

\begin{proof}
All items rest on the field laws of the states of the net.
In the Fock vacuum state, the field of $f$ has the Gaussian law with mean $0$
and variance $\frac{1}{2}\norm{f}^{2}$, and $\Theta_{\ell_{\iota}}$ adds
$\fun{\ell_{\iota}}{f}$ to the field, so that $\msrprb_{\psi_{\iota},f}$ is the
Gaussian law with mean $\fun{\ell_{\iota}}{f}$ and the same variance.

(1): Limits of convergent nets are additive and homogeneous, so that
$\sphilbfrak{d}$ is a real linear subspace and $\ell_{\sphilbfrak{d}}$ is real
linear.
For $f\in\sphilbfrak{t}$, the shifted laws converge to $\msrdirac_{\infty}$
against every $h\in\conti(\gtoneptcpt{\fldreal})$, which gives
$\sphilbfrak{t}\subset X_{\infty}$, and
Proposition \ref{prop:infinite-field-reduction}(1) gives the inclusion of the
ideal.

(2): For $f\in\sphilbfrak{d}\setminus\setone{0}$, the shifted laws converge
weakly to the Gaussian law with mean $\fun{\ell_{\sphilbfrak{d}}}{f}$ and
variance $\frac{1}{2}\norm{f}^{2}$, which is
$\msrprb_{\psi,f}$ as in the proof of
Proposition \ref{prop:infinite-field-reduction}(1).

(3): For $\lambda\in\fldreal\setminus\setone{0}$ and
$f\in\sphilbfrak{d}_{0}$,
$$\begin{aligned}
\fun{\oastate[\psi_{\iota}]}{\fun{\oaresolvent}{\lambda,f}}
&=
\fun{\oastate[\psi_{\txtfock,\txtvac}]}{\fun{\oarepn_{\txtfock}}{\fun{\oaresolvent}{\lambda+\imunit\fun{\ell_{\iota}}{f},f}}},
\end{aligned}$$
which converges to the value with $\fun{\ell_{\sphilbfrak{d}}}{f}$ by the norm
continuity of $z\mapsto\fun{\oaresolvent}{z,f}$; the accumulation state
$\oastate[\psi]$ takes the limit value, and, since the resolvents of the vectors
of $\sphilbfrak{d}_{0}$ span a dense subspace of
$\oaresolventalgebra(\sphilbfrak{d}_{0})$ (Section \ref{sec:setting}), this
gives (3).

(4): By (3) for $\sphilbfrak{d}_{0}=\sphilbfrak{h}$, every weak-$\ast$
accumulation state of the net is
$\oastate[\psi_{\txtfock,\txtvac}]\circ\oarepn_{\txtfock}\circ\Theta_{\ell_{\sphilbfrak{d}}}$;
by weak-$\ast$ compactness, the net converges to it.
\end{proof}

\subsection{Characters and the commutator ideal}\label{sec:partial-configurations}

Let \(\Delta_X\) be the set of characters of Section \ref{sec:characters}, and let \(\oaideal{J}_{\mathrm{c}}\) be the commutator ideal \begin{equation}\label{eq:commutator-ideal}
\begin{aligned}
\oaideal{J}_{\mathrm{c}}
&=
\gtclos{\splinspan_{\fldcmp}\set{A\commutator{B}{C}D}{A,B,C,D\in\fun{\oaresolventalgebra}{X,\opform{\sigma}}}}^{\norm{\mathord{\cdot}}}.
\end{aligned}
\end{equation} By Proposition \ref{prop:character-classification}, a character \(\chi\in\Delta_X\) assigns a real value to the fields of an isotropic subspace and the value \(\infty\) to the others. The states of interest in Proposition \ref{prop:partial-configurations}(3) are the mixtures of characters \begin{equation}\label{eq:classical-random-configuration}
\begin{aligned}
\fun{\oastate[\psi]}{A}
&=
\sqfun{\prbexp_{\msrprb_{\psi}(\chi)}}{\fun{\chi}{A}}
&&\rbk{A\in\fun{\oaresolventalgebra}{X,\opform{\sigma}}}
\end{aligned}
\end{equation} with a Radon probability measure \(\msrprb_{\psi}\) on \(\Delta_X\).

\begin{prop}[Characters and the commutator ideal]\label{prop:partial-configurations}
\begin{enumerate}
\item The space $\Delta_X$ is a nonempty compact Hausdorff space.
The map $A\mapsto\widehat{A}$ with $\fun{\widehat{A}}{\chi}=\fun{\chi}{A}$ is a
$\ast$-homomorphism of $\fun{\oaresolventalgebra}{X,\opform{\sigma}}$ onto
$\fun{\conti}{\Delta_X}$, its kernel is $\oaideal{J}_{\mathrm{c}}$, and
$\idone\notin\oaideal{J}_{\mathrm{c}}$.
\item A state belongs to $\Delta_X$ if and only if all its field laws are
Dirac measures.
\item A state vanishes on $\oaideal{J}_{\mathrm{c}}$ if and only if it is of the form
\eqref{eq:classical-random-configuration}, and the measure is then unique.
\end{enumerate}
\end{prop}

\begin{proof}
(1): $\chi_{\setone{0},0}\in\Delta_X$, and $\Delta_X$ is weak-$\ast$ closed in the
unit ball of the dual.
The elements of $\Delta_X$ vanish on $\oaideal{J}_{\mathrm{c}}$ and are the
characters of the commutative quotient by $\oaideal{J}_{\mathrm{c}}$, so the
Gelfand transform of
the quotient gives the $\ast$-homomorphism onto $\fun{\conti}{\Delta_X}$, whose
kernel is $\oaideal{J}_{\mathrm{c}}$.

(2): The field laws of a character $\chi\in\Delta_X$ are Dirac measures, since
$h\mapsto\fun{\chi}{\fun{\gamma_f}{h}}$ is a character of
$\conti(\gtoneptcpt{\fldreal})$.
Conversely, if all field laws are Dirac measures, the proof of
Proposition \ref{prop:principal-ideal-sharp-value}(1) shows that every resolvent
is represented by a scalar in $\oarepn_{\psi}$, hence every element is, and
$\oastate[\psi]$ is multiplicative.

(3): A state vanishing on $\oaideal{J}_{\mathrm{c}}$ is a state of
$\fun{\conti}{\Delta_X}$ by (1), and the Riesz representation theorem gives the
unique measure; the converse holds because characters vanish on commutators.
\end{proof}

For a character \(\chi_{Z,\ell}\in\Delta_X\) and \(f\in X\setminus\setone{0}\), the field law of \(f\) is \(\msrdirac_{\fun{\ell}{f}}\) for \(f\in Z\) and \(\msrdirac_{\infty}\) for \(f\notin Z\).

\section{States of commuting random variables}\label{sec:random-variables}

In this section, the state given by the expectations of a family of commuting random variables on an isotropic subspace is constructed, its Gelfand--Naimark--Segal representation is identified with the multiplication representation on the \(\lp^{2}\) space of the \(\sigma\)-algebra of the fields, and the quotient by \(\oaideal{J}_{W\setminus Y}\) is identified with the restriction to a sub-\(\sigma\)-algebra.

Let \(W\subset X\) be isotropic, let \(\pairbk{\Omega,\mblfmlfrak{S},\msrprb}\) be a probability space, and let \(\xi=\fml{\xi(f)}{f\in W}\) be a family of real random variables with \begin{equation}\label{eq:random-variable-linearity}
\begin{aligned}
\xi(f+g)
=
\xi(f)+\xi(g),
\quad
\xi(cf)
=
c\, \xi(f)
\quad
\msras{\msrprb}
\end{aligned}
\end{equation} for all \(f,g\in W\) and \(c\in\fldreal\). For a set \(F\subset W\), let \(\mblfmlfrak{S}_{\xi}(F)\) be the \(\sigma\)-algebra generated by the random variables \(\xi(f)\), \(f\in F\). For a real linear subspace \(Y\subset W\), let \begin{equation}\label{eq:infinite-outside}
\begin{aligned}
\xi_Y(f)
&=
\begin{cases}
\xi(f),&f\in Y,\\
\infty,&f\in W\setminus Y,
\end{cases}
\end{aligned}
\end{equation} with values in \(\gtoneptcpt{\fldreal}\).

\begin{thm}[States of commuting random variables and the ideals of subspaces]\label{thm:random-variable-ideals}
In the setting of Section \ref{sec:random-variables}, with $\xi$ satisfying
\eqref{eq:random-variable-linearity} and $\xi_Y$ defined by
\eqref{eq:infinite-outside}, the following hold.

\begin{enumerate}
\item There is exactly one state $\oastate[\psi_{\xi}]$ of
$\oaresolventalgebra(W)$ with
\begin{equation}\label{eq:random-variable-state}
\begin{aligned}
\fun{\oastate[\psi_{\xi}]}{\fun{\oaresolvent}{\lambda_1,f_1}\cdots\fun{\oaresolvent}{\lambda_n,f_n}}
&=
\sqfun{\prbexp_{\msrprb}}{\prod_{j=1}^{n}\fun{r_{\lambda_j}}{\xi(f_j)}}
\end{aligned}
\end{equation}
for $n\in\semigrposint$, $\lambda_1,\ldots,\lambda_n\in\fldreal\setminus\setone{0}$,
and $f_1,\ldots,f_n\in W$.
\item For $F\subset W$, $\mblfmlfrak{S}_{\xi}(F)$ is generated by the random
variables $\fun{r_{\lambda}}{\xi(f)}$ with $\lambda\in\fldreal\setminus\setone{0}$
and $f\in F$, the random variables of the generators of
$\oaresolventalgebra(F)$ in \eqref{eq:random-variable-state} when $F$ is a real
linear subspace.
\item For a real linear subspace $Y\subset W$, there is exactly one state
$\oastate[\psi_{\xi,Y}]$ of $\oaresolventalgebra(W)$ that agrees with
$\oastate[\psi_{\xi}]$ on $\oaresolventalgebra(Y)$ and vanishes on
$\oaideal{J}_{W\setminus Y}$, and
\begin{equation}\label{eq:random-variable-quotient-state}
\begin{aligned}
\fun{\oastate[\psi_{\xi,Y}]}{\fun{\oaresolvent}{\lambda_1,f_1}\cdots\fun{\oaresolvent}{\lambda_n,f_n}}
&=
\sqfun{\prbexp_{\msrprb}}
{\prod_{j=1}^{n}\fun{r_{\lambda_j}}{\fun{\xi_Y}{f_j}}}
\end{aligned}
\end{equation}
for $\lambda_j$ and $f_j$ as in (1); the integrand on the right-hand side is
$\mblfmlfrak{S}_{\xi}(Y)$-measurable.
\item For $\lambda\in\fldreal\setminus\setone{0}$ and $f\in W$,
$\fun{\oastate[\psi_{\xi}]}{\fun{\oaresolvent}{\lambda,f}}\neq0$.
Hence $\oastate[\psi_{\xi}]$ vanishes on $\oaideal{J}_{W\setminus Y}$ if and
only if $Y=W$.
\end{enumerate}
\end{thm}

\begin{proof}
(1): The resolvents span a dense subspace of $\oaresolventalgebra(W)$
(Section \ref{sec:setting}), so a state is determined by the values
\eqref{eq:random-variable-state} with $n=1$, which gives the uniqueness.
In the commutative $\oacstar$-algebra
$\fun{\lp^{\infty}}{\Omega,\mblfmlfrak{S},\msrprb}$, the classes of
$\fun{r_{\lambda}}{\xi(f)}$, $f\in W$, satisfy the relations for $f,g\in W$,
since outside the null sets of \eqref{eq:random-variable-linearity} for $\nu f$
and $f+g$ they reduce, as in Lemma \ref{lem:zero-extension}(2), to
\eqref{eq:value-function-relations} and \eqref{eq:value-function-sum} with
$a=\fun{\xi(f)}{\omega}$ and $b=\fun{\xi(g)}{\omega}$.
Lemma \ref{lem:zero-extension}(1) with $Y=W$ gives a $\ast$-homomorphism
$\oarepn[\rho]_{\xi}\colon\fun{\oaresolventalgebra}{X,\opform{\sigma}}\to\fun{\lp^{\infty}}{\Omega,\mblfmlfrak{S},\msrprb}$,
and $A\mapsto\sqfun{\prbexp_{\msrprb}}{\fun{\oarepn[\rho]_{\xi}}{A}}$ is positive, since
$\fun{\oarepn[\rho]_{\xi}}{\faadj{A}A}=\abs{\fun{\oarepn[\rho]_{\xi}}{A}}^{2}$, and takes the value
$1$ at $\idone$.
Its restriction to $\oaresolventalgebra(W)$ satisfies
\eqref{eq:random-variable-state}.

(2): By \eqref{eq:resolvent-value-function}, $\xi(f)$ and
$\fun{r_{\lambda}}{\xi(f)}$ are Borel functions of each other.

(3): The family $\fml{\xi(f)}{f\in Y}$ on the isotropic subspace $Y$ gives,
by the construction of (1), the $\ast$-homomorphism $\oarepn[\rho]_{\xi,Y}$ with
$\fun{\oarepn[\rho]_{\xi,Y}}{\fun{\oaresolvent}{\lambda,f}}$ the class of
$\fun{r_{\lambda}}{\xi(f)}$ for $f\in Y$ and $0$ for $f\notin Y$, that is, the
class of $\fun{r_{\lambda}}{\fun{\xi_Y}{f}}$ for $f\in W$.
Acting by multiplication on $\fun{\lp^{2}}{\Omega,\mblfmlfrak{S},\msrprb}$, it is
the representation of Theorem \ref{thm:subspace-complement-quotient}(1) for the
restriction of the multiplication representation of $\oarepn[\rho]_{\xi}$ to
$\oaresolventalgebra(Y)$, since the two agree on the generators.
It vanishes on $\oaideal{J}_{W\setminus Y}$, and so does the state
$A\mapsto\sqfun{\prbexp_{\msrprb}}{\fun{\oarepn[\rho]_{\xi,Y}}{A}}$, which
agrees with $\oastate[\psi_{\xi}]$ on $\oaresolventalgebra(Y)$.
Theorem \ref{thm:subspace-complement-quotient}(3) shows that it is the only
such state $\oastate[\psi_{\xi,Y}]$.
Since $\fun{r_{\lambda}}{\infty}=0$, the value of $\oarepn[\rho]_{\xi,Y}$ at a
product of resolvents of vectors of $W$ is the class of the integrand of
\eqref{eq:random-variable-quotient-state}, which is $0$ or a product of the
random variables $\fun{r_{\lambda_j}}{\xi(f_j)}$ with $f_j\in Y$.
This gives \eqref{eq:random-variable-quotient-state} and the measurability.

(4): $\opimag\fun{r_{\lambda}}{u}=-\lambda\rbk{\lambda^{2}+u^{2}}^{-1}$ has the
sign of $-\lambda$ for $u\in\fldreal$, so
$\opimag\fun{\oastate[\psi_{\xi}]}{\fun{\oaresolvent}{\lambda,f}}\neq0$.
If $Y\neq W$, a vector $f\in W\setminus Y$ gives a generator of
$\oaideal{J}_{W\setminus Y}$ at which the state does not vanish; if $Y=W$, the
ideal is $\setone{0}$.
\end{proof}

By Theorem \ref{thm:random-variable-ideals}(3) and (4), on the side of the random variables, the passage to the quotient by \(\oaideal{J}_{W\setminus Y}\) with the state that agrees on \(\oaresolventalgebra(Y)\) is the restriction to the sub-\(\sigma\)-algebra \(\mblfmlfrak{S}_{\xi}(Y)\) together with the value \(\infty\) for the random variables of the vectors outside \(Y\); a state given by real-valued random variables, which are finite everywhere, vanishes on no nonzero ideal of this form.

The \(\ast\)-homomorphism \(\oarepn[\rho]_{\xi}\) of the proof of Theorem \ref{thm:random-variable-ideals}(1), used in Theorem \ref{thm:random-variable-gns}, maps \(\fun{\oaresolvent}{\lambda,f}\) to the class of \(\fun{r_{\lambda}}{\xi(f)}\) for \(f\in W\). The next theorem identifies the Gelfand--Naimark--Segal representation of \(\oastate[\psi_{\xi}]\), and with it the correspondence between resolvent subalgebras and sub-\(\sigma\)-algebras on the level of von Neumann algebras.

\begin{thm}[Gelfand--Naimark--Segal representation of a state of commuting random variables]\label{thm:random-variable-gns}
In the setting of Theorem \ref{thm:random-variable-ideals}, let
$\sphilbfrak{H}_{\xi}=\fun{\lp^{2}}{\Omega,\mblfmlfrak{S}_{\xi}(W),\msrprb}$, and let
$\fun{\oarepn_{\xi}}{A}$ be the multiplication by $\fun{\oarepn[\rho]_{\xi}}{A}$ on
$\sphilbfrak{H}_{\xi}$ for $A\in\oaresolventalgebra(W)$.
\begin{enumerate}
\item With the constant function $\oagnsvector[\Psi_{\xi}]=1$, the triple
$\pairbk{\sphilbfrak{H}_{\xi},\oarepn_{\xi},\oagnsvector[\Psi_{\xi}]}$ is a
Gelfand--Naimark--Segal triple of $\oastate[\psi_{\xi}]$, and
$\Ker\oarepn_{\xi}$ is the set of $A\in\oaresolventalgebra(W)$ with
$\fun{\oarepn[\rho]_{\xi}}{A}=0$ $\msrprb$-almost everywhere.
\item For every real linear subspace $Y\subset W$, the von Neumann algebra
generated by $\fun{\oarepn_{\xi}}{\oaresolventalgebra(Y)}$ is the algebra of the
multiplications by the bounded $\mblfmlfrak{S}_{\xi}(Y)$-measurable functions;
for $Y=W$, it is maximal abelian.
\item For $f\in W\setminus\setone{0}$, the field law $\msrprb_{\psi_{\xi},f}$ is
the law of $\xi(f)$, regarded as a probability measure on
$\gtoneptcpt{\fldreal}$ without mass at $\infty$.
\end{enumerate}
\end{thm}

\begin{proof}
(2): The proof of the implication from (1) to (3) in
\cite[Lemma I.5, p.~14]{BarrySimon004} applies: the projections of the von
Neumann algebra generated by $\fun{\oarepn_{\xi}}{\oaresolventalgebra(Y)}$ are
multiplications by indicator functions, and their events form a
$\sigma$-algebra with respect to which the fields of $Y$ are measurable, hence
containing $\mblfmlfrak{S}_{\xi}(Y)$.
These events belong to $\mblfmlfrak{S}_{\xi}(Y)$, since the von Neumann algebra
is the strong closure of multiplications by $\mblfmlfrak{S}_{\xi}(Y)$-measurable
functions, by Theorem \ref{thm:random-variable-ideals}(2), and
$\fun{\lp^{2}}{\Omega,\mblfmlfrak{S}_{\xi}(Y),\msrprb}$ is closed in
$\sphilbfrak{H}_{\xi}$.
For $Y=W$, the von Neumann algebra is therefore that of all multiplications by
bounded $\mblfmlfrak{S}_{\xi}(W)$-measurable functions, which is maximal
abelian \cite[Lemma I.5(3), p.~14]{BarrySimon004}.

(1): By (2) with $Y=W$, the multiplication by $\fndef{B}$ for
$B\in\mblfmlfrak{S}_{\xi}(W)$ belongs to the strong closure of
$\fun{\oarepn_{\xi}}{\oaresolventalgebra(W)}$, so that
$\fndef{B}=\fndef{B}\oagnsvector[\Psi_{\xi}]$ lies in the closure of
$\fun{\oarepn_{\xi}}{\oaresolventalgebra(W)}\oagnsvector[\Psi_{\xi}]$.
The linear combinations of these indicator functions are dense in
$\sphilbfrak{H}_{\xi}$, and $\oagnsvector[\Psi_{\xi}]$ is cyclic.
By the construction in the proof of
Theorem \ref{thm:random-variable-ideals}(1),
$\bkt{\oagnsvector[\Psi_{\xi}]}{\fun{\oarepn_{\xi}}{A}\oagnsvector[\Psi_{\xi}]}=\sqfun{\prbexp_{\msrprb}}{\fun{\oarepn[\rho]_{\xi}}{A}}=\fun{\oastate[\psi_{\xi}]}{A}$,
and the uniqueness of the Gelfand--Naimark--Segal representation up to unitary
equivalence \cite[Theorem 2.3.16]{BratteliRobinson003} gives the first
statement.
A multiplication operator vanishes exactly when its function vanishes almost
everywhere, since it maps $\oagnsvector[\Psi_{\xi}]=1$ to the function; this
gives the kernel.

(3): By \eqref{eq:random-variable-state},
$\fun{\oastate[\psi_{\xi}]}{\fun{\gamma_f}{h}}=\sqfun{\prbexp_{\msrprb}}{\fun{h}{\xi(f)}}$
for $h=r_{\lambda}$, hence for all $h\in\conti(\gtoneptcpt{\fldreal})$ by
Lemma \ref{lem:single-field-algebra}(2).
\end{proof}

On the level of the von Neumann algebras of the representation, the correspondence between resolvent subalgebras and sub-\(\sigma\)-algebras takes the following form.

\begin{cor}[Resolvent subalgebras and sub-$\sigma$-algebras in the Gelfand--Naimark--Segal representation]\label{cor:random-variable-subalgebras}
In the setting of Theorem \ref{thm:random-variable-gns}, for a real linear
subspace $Y\subset W$, let $\oa{M}_Y$ be the von Neumann algebra generated by
$\fun{\oarepn_{\xi}}{\oaresolventalgebra(Y)}$.
\begin{enumerate}
\item For real linear subspaces $Y_1,Y_2\subset W$,
$\oa{M}_{Y_1}\subset\oa{M}_{Y_2}$ if and only if every event of
$\mblfmlfrak{S}_{\xi}(Y_1)$ differs from an event of
$\mblfmlfrak{S}_{\xi}(Y_2)$ by a $\msrprb$-null set.
In particular, $Y_1\subset Y_2$ implies
$\mblfmlfrak{S}_{\xi}(Y_1)\subset\mblfmlfrak{S}_{\xi}(Y_2)$ and
$\oa{M}_{Y_1}\subset\oa{M}_{Y_2}$.
\item For a real linear subspace $Y\subset W$,
\begin{equation}\label{eq:random-variable-cyclic-subspace}
\begin{aligned}
\gtclos{\fun{\oarepn_{\xi}}{\oaresolventalgebra(Y)}\oagnsvector[\Psi_{\xi}]}
&=
\fun{\lp^{2}}{\Omega,\mblfmlfrak{S}_{\xi}(Y),\msrprb},
\end{aligned}
\end{equation}
the subspace of the classes in $\sphilbfrak{H}_{\xi}$ with an
$\mblfmlfrak{S}_{\xi}(Y)$-measurable representative, and the orthogonal
projection of $\sphilbfrak{H}_{\xi}$ onto it
maps every $Z\in\sphilbfrak{H}_{\xi}$ to a version of the conditional
expectation $\sqfuncond{\prbexp_{\msrprb}}{Z}{\mblfmlfrak{S}_{\xi}(Y)}$.
\end{enumerate}
\end{cor}

\begin{proof}
Both items are read off from the description of $\oa{M}_Y$ in
Theorem \ref{thm:random-variable-gns}(2).

(1): Let $\oa{M}_{Y_1}\subset\oa{M}_{Y_2}$ and $B\in\mblfmlfrak{S}_{\xi}(Y_1)$.
The multiplication by $\fndef{B}$ belongs to $\oa{M}_{Y_2}$ and is the
multiplication by a bounded $\mblfmlfrak{S}_{\xi}(Y_2)$-measurable function $g$;
applied to $\oagnsvector[\Psi_{\xi}]=1$, it gives $\fndef{B}=g$ almost
everywhere, and the event $B'\in\mblfmlfrak{S}_{\xi}(Y_2)$ on which $g=1$
differs from $B$ by a null set.
Conversely, if every event of $\mblfmlfrak{S}_{\xi}(Y_1)$ differs from an event
of $\mblfmlfrak{S}_{\xi}(Y_2)$ by a null set, every bounded
$\mblfmlfrak{S}_{\xi}(Y_1)$-measurable function is almost everywhere equal to a
bounded $\mblfmlfrak{S}_{\xi}(Y_2)$-measurable one, and
Theorem \ref{thm:random-variable-gns}(2) gives $\oa{M}_{Y_1}\subset\oa{M}_{Y_2}$.
If $Y_1\subset Y_2$, the generators $\xi(f)$, $f\in Y_1$, of
$\mblfmlfrak{S}_{\xi}(Y_1)$ are among those of $\mblfmlfrak{S}_{\xi}(Y_2)$.

(2): By Theorem \ref{thm:random-variable-gns}(2), $\oa{M}_Y$ maps
$\oagnsvector[\Psi_{\xi}]=1$ onto the bounded $\mblfmlfrak{S}_{\xi}(Y)$-measurable
functions, which are dense in
$\fun{\lp^{2}}{\Omega,\mblfmlfrak{S}_{\xi}(Y),\msrprb}$; this gives
\eqref{eq:random-variable-cyclic-subspace}, and the orthogonal projection onto
this subspace is the conditional expectation given $\mblfmlfrak{S}_{\xi}(Y)$.
\end{proof}

\section{Inclusion of the ideals and independence}\label{sec:inclusion-independence}

In this section, the inclusion of the ideals of subspaces is compared with the inclusion of the sub-\(\sigma\)-algebras, independence from a given sub-\(\sigma\)-algebra is characterized by conditional expectations, and the reading of these statements as statements on information is explained. Throughout the section, \(\mblfmlfrak{S}_0\) is an arbitrary sub-\(\sigma\)-algebra of \(\mblfmlfrak{S}\), the information with which the fields are compared; the subscript \(0\) does not refer to a time, and \(\mblfmlfrak{S}_0\) is not the \(\sigma\)-algebra of the fields at the time zero of a path space. In the applications of Section \ref{sec:shifted-fields}, it is the \(\sigma\)-algebra of the order coordinate of the free Bose gas or of the paths of the particles in the Nelson model.

\begin{thm}[Inclusion of the ideals and independence of the sub-$\sigma$-algebras]\label{thm:ideal-inclusion-sigma-independence}
In the setting of Theorem \ref{thm:random-variable-ideals}, let
$\mblfmlfrak{S}_0\subset\mblfmlfrak{S}$ be a sub-$\sigma$-algebra.
\begin{enumerate}
\item For real linear subspaces $Y_1,Y_2\subset W$,
\begin{equation}\label{eq:ideal-inclusion-sigma-inclusion}
\begin{aligned}
\oaideal{J}_{W\setminus Y_2}\subset\oaideal{J}_{W\setminus Y_1}
&\implies
\mblfmlfrak{S}_{\xi}(Y_1)\subset\mblfmlfrak{S}_{\xi}(Y_2).
\end{aligned}
\end{equation}
\item For a real linear subspace $Y\subset W$, the $\sigma$-algebras
$\mblfmlfrak{S}_{\xi}(Y)$ and $\mblfmlfrak{S}_0$ are independent under
$\msrprb$ if and only if
\begin{equation}\label{eq:conditional-expectation-constant}
\begin{aligned}
\sqfuncond{\prbexp_{\msrprb}}{\fun{r_{\lambda}}{\xi(f)}}{\mblfmlfrak{S}_0}
&=
\fun{\oastate[\psi_{\xi}]}{\fun{\oaresolvent}{\lambda,f}}
\quad
\msras{\msrprb}
&&\rbk{\lambda\in\fldreal\setminus\setone{0},\ f\in Y}.
\end{aligned}
\end{equation}
\item Let $Y\subset W$ be a real linear subspace such that
$\mblfmlfrak{S}_{\xi}(Y)$ and $\mblfmlfrak{S}_0$ are independent under
$\msrprb$.
Then
\begin{equation}\label{eq:quotient-conditional-expectation}
\begin{aligned}
\sqfuncond{\prbexp_{\msrprb}}{\fun{r_{\lambda}}{\fun{\xi_Y}{f}}}{\mblfmlfrak{S}_0}
&=
\fun{\oastate[\psi_{\xi,Y}]}{\fun{\oaresolvent}{\lambda,f}}
\quad
\msras{\msrprb}
&&\rbk{\lambda\in\fldreal\setminus\setone{0},\ f\in W}.
\end{aligned}
\end{equation}
\item Let $Y\subset W$ be as in (3), and let $f\in W$ be a vector such that
$\mblfmlfrak{S}_{\xi}(\fldreal f)$ and $\mblfmlfrak{S}_0$ are not independent
under $\msrprb$.
Then $f\notin Y$, and
$\fun{\oaresolvent}{\lambda,f}\in\oaideal{J}_{W\setminus Y}$ for every
$\lambda\in\fldreal\setminus\setone{0}$.
\end{enumerate}
\end{thm}

\begin{proof}
(1): By Theorem \ref{thm:subspace-complement-quotient}(4), $Y_1\subset Y_2$,
and the generators of $\mblfmlfrak{S}_{\xi}(Y_1)$ are among those of
$\mblfmlfrak{S}_{\xi}(Y_2)$.

(2): If the two $\sigma$-algebras are independent, then, for $f\in Y$, the
bounded $\mblfmlfrak{S}_{\xi}(Y)$-measurable random variable
$Z=\fun{r_{\lambda}}{\xi(f)}$ has the constant
$\sqfun{\prbexp_{\msrprb}}{Z}$ as a version of its conditional expectation given
$\mblfmlfrak{S}_0$, and
\eqref{eq:random-variable-state} gives
\eqref{eq:conditional-expectation-constant}.
Conversely, let \eqref{eq:conditional-expectation-constant} hold, and let
$B\in\mblfmlfrak{S}_0$.
For $f\in Y$, the finite Borel measures
\begin{equation}\label{eq:conditioned-field-laws}
\begin{aligned}
\fun{\msr{\mu}_f}{D}
&=
\fun{\msrprb}{B\cap\fun{\inv{\xi(f)}}{D}},
\\ %%%%%%%%%%%%%%%%
\fun{\msr{\nu}_f}{D}
&=
\fun{\msrprb}{B}\fun{\msrprb}{\fun{\inv{\xi(f)}}{D}}
\end{aligned}
\end{equation}
on $\gtoneptcpt{\fldreal}$ give the same integral to the constant $1$ and, by
the defining property of the conditional expectation,
\eqref{eq:conditional-expectation-constant}, and
\eqref{eq:random-variable-state}, to every $r_{\lambda}$.
By \eqref{eq:value-function-relations}, the norm closure of
the linear span of $1$ and the functions $r_{\lambda}$ contains the products
$r_{\lambda}r_{\mu}=\frac{r_{\lambda}-r_{\mu}}{\imunit\rbk{\mu-\lambda}}$ for
$\lambda\neq\mu$ and their uniform limits $r_{\lambda}^{2}$ as $\mu\to\lambda$,
so it is the closed subalgebra generated by the functions $r_{\lambda}$, which is
$\conti(\gtoneptcpt{\fldreal})$ by Lemma \ref{lem:single-field-algebra}(2).
Hence $\msr{\mu}_f$ and $\msr{\nu}_f$ agree on
$\conti(\gtoneptcpt{\fldreal})$, and so, by the regularity of finite Borel
measures on the compact metrizable space $\gtoneptcpt{\fldreal}$, on the Borel
sets.
Let $m\in\semigrposint$, let $g_1,\ldots,g_m\in Y$, let
$\Xi=\rbk{\xi(g_1),\ldots,\xi(g_m)}\colon\Omega\to\fldreal^{m}$, and let
$t\in\fldreal^{m}$.
For $f=\sum_kt_kg_k\in Y$, \eqref{eq:random-variable-linearity} gives
$t\vainnprod\Xi=\xi(f)$ almost everywhere, and the equality
$\msr{\mu}_f=\msr{\nu}_f$ integrated against $u\mapsto\napiernum^{\imunit u}$
on $\fldreal$, which carries the whole mass of both measures, gives
$\sqfun{\prbexp_{\msrprb}}{\fndef{B}\napiernum^{\imunit t\vainnprod\Xi}}=\fun{\msrprb}{B}\sqfun{\prbexp_{\msrprb}}{\napiernum^{\imunit t\vainnprod\Xi}}$.
If $\fun{\msrprb}{B}>0$, the probability measures
$A\mapsto\frac{\fun{\msrprb}{B\cap\fun{\inv{\Xi}}{A}}}{\fun{\msrprb}{B}}$ and
$A\mapsto\fun{\msrprb}{\fun{\inv{\Xi}}{A}}$ on $\fldreal^{m}$ have the same
characteristic function and coincide;
if $\fun{\msrprb}{B}=0$, both sides of
$\fun{\msrprb}{B\cap\fun{\inv{\Xi}}{A}}=\fun{\msrprb}{B}\fun{\msrprb}{\fun{\inv{\Xi}}{A}}$
vanish.
The sets $\fun{\inv{\Xi}}{A}$ with $m\in\semigrposint$, $g_j\in Y$, and Borel
$A\subset\fldreal^{m}$ form a $\pi$-system generating
$\mblfmlfrak{S}_{\xi}(Y)$, and the $\pi$-$\lambda$ theorem gives
$\fun{\msrprb}{B\cap A'}=\fun{\msrprb}{B}\fun{\msrprb}{A'}$ for every
$A'\in\mblfmlfrak{S}_{\xi}(Y)$.

(3): If $f\notin Y$, both sides of \eqref{eq:quotient-conditional-expectation}
vanish, the left-hand side by $\fun{r_{\lambda}}{\infty}=0$ and the right-hand
side because $\fun{\oaresolvent}{\lambda,f}\in\oaideal{J}_{W\setminus Y}$.
If $f\in Y$, then
$\fun{\oastate[\psi_{\xi,Y}]}{\fun{\oaresolvent}{\lambda,f}}=\fun{\oastate[\psi_{\xi}]}{\fun{\oaresolvent}{\lambda,f}}$
and $\fun{\xi_Y}{f}=\xi(f)$, and (2) applies.

(4): For $f\in Y$,
$\mblfmlfrak{S}_{\xi}(\fldreal f)\subset\mblfmlfrak{S}_{\xi}(Y)$ is independent
of $\mblfmlfrak{S}_0$; hence the vector $f$ of (4) lies outside $Y$, and
Theorem \ref{thm:subspace-complement-quotient}(4) gives
$\fun{\oaresolvent}{\lambda,f}\in\oaideal{J}_{W\setminus Y}$.
\end{proof}

\begin{rem}[Sub-$\sigma$-algebras and ideals as carriers of information]\label{rem:information}
A sub-$\sigma$-algebra $\mblfmlfrak{S}_0$ represents information, and
$\sqfuncond{\prbexp_{\msrprb}}{Z}{\mblfmlfrak{S}_0}$ is the best prediction in
$\lp^{2}$ of $Z$ from it.

Through Theorems \ref{thm:subspace-complement-quotient},
\ref{thm:random-variable-ideals}, and
\ref{thm:ideal-inclusion-sigma-independence}, the same reading attaches physical
information to the ideals $\oaideal{J}_{W\setminus Y}$.

\begin{itemize}
\item By Theorem \ref{thm:random-variable-ideals}(2), $\mblfmlfrak{S}_{\xi}(Y)$
is the information supplied by the observables of $\oaresolventalgebra(Y)$.
\item By Theorem \ref{thm:ideal-inclusion-sigma-independence}(1), a larger
ideal corresponds to a smaller sub-$\sigma$-algebra.
For an increasing family $\fml{Y_t}{t\in I}$ of real linear subspaces of $W$
indexed by a totally ordered set $I$, the ideals $\oaideal{J}_{W\setminus Y_t}$
decrease and the sub-$\sigma$-algebras $\mblfmlfrak{S}_{\xi}(Y_t)$ increase in
$t$, as a filtration does.
\item By Theorems \ref{thm:subspace-complement-quotient}(2) and
\ref{thm:random-variable-ideals}(3), the quotient by
$\oaideal{J}_{W\setminus Y}$ keeps the information of the directions of $Y$ and
discards that of the other directions, whose fields are given the value
$\infty$, at which every $r_{\lambda}$ vanishes.
\item Independence means that one piece of information says nothing about
another.
By Theorem \ref{thm:ideal-inclusion-sigma-independence}(2),
$\mblfmlfrak{S}_{\xi}(Y)$ is independent of $\mblfmlfrak{S}_0$ exactly when
the conditional expectations \eqref{eq:conditional-expectation-constant} do not
depend on the information in $\mblfmlfrak{S}_0$.
If $\mblfmlfrak{S}_0$ is generated by macroscopic variables, such as the
coordinates of an order parameter, the observables of $\oaresolventalgebra(Y)$
then do not detect these variables.
Under this independence, the fields of $Y$ carry no information on these
variables, and no nonconstant one among them is almost everywhere a function of
the fields of $Y$, since a random variable independent of itself is almost
everywhere constant.
This reproduces, within the present framework, the view of the resolvent algebra
as a $\oacstar$-algebra of purely quantum observables, generated by the
resolvents of the fields, that contains no macroscopic variables: in the
applications of Section \ref{sec:shifted-fields}, the macroscopic variables
enter through the state, as the parameters of the mixture in
Theorem \ref{thm:shifted-fields-correlated-ideal}(1).
\item By Theorem \ref{thm:ideal-inclusion-sigma-independence}(3) and (4), the
state $\oastate[\psi_{\xi,Y}]$ of the quotient then carries no information on
$\mblfmlfrak{S}_0$, and every direction whose field is not independent of
$\mblfmlfrak{S}_0$ has its resolvents in $\oaideal{J}_{W\setminus Y}$.
\end{itemize}

This reading rests on an algebraic property of the resolvent algebra: it has
nontrivial closed two-sided ideals, such as the ideals
$\oaideal{J}_{X\setminus Y}$ of Theorem \ref{thm:subspace-complement-quotient}
for proper subspaces $Y\subset X$ and the principal ideals of
Section \ref{sec:field-laws}, whereas the Weyl algebra of a nondegenerate
symplectic space is simple \cite[Theorem 5.2.8(3)]{BratteliRobinson004},
\cite[Section 1]{DetlevBuchholz001}.
An ideal that carries physical information, such as the ideal of the condensate
directions or that of the infrared-singular directions, gives a quotient
$\oacstar$-algebra, which again has a physical meaning: by
Theorem \ref{thm:subspace-complement-quotient}(2) and (3), the quotient by
$\oaideal{J}_{W\setminus Y}$ is the resolvent algebra of the retained directions
$Y$, it is faithfully represented by the representations obtained from the
faithful representations of $\oaresolventalgebra(Y)$ by letting the resolvents of
the discarded directions vanish, so that the fields of these directions are
infinite, and the states that vanish on the ideal are the vector states
\eqref{eq:subspace-extension-state} in such representations.
For the simple Weyl algebra, every nonzero representation is faithful, and the
passage to the retained directions is not expressed by a quotient.

The commutation relations are not seen by the $\sigma$-algebras; the comparison is made on an isotropic subspace.
\end{rem}

\section{Shifted fields and the ideal of the correlated directions}\label{sec:shifted-fields}

In this section, the fields are sums of fields independent of a given sub-\(\sigma\)-algebra and of shifts measurable with respect to it, the abstract form of the mean shifts of the fields in the models of Section \ref{sec:applications}. The directions in which the shift is not almost surely constant generate an ideal, and a subspace of directions has fields independent of the sub-\(\sigma\)-algebra exactly when this ideal is contained in the ideal of the complement of the subspace.

The shifts are modeled on the mean shifts of the fields in the models with a linear coupling or with a condensate. In the van Hove model with an infrared-regular source, the ground state is the Fock vacuum transformed by a Weyl operator \cite[Theorem 13.9(iii)]{AsaoArai026}, so that every field is shifted by a deterministic real number depending linearly on the test function; this is the case of a constant shift in \eqref{eq:constant-shift-directions}, with \(Y_0=W\). In the free Bose gas above the critical density, the BEC state is a mixture of states whose fields are shifted by the condensate parts, which are functions of the random order coordinate \cite{YoshitsuguSekine004}; there \(\mblfmlfrak{S}_0\) is the \(\sigma\)-algebra of the order coordinate. In the Nelson model, the Gibbs measure of the particle path and the field is a mixture of Gaussian measures: conditioned on the path, the field is Gaussian with the covariance of the free field and with a mean that depends linearly on the test function and on the path \cite[Section 4, after Theorem 4.1]{BetzHiroshimaLorincziMinlosSpohn001}; there \(\mblfmlfrak{S}_0\) is the \(\sigma\)-algebra of the path, \(\eta(f)\) is the centered field, and \(\ell(f)\) is the path-dependent mean.

Let \(W\subset X\) be isotropic, let \(\pairbk{\Omega,\mblfmlfrak{S},\msrprb}\) be a probability space with a sub-\(\sigma\)-algebra \(\mblfmlfrak{S}_0\), and let \(\eta=\fml{\eta(f)}{f\in W}\) and \(\ell=\fml{\ell(f)}{f\in W}\) be families of real random variables, both satisfying \eqref{eq:random-variable-linearity}, such that every \(\ell(f)\) is \(\mblfmlfrak{S}_0\)-measurable and \(\mblfmlfrak{S}_{\eta}(W)\) is independent of \(\mblfmlfrak{S}_0\). Let \(\xi(f)=\eta(f)+\ell(f)\), and let \begin{equation}\label{eq:constant-shift-directions}
\begin{aligned}
Y_0
&=
\set{f\in W}{\ell(f)\ \text{is}\ \msrprb\text{-almost everywhere equal to a constant}\ \fun{c}{f}}.
\end{aligned}
\end{equation}

\begin{thm}[Shifted fields and the ideal of the correlated directions]\label{thm:shifted-fields-correlated-ideal}
In the setting of Section \ref{sec:shifted-fields}, with
$\xi(f)=\eta(f)+\ell(f)$ and $Y_0$ defined by
\eqref{eq:constant-shift-directions}, the following hold.

\begin{enumerate}
\item For $n\in\semigrposint$, $\lambda_j\in\fldreal\setminus\setone{0}$, and
$f_j\in W$,
\begin{equation}\label{eq:shifted-conditional-expectation}
\begin{aligned}
\sqfuncond{\prbexp_{\msrprb}}{\prod_{j=1}^{n}\fun{r_{\lambda_j}}{\xi(f_j)}}{\mblfmlfrak{S}_0}
&=
\fun{\oastate[\psi_{\eta}]}{\prod_{j=1}^{n}\fun{\oaresolvent}{\lambda_j+\imunit\ell(f_j),f_j}}
\end{aligned}
\end{equation}
$\msrprb$-almost everywhere, with the resolvents of complex parameters of
Section \ref{sec:setting}, the state $\oastate[\psi_{\eta}]$ extended to them by
continuity, and the random numbers $\ell(f_j)$ inserted.
If $\ell(\cdot)(\omega)$ is real linear on $W$ for every $\omega$, the right-hand
side is $\fun{\oastate[\psi_{\eta}]\circ\Theta_{\ell(\cdot)(\omega)}}{\prod_j\fun{\oaresolvent}{\lambda_j,f_j}}$,
and $\oastate[\psi_{\xi}]$ is the mixture of the shifted states
$\oastate[\psi_{\eta}]\circ\Theta_{\ell(\cdot)(\omega)}$ under $\msrprb$.
\item The set $Y_0$ is a real linear subspace of $W$, and $c$ is real linear on
$Y_0$.
\item For a real linear subspace $Y\subset W$, the $\sigma$-algebras
$\mblfmlfrak{S}_{\xi}(Y)$ and $\mblfmlfrak{S}_0$ are independent if and only if
$Y\subset Y_0$, that is, if and only if
$\oaideal{J}_{W\setminus Y_0}\subset\oaideal{J}_{W\setminus Y}$.
\item For $f\in W$ and $\lambda\in\fldreal\setminus\setone{0}$,
$\fun{\oaresolvent}{\lambda,f}\in\oaideal{J}_{W\setminus Y_0}$ if and only if
$\mblfmlfrak{S}_{\xi}(\fldreal f)$ and $\mblfmlfrak{S}_0$ are not independent.
On $\oaresolventalgebra(Y_0)$,
$\oastate[\psi_{\xi}]=\oastate[\psi_{\eta}]\circ\Theta_{\tilde{c}}$ for every
real-linear extension $\tilde{c}$ of $c$ to $X$, and
the state $\oastate[\psi_{\xi,Y_0}]$ of
Theorem \ref{thm:random-variable-ideals}(3), the only state that agrees with
$\oastate[\psi_{\xi}]$ on $\oaresolventalgebra(Y_0)$ and vanishes on
$\oaideal{J}_{W\setminus Y_0}$, takes at the resolvents the constant conditional
expectations of Theorem \ref{thm:ideal-inclusion-sigma-independence}(3) for
$Y=Y_0$.
\end{enumerate}
\end{thm}

\begin{proof}
(1): For $B\in\mblfmlfrak{S}_0$, the joint law of $\rbk{\eta(f_j)}_j$ is
independent of that of $\rbk{\ell(f_j)}_j$ and $\fndef{B}$, and Fubini's theorem
for the product of the two laws gives
$$\sqfun{\prbexp_{\msrprb}}{\fndef{B}\prod_j\fun{r_{\lambda_j}}{\eta(f_j)+\ell(f_j)}}
=
\sqfun{\prbexp_{\msrprb}}{\fndef{B}\fun{\Phi}{\ell(f_1),\ldots,\ell(f_n)}}$$
with
$\fun{\Phi}{c_1,\ldots,c_n}=\sqfun{\prbexp_{\msrprb}}{\prod_j\fun{r_{\lambda_j}}{\eta(f_j)+c_j}}$.
Since $\fun{r_{\lambda}}{u+c}=\rbk{\imunit\rbk{\lambda+\imunit c}-u}^{-1}$ and
$\oarepn[\rho]_{\eta}$ of the proof of Theorem \ref{thm:random-variable-ideals}(1) maps
$\fun{\oaresolvent}{z,f}$, defined by a norm-convergent Neumann series in a
resolvent of $f$, to the class of $\rbk{\imunit z-\eta(f)}^{-1}$, the function
$\Phi$ is the right-hand side of \eqref{eq:shifted-conditional-expectation}.
If $\ell(\cdot)(\omega)$ is linear, the automorphism of
\cite[Proposition 3.7]{BuchholzGrundling002} for a linear extension of it gives
the second statement, and taking expectations gives the mixture.

(2): For $f,g\in Y_0$ and $a\in\fldreal$, \eqref{eq:random-variable-linearity}
for $\ell$ gives $\ell(f+ag)=\fun{c}{f}+a\fun{c}{g}$ almost everywhere.

(3): If $Y\subset Y_0$, then $\xi(f)=\eta(f)+\fun{c}{f}$ almost everywhere for
$f\in Y$, so every event of $\mblfmlfrak{S}_{\xi}(Y)$ differs from an event of
$\mblfmlfrak{S}_{\eta}(W)$ by a null set, and the independence of
$\mblfmlfrak{S}_{\eta}(W)$ and $\mblfmlfrak{S}_0$ gives that of
$\mblfmlfrak{S}_{\xi}(Y)$ and $\mblfmlfrak{S}_0$.
Conversely, let the latter hold and $f\in Y$.
For $u\in\fldreal$, the independence gives
$\sqfuncond{\prbexp_{\msrprb}}{\napiernum^{\imunit u\xi(f)}}{\mblfmlfrak{S}_0}=\sqfun{\prbexp_{\msrprb}}{\napiernum^{\imunit u\xi(f)}}$,
and the argument of (1) gives
$\sqfuncond{\prbexp_{\msrprb}}{\napiernum^{\imunit u\xi(f)}}{\mblfmlfrak{S}_0}=\napiernum^{\imunit u\ell(f)}\sqfun{\prbexp_{\msrprb}}{\napiernum^{\imunit u\eta(f)}}$
almost everywhere.
The characteristic function of $\eta(f)$ is continuous with the value $1$ at
$u=0$, hence nonzero for $\abs{u}<\varepsilon$ with some $\varepsilon>0$, and
$\napiernum^{\imunit u\ell(f)}$ is almost everywhere equal to a constant for every
rational $u$ with $\abs{u}<\varepsilon$.
Outside a null set $N$, these equalities hold simultaneously; for
$\omega,\omega'\notin N$, $\napiernum^{\imunit u\rbk{\fun{\ell(f)}{\omega}-\fun{\ell(f)}{\omega'}}}=1$
for these $u$, and by continuity in $u$ for all $\abs{u}<\varepsilon$, which
gives $\fun{\ell(f)}{\omega}=\fun{\ell(f)}{\omega'}$.
Hence $f\in Y_0$.
The equivalence with the inclusion of the ideals is
Theorem \ref{thm:subspace-complement-quotient}(4).

(4): By (3) for $Y=\fldreal f$ and (2), $\mblfmlfrak{S}_{\xi}(\fldreal f)$ and
$\mblfmlfrak{S}_0$ are independent exactly when $f\in Y_0$, and
Theorem \ref{thm:subspace-complement-quotient}(4) gives the first statement.
On $\oaresolventalgebra(Y_0)$, the field of $f$ is $\eta(f)+\fun{c}{f}$ almost
everywhere, and the computation of (1) with constant shifts gives the shifted
state; the remaining statements are
Theorem \ref{thm:random-variable-ideals}(3) and
Theorem \ref{thm:ideal-inclusion-sigma-independence}(3) for $Y=Y_0$.
\end{proof}

The same mechanism appears on the side of representations, where the mixture of Theorem \ref{thm:shifted-fields-correlated-ideal}(1) becomes a direct integral of shifted representations. Let \(\Gamma\) be a locally compact space with a Radon probability measure \(\msrprb_{\Gamma}\), let \(\sphilbfrak{H}_0\neq\setone{0}\) be a separable Hilbert space, and let \(\oarepn_0\) be a representation of \(\fun{\oaresolventalgebra}{X,\opform{\sigma}}\) on \(\sphilbfrak{H}_0\) that is regular on \(X\), with the fields \(\fun{\phi_{\oarepn_0}}{f}\) of Section \ref{sec:setting}. For \(\gamma\in\Gamma\), let \(\ell_{\gamma}\colon X\to\fldreal\) be real linear, with \(\gamma\mapsto\ell_{\gamma}(f)\) continuous for every \(f\in X\), and put \(\oarepn_{\gamma}=\oarepn_0\circ\Theta_{\ell_{\gamma}}\), that is, \begin{equation}\label{eq:shifted-representation}
\begin{aligned}
\fun{\oarepn_{\gamma}}{\fun{\oaresolvent}{\lambda,f}}
&=
\rbk{\imunit\lambda-\fun{\phi_{\oarepn_0}}{f}-\ell_{\gamma}(f)}^{-1}
\quad\rbk{\lambda\in\fldreal\setminus\setone{0},\ f\in X}.
\end{aligned}
\end{equation} Let \begin{equation}\label{eq:constant-shift-directions-representation}
\begin{aligned}
Y_{\Gamma}
&=
\set{f\in X}{\ell_{\gamma}(f)=\fun{c_{\Gamma}}{f}\ \text{for}\ \msrprb_{\Gamma}\text{-almost every}\ \gamma\ \text{with a constant}\ \fun{c_{\Gamma}}{f}}.
\end{aligned}
\end{equation} The space \(\fun{\lp^{2}}{\Gamma,\msrprb_{\Gamma};\sphilbfrak{H}_0}\) of square-integrable \(\msrprb_{\Gamma}\)-measurable \(\sphilbfrak{H}_0\)-valued functions is the direct integral of the constant field \(\sphilbfrak{H}_0\), and the map \begin{equation}\label{eq:direct-integral-shifted-unitary}
g\otimes\eta\mapsto g\eta
\end{equation} extends to a unitary operator from \(\fun{\lp^{2}}{\Gamma,\msrprb_{\Gamma}}\otimes\sphilbfrak{H}_0\) onto it.

\begin{prop}[Direct integral of shifted representations and the quotient]\label{prop:direct-integral-shifted-quotient}
Under the hypotheses of Section \ref{sec:shifted-fields} for the family
$\oarepn_{\gamma}$ in \eqref{eq:shifted-representation}, with $Y_{\Gamma}$
defined by \eqref{eq:constant-shift-directions-representation}, the following hold.

\begin{enumerate}
\item For $A\in\fun{\oaresolventalgebra}{X,\opform{\sigma}}$, the map
$\gamma\mapsto\fun{\oarepn_{\gamma}}{A}$ is norm continuous, and
\begin{equation}\label{eq:direct-integral-shifted-representation}
\begin{aligned}
\funrbk{\fun{\oarepn}{A}\xi}{\gamma}
&=
\fun{\oarepn_{\gamma}}{A}\fun{\xi}{\gamma}
\quad\rbk{\xi\in\fun{\lp^{2}}{\Gamma,\msrprb_{\Gamma};\sphilbfrak{H}_0}}
\end{aligned}
\end{equation}
defines a representation
$\oarepn=\int_{\Gamma}^{\oplus}\oarepn_{\gamma}\opdmsr{\msrprb_{\Gamma}(\gamma)}$
of $\fun{\oaresolventalgebra}{X,\opform{\sigma}}$.
\item The set $Y_{\Gamma}$ is a real linear subspace of $X$, and $c_{\Gamma}$ is
real linear on it.
For every real-linear extension $\tilde{c}_{\Gamma}$ of $c_{\Gamma}$ to $X$ and
every $A\in\oaresolventalgebra(Y_{\Gamma})$, the operator $\fun{\oarepn}{A}$
corresponds to $\idone\otimes\fun{\oarepn_0}{\fun{\Theta_{\tilde{c}_{\Gamma}}}{A}}$
under the unitary operator defined by \eqref{eq:direct-integral-shifted-unitary}.
\item For $f\in X$ and $\lambda\in\fldreal\setminus\setone{0}$, the operator
$\fun{\oarepn}{\fun{\oaresolvent}{\lambda,f}}$ corresponds to an operator of
the form $\idone\otimes T$ if and only if $f\in Y_{\Gamma}$, that is, if and only
if $\fun{\oaresolvent}{\lambda,f}\notin\oaideal{J}_{X\setminus Y_{\Gamma}}$.
\item For a unit vector $\Upsilon_0\in\sphilbfrak{H}_0$, the vector state
$\oastate[\psi]$ of the constant section $\gamma\mapsto\Upsilon_0$ in $\oarepn$ is
$\fun{\oastate[\psi]}{A}=\int_{\Gamma}\bkt{\Upsilon_0}{\fun{\oarepn_{\gamma}}{A}\Upsilon_0}\opdmsr{\msrprb_{\Gamma}(\gamma)}$.
Let $\oarepn[\rho]_{\Gamma}$ be the restriction of
$\oarepn_0\circ\Theta_{\tilde{c}_{\Gamma}}$ to $\oaresolventalgebra(Y_{\Gamma})$,
which does not depend on the extension $\tilde{c}_{\Gamma}$, and let
$\oarepn[\rho]_{\Gamma,\infty}$ be the representation of
Theorem \ref{thm:subspace-complement-quotient}(1) for it, with $Y=Y_{\Gamma}$.
Then
$A+\oaideal{J}_{X\setminus Y_{\Gamma}}\mapsto\fun{\oarepn[\rho]_{\Gamma,\infty}}{A}$
is a well-defined representation of the quotient
$\setquot{\fun{\oaresolventalgebra}{X,\opform{\sigma}}}{\oaideal{J}_{X\setminus Y_{\Gamma}}}$,
and the vector state of $\Upsilon_0$ in $\oarepn[\rho]_{\Gamma,\infty}$ is the
only state that agrees with $\oastate[\psi]$ on
$\oaresolventalgebra(Y_{\Gamma})$ and vanishes on
$\oaideal{J}_{X\setminus Y_{\Gamma}}$.
\end{enumerate}
\end{prop}

\begin{proof}
(1): For $a,b\in\fldreal$, the operators
$\rbk{\imunit\lambda-\fun{\phi_{\oarepn_0}}{f}-a}^{-1}$ have norm at most
$\abs{\lambda}^{-1}$, and the difference of the two inverses for $a$ and $b$ is
$\rbk{a-b}\rbk{\imunit\lambda-\fun{\phi_{\oarepn_0}}{f}-a}^{-1}\rbk{\imunit\lambda-\fun{\phi_{\oarepn_0}}{f}-b}^{-1}$,
of norm at most $\lambda^{-2}\abs{a-b}$.
Hence $\fun{\oarepn_{\gamma}}{A}$ is norm continuous in $\gamma$ for every
linear combination $A$ of resolvents.
These linear combinations are dense (Section \ref{sec:setting}), and
$\norm{\fun{\oarepn_{\gamma}}{A}-\fun{\oarepn_{\gamma}}{B}}\leq\norm{A-B}$
uniformly in $\gamma$, so the continuity holds for every $A$.
For a $\msrprb_{\Gamma}$-measurable $\xi$, a compact set $K_0$ on which $\xi$ is
continuous is also one on which $\gamma\mapsto\fun{\oarepn_{\gamma}}{A}\fun{\xi}{\gamma}$
is continuous, and $\norm{\fun{\oarepn_{\gamma}}{A}\fun{\xi}{\gamma}}\leq\norm{A}\norm{\fun{\xi}{\gamma}}$;
hence \eqref{eq:direct-integral-shifted-representation} defines bounded
operators, and the algebraic properties hold pointwise in $\gamma$.

(2): For $f,g\in Y_{\Gamma}$ and $a\in\fldreal$,
$\ell_{\gamma}(f+ag)=\fun{c_{\Gamma}}{f}+a\fun{c_{\Gamma}}{g}$ for almost every
$\gamma$.
For $f\in Y_{\Gamma}$, \eqref{eq:shifted-representation} gives
$\fun{\oarepn_{\gamma}}{\fun{\oaresolvent}{\lambda,f}}=\fun{\oarepn_0}{\fun{\Theta_{\tilde{c}_{\Gamma}}}{\fun{\oaresolvent}{\lambda,f}}}$
for $\gamma$ outside a null set.
For a sequence $A_n$ of linear combinations of resolvents of vectors of
$Y_{\Gamma}$ converging to $A\in\oaresolventalgebra(Y_{\Gamma})$, the union of the countably
many null sets for the vectors in the $A_n$ is null, and outside it
$\fun{\oarepn_{\gamma}}{A}=\fun{\oarepn_0}{\fun{\Theta_{\tilde{c}_{\Gamma}}}{A}}$.
Hence $\fun{\oarepn}{A}$ multiplies every section pointwise by the constant
operator $T=\fun{\oarepn_0}{\fun{\Theta_{\tilde{c}_{\Gamma}}}{A}}$, which
corresponds to $\idone\otimes T$, since $g\otimes\eta$ and $g\otimes T\eta$
correspond to $g\eta$ and $gT\eta$.

(3): If $f\in Y_{\Gamma}$, (2) applies.
Conversely, let $\fun{\oarepn}{\fun{\oaresolvent}{\lambda,f}}$ correspond to
$\idone\otimes T$.
Applied to the sections $\gamma\mapsto\eta$ for $\eta$ in a countable dense
subset of $\sphilbfrak{H}_0$, this gives
$\fun{\oarepn_{\gamma}}{\fun{\oaresolvent}{\lambda,f}}=T$ for $\gamma$ outside a
null set $N$.
For $\gamma_0,\gamma\notin N$, the inverses in \eqref{eq:shifted-representation}
coincide, hence so do the operators
$\imunit\lambda-\fun{\phi_{\oarepn_0}}{f}-\ell_{\gamma}(f)$ and
$\imunit\lambda-\fun{\phi_{\oarepn_0}}{f}-\ell_{\gamma_0}(f)$ on the nonzero
domain of $\fun{\phi_{\oarepn_0}}{f}$, which gives
$\ell_{\gamma}(f)=\ell_{\gamma_0}(f)$.
Hence $f\in Y_{\Gamma}$.
The equivalence with the ideal is
Theorem \ref{thm:subspace-complement-quotient}(4) with $W=X$.

(4): The first statement is the definition of the inner product of
$\fun{\lp^{2}}{\Gamma,\msrprb_{\Gamma};\sphilbfrak{H}_0}$.
For $f\in Y_{\Gamma}$, $\fun{\Theta_{\tilde{c}_{\Gamma}}}{\fun{\oaresolvent}{\lambda,f}}=\fun{\oaresolvent}{\lambda+\imunit\fun{c_{\Gamma}}{f},f}$
depends only on $c_{\Gamma}$, and so does $\oarepn[\rho]_{\Gamma}$.
By Theorem \ref{thm:subspace-complement-quotient}(2) with $W=X$, the kernel of
$\oarepn[\rho]_{\Gamma,\infty}$ contains $\oaideal{J}_{X\setminus Y_{\Gamma}}$,
which gives the representation of the quotient, and the vector state of
$\Upsilon_0$ in it vanishes on $\oaideal{J}_{X\setminus Y_{\Gamma}}$.
For $A\in\oaresolventalgebra(Y_{\Gamma})$, (2) and the first statement give
$\fun{\oastate[\psi]}{A}=\bkt{\Upsilon_0}{\fun{\oarepn[\rho]_{\Gamma}}{A}\Upsilon_0}=\bkt{\Upsilon_0}{\fun{\oarepn[\rho]_{\Gamma,\infty}}{A}\Upsilon_0}$,
and Theorem \ref{thm:subspace-complement-quotient}(3) gives the uniqueness.
\end{proof}

By Proposition \ref{prop:direct-integral-shifted-quotient}, the generators of \(\oaideal{J}_{X\setminus Y_{\Gamma}}\) are the resolvents whose representatives vary over the integration variable \(\gamma\), and on the algebra \(\oaresolventalgebra(Y_{\Gamma})\) of the remaining directions, which the quotient identifies with \(\setquot{\fun{\oaresolventalgebra}{X,\opform{\sigma}}}{\oaideal{J}_{X\setminus Y_{\Gamma}}}\), the direct integral reduces to one integrand \(\oarepn[\rho]_{\Gamma}\) with the multiplicity space \(\fun{\lp^{2}}{\Gamma,\msrprb_{\Gamma}}\). If \(Y_{\Gamma}\neq X\), the direct integral itself does not pass to the quotient, since by \eqref{eq:shifted-representation} it maps every resolvent to a nonzero operator, whereas the resolvents of the vectors outside \(Y_{\Gamma}\) belong to \(\oaideal{J}_{X\setminus Y_{\Gamma}}\); the quotient is represented by \(\oarepn[\rho]_{\Gamma,\infty}\), which agrees at each element of \(\oaresolventalgebra(Y_{\Gamma})\) with almost every integrand, by (2), and in which the resolvents of the other directions vanish. The passage to the quotient thus keeps the integrand representation on the retained directions, and the integration variable, which labels the integrands, is carried only by the ideal.

The direct integral itself sees the ideal \(\oaideal{J}_{X\setminus Y_{\Gamma}}\) in a way opposite to the kernels of Section \ref{sec:infinite-field-reduction}. The following proposition shows that the direct integral is faithful, so that the ideal is not visible on the level of the \(\oacstar\)-algebra, and that, along sequences of directions in which the unshifted fields vanish in the limit and the shifts converge, the strong limits of the images of generators of the ideal are central and are functions of the integration variable.

\begin{prop}[Faithfulness of the direct integral and the center of its weak closure]\label{prop:direct-integral-faithful-center}
In the setting of Proposition \ref{prop:direct-integral-shifted-quotient}, let
$\oa{M}=\oadoublecommutant{\fun{\oarepn}{\fun{\oaresolventalgebra}{X,\opform{\sigma}}}}$.
\begin{enumerate}
\item The representation $\oarepn$ is regular and faithful,
$\fun{\oastate[\psi]}{\fun{\oaresolvent}{\lambda,f}}\neq0$ for every $f\in X$
and $\lambda\in\fldreal\setminus\setone{0}$, and
\begin{equation}\label{eq:direct-integral-image-center}
\begin{aligned}
\fun{\oarepn}{\fun{\oaresolventalgebra}{X,\opform{\sigma}}}\cap\fun{\oacenter}{\oa{M}}
&=
\fldcmp.
\end{aligned}
\end{equation}
\item Let $\seq{f_n}{n\in\semigrposint}$ be a sequence in $X$ and
$m\colon\Gamma\to\fldreal$ a function with
\begin{equation}\label{eq:direct-integral-central-probes}
\begin{aligned}
\slim_{n\to\infty}\fun{\oarepn_0}{\fun{\oaresolvent}{1,f_n}}
&=
-\imunit,
\quad
\lim_{n\to\infty}\ell_{\gamma}(f_n)
=
\fun{m}{\gamma}
\quad\rbk{\gamma\in\Gamma}.
\end{aligned}
\end{equation}
Then, for $\lambda\in\fldreal\setminus\setone{0}$, the operators
$\fun{\oarepn}{\fun{\oaresolvent}{\lambda,f_n}}$ converge strongly to the
operator $C_{\lambda}$ defined by
\begin{equation}\label{eq:direct-integral-central-limit}
\begin{aligned}
\funrbk{C_{\lambda}\zeta}{\gamma}
&=
\frac{\fun{\zeta}{\gamma}}{\imunit\lambda-\fun{m}{\gamma}}
\quad\rbk{\zeta\in\fun{\lp^{2}}{\Gamma,\msrprb_{\Gamma};\sphilbfrak{H}_0}},
\end{aligned}
\end{equation}
and $C_{\lambda}\in\fun{\oacenter}{\oa{M}}$.
\item If, in (2), $m$ is not $\msrprb_{\Gamma}$-almost everywhere constant, then
$f_n\notin Y_{\Gamma}$ and
$\fun{\oaresolvent}{\lambda,f_n}\in\oaideal{J}_{X\setminus Y_{\Gamma}}$ for all
but finitely many $n$, and $C_{\lambda}$ is not a scalar; in particular,
$\oa{M}$ is not a factor.
\end{enumerate}
\end{prop}

\begin{proof}
(1): Let $f\in X$, $\lambda\in\fldreal\setminus\setone{0}$, and
$\fun{\oarepn}{\fun{\oaresolvent}{\lambda,f}}\zeta=0$.
By \eqref{eq:direct-integral-shifted-representation},
$\fun{\oarepn_{\gamma}}{\fun{\oaresolvent}{\lambda,f}}\fun{\zeta}{\gamma}=0$ for
almost every $\gamma$, and the operator \eqref{eq:shifted-representation} is
injective as an inverse; hence $\zeta=0$, and $\oarepn$ is regular.
Every regular representation of $\fun{\oaresolventalgebra}{X,\opform{\sigma}}$
is faithful \cite[Theorem 4.10]{BuchholzGrundling002}.
The relation
$\Upsilon_0=\rbk{\imunit\lambda-\fun{\phi_{\oarepn_0}}{f}-\ell_{\gamma}(f)}\fun{\oarepn_{\gamma}}{\fun{\oaresolvent}{\lambda,f}}\Upsilon_0$
and the self-adjointness of the field give
\begin{equation}\label{eq:direct-integral-resolvent-imaginary-part}
\begin{aligned}
\opimag\bkt{\Upsilon_0}{\fun{\oarepn_{\gamma}}{\fun{\oaresolvent}{\lambda,f}}\Upsilon_0}
&=
-\lambda\norm{\fun{\oarepn_{\gamma}}{\fun{\oaresolvent}{\lambda,f}}\Upsilon_0}^{2},
\end{aligned}
\end{equation}
and the vector $\fun{\oarepn_{\gamma}}{\fun{\oaresolvent}{\lambda,f}}\Upsilon_0$ is nonzero by injectivity.
By Proposition \ref{prop:direct-integral-shifted-quotient}(4),
$$\begin{aligned}
\opimag\fun{\oastate[\psi]}{\fun{\oaresolvent}{\lambda,f}}
&=
-\lambda\sqfun{\prbexp_{\msrprb_{\Gamma}}}{\norm{\fun{\oarepn_{\gamma}}{\fun{\oaresolvent}{\lambda,f}}\Upsilon_0}^{2}}
\neq
0.
\end{aligned}$$
The representation $\oarepn_0$ is unital, since its regularity at the zero
vector makes the projection
$\fun{\oarepn_0}{\idone}=\imunit\fun{\oarepn_0}{\fun{\oaresolvent}{1,0}}$
injective, and so is $\oarepn$; hence the scalars belong to the left-hand side
of \eqref{eq:direct-integral-image-center}.
Conversely, if $\fun{\oarepn}{A}$ lies in the center of $\oa{M}$, it commutes
with $\fun{\oarepn}{B}$ for every $B$, and the faithfulness gives $AB=BA$; the
center of $\fun{\oaresolventalgebra}{X,\opform{\sigma}}$ consists of the
scalars \cite[remark after Theorem 4.10]{BuchholzGrundling002}.

(2): Fix $\lambda$ and $\gamma$.
The first relation of \eqref{eq:direct-integral-central-probes} is the
convergence of $\fun{\phi_{\oarepn_0}}{f_n}$ to $0$ in the strong resolvent
sense, so that the operators
$\rbk{\imunit\lambda-\fun{\phi_{\oarepn_0}}{f_n}-\fun{m}{\gamma}}^{-1}$ converge strongly to
$\rbk{\imunit\lambda-\fun{m}{\gamma}}^{-1}$.
The proof of Proposition \ref{prop:direct-integral-shifted-quotient}(1) gives
$$\begin{aligned}
\norm{\rbk{\imunit\lambda-\fun{\phi_{\oarepn_0}}{f_n}-\ell_{\gamma}(f_n)}^{-1}-\rbk{\imunit\lambda-\fun{\phi_{\oarepn_0}}{f_n}-\fun{m}{\gamma}}^{-1}}
&\leq
\frac{\abs{\ell_{\gamma}(f_n)-\fun{m}{\gamma}}}{\lambda^{2}},
\end{aligned}$$
which tends to $0$.
Hence, for every $\gamma$,
$$\begin{aligned}
\lim_{n\to\infty}\fun{\oarepn_{\gamma}}{\fun{\oaresolvent}{\lambda,f_n}}\fun{\zeta}{\gamma}
&=
\frac{\fun{\zeta}{\gamma}}{\imunit\lambda-\fun{m}{\gamma}}.
\end{aligned}$$
The function $m$ is a pointwise limit of continuous functions, hence Borel, so
that $C_{\lambda}$ is a bounded operator of norm at most $\abs{\lambda}^{-1}$;
the differences are bounded by $2\abs{\lambda}^{-1}\norm{\fun{\zeta}{\gamma}}$,
and dominated convergence gives the strong convergence.
The von Neumann algebra $\oa{M}$ is strongly closed and contains the operators
$\fun{\oarepn}{\fun{\oaresolvent}{\lambda,f_n}}$, so that
$C_{\lambda}\in\oa{M}$.
Every $\fun{\oarepn}{B}$ acts fiberwise by
\eqref{eq:direct-integral-shifted-representation}, and $C_{\lambda}$ multiplies
each fiber by a scalar, so that $C_{\lambda}$ commutes with $\oarepn$ and belongs
to $\oacommutant{\oa{M}}$.

(3): For the $n$ with $f_n\in Y_{\Gamma}$, outside the union of countably many
null sets, $\ell_{\gamma}(f_n)=\fun{c_{\Gamma}}{f_n}$, so along an infinite set of
such $n$ the function $m$ is almost everywhere equal to the limit of the
constants $\fun{c_{\Gamma}}{f_n}$.
Since $m$ is not almost everywhere constant, the set of such $n$ is finite, and
Theorem \ref{thm:subspace-complement-quotient}(4) with $W=X$ gives
$\fun{\oaresolvent}{\lambda,f_n}\in\oaideal{J}_{X\setminus Y_{\Gamma}}$ for the
other $n$.
Let $c\in\fldcmp$ and let $\zeta_0\in\sphilbfrak{H}_0\setminus\setone{0}$.
The function $\gamma\mapsto\rbk{\imunit\lambda-\fun{m}{\gamma}}^{-1}$ takes the
value $c$ only where $c\neq0$ and $\fun{m}{\gamma}=\imunit\lambda-c^{-1}$, and,
since $m$ is not almost everywhere constant, the set of the other $\gamma$ has
positive $\msrprb_{\Gamma}$-measure.
For the constant section $\fun{\zeta}{\gamma}=\zeta_0$, the section
$\funrbk{\rbk{C_{\lambda}-c}\zeta}{\gamma}=\rbk{\rbk{\imunit\lambda-\fun{m}{\gamma}}^{-1}-c}\zeta_0$
is therefore nonzero on a set of positive measure, so that $C_{\lambda}\neq c$.
Hence $C_{\lambda}$ is a nonscalar element of the center of $\oa{M}$.
\end{proof}

By Proposition \ref{prop:direct-integral-faithful-center}(1), the ideal \(\oaideal{J}_{X\setminus Y_{\Gamma}}\) is not visible on the level of the \(\oacstar\)-algebra: the kernel of \(\oarepn\) is \(\setone{0}\), the state \(\oastate[\psi]\) vanishes at none of the generators of the ideal, and no element of \(\fun{\oarepn}{\fun{\oaresolventalgebra}{X,\opform{\sigma}}}\) other than the scalars lies in the center of \(\oa{M}\). By (2) and (3), the ideal appears in the weak closure when there are directions in which the fields of \(\oarepn_0\) vanish in the limit and the shifts converge to a function of \(\gamma\) that is not \(\msrprb_{\Gamma}\)-almost everywhere constant: the images of generators of the ideal converge to nonscalar central elements, functions of the integration variable. The existence of such directions is a hypothesis on \(\oarepn_0\) and the shifts, not a consequence of \(Y_{\Gamma}\neq X\). In the free Bose gas, such directions are given by the zero-mode probe test functions. This is the opposite of the escape of fields to infinity in Section \ref{sec:escape-to-infinity}, where the limit state vanishes on the ideal of the escaping directions, the ideal lies in the kernel of the GNS representation, and the represented algebra is reduced to the resolvent algebra of the remaining directions (Theorem \ref{thm:escape-to-infinity}).

In the free Bose gas above the critical density \cite{YoshitsuguSekine004}, the fields of the real-valued test functions are of the form of Theorem \ref{thm:shifted-fields-correlated-ideal}, with \(\eta\) the centered nonzero-mode fields, \(\ell\) the condensate parts given by the order coordinate, and \(\mblfmlfrak{S}_0\) the \(\sigma\)-algebra of the order coordinate; the ideal \(\oaideal{J}_{W\setminus Y_0}\) of Theorem \ref{thm:shifted-fields-correlated-ideal} is then the ideal of the condensate directions. On the side of representations, the GNS representation of the BEC state there is the direct integral, over the order coordinate with the standard complex Gaussian measure, of the left Araki--Woods representation shifted by the condensate parts, which is the situation of Proposition \ref{prop:direct-integral-shifted-quotient} with \(\oarepn_0\) the left Araki--Woods representation, \(Y_{\Gamma}\) the test functions on which the zero-mode form vanishes, and \(c_{\Gamma}=0\). The quotient by the ideal of the condensate directions is therefore represented by the representation that agrees with the unshifted Araki--Woods representation, the integrand at the vanishing order coordinate, on the retained directions and sends the resolvents of the condensate directions to \(0\). By Proposition \ref{prop:direct-integral-faithful-center}(1), the GNS representation of the BEC state is faithful, and the ideal of the condensate directions is not visible on the level of the \(\oacstar\)-algebra. The analogue of Proposition \ref{prop:direct-integral-faithful-center}(2)--(3) is proved there for the zero-mode probe test functions, which are condensate directions: their resolvents converge strongly to elements of the center of the weak closure that are functions of the order coordinate, and, under the hypotheses of the central decomposition there, these limits generate the center. The order coordinate, which labels the integrands, thus appears as the center of the weak closure through the generators of the ideal.

\section{Escape of fields to infinity}\label{sec:escape-to-infinity}

In this section, the escape of the fields of some directions to infinity is shown to give in the limit the state that extends the state of the remaining fields and vanishes on the ideal of the escaping directions, the state at which the escaping fields take the value \(\infty\) as in Theorem \ref{thm:random-variable-ideals}(3).

\begin{thm}[Escape to infinity and the limit state]\label{thm:escape-to-infinity}
Let $W\subset X$ be isotropic and $Y\subset W$ a real linear subspace.
For $n$ in a directed set $I$, let $\xi_n$ be a family of real random variables
on $W$ on a probability space $\pairbk{\Omega_n,\mblfmlfrak{S}_n,\msrprb_n}$ with
\eqref{eq:random-variable-linearity}, and let $\xi$ be such a family on $Y$ on
$\pairbk{\Omega,\mblfmlfrak{S},\msrprb}$.
Assume that
\begin{enumerate}
\item $\lim_{n}\fun{\msrprb_n}{\abs{\xi_n(f)}\leq M}=0$ for every
$f\in W\setminus Y$ and every $M>0$;
\item for $g_1,\ldots,g_m\in Y$, the joint laws of
$\rbk{\xi_n(g_1),\ldots,\xi_n(g_m)}$ converge weakly to that of
$\rbk{\xi(g_1),\ldots,\xi(g_m)}$.
\end{enumerate}
Then the states $\oastate[\psi_{\xi_n}]$ of $\oaresolventalgebra(W)$ converge in the
weak-$\ast$ topology to the only state $\oastate[\psi_{\xi,\infty}]$ of
$\oaresolventalgebra(W)$ that extends
the state $\oastate[\psi_{\xi}]$ of $\oaresolventalgebra(Y)$ and vanishes on
$\oaideal{J}_{W\setminus Y}$; its values at the products of resolvents are the
expectations \eqref{eq:random-variable-quotient-state} with $\xi_Y$ of
\eqref{eq:infinite-outside}, and the field laws of the
directions $f\in W\setminus Y$ converge to $\msrdirac_{\infty}$.
The GNS representation $\oarepn_{\psi_{\xi,\infty}}$ of $\oastate[\psi_{\xi,\infty}]$ satisfies
\begin{equation}\label{eq:escape-limit-algebra-reduction}
\begin{aligned}
\oaideal{J}_{W\setminus Y}
&\subset
\Ker\oarepn_{\psi_{\xi,\infty}},
\quad
\fun{\oarepn_{\psi_{\xi,\infty}}}{\oaresolventalgebra(W)}
=
\fun{\oarepn_{\psi_{\xi,\infty}}}{\oaresolventalgebra(Y)}.
\end{aligned}
\end{equation}
\end{thm}

\begin{proof}
For $f\in W\setminus Y$, the bound
$\abs{\fun{r_{\lambda}}{u}}\leq\rbk{\lambda^{2}+u^{2}}^{-\frac{1}{2}}$ gives
$$\begin{aligned}
\abs{\sqfun{\prbexp_{\msrprb_n}}{\fun{r_{\lambda}}{\xi_n(f)}}}
&\leq
\rbk{\lambda^{2}+M^{2}}^{-\frac{1}{2}}+\abs{\lambda}^{-1}\fun{\msrprb_n}{\abs{\xi_n(f)}\leq M}
\end{aligned}$$
for every $M>0$, which tends to $0$ by (1) and the choice of large $M$.
For $f\in Y$, $r_{\lambda}$ is a bounded continuous function, and (2) gives the
convergence to the value of $\oastate[\psi_{\xi}]$.
The state $\oastate[\psi_{\xi,\infty}]$ is the state of
Theorem \ref{thm:subspace-complement-quotient}(3) for $\oastate[\psi_{\xi}]$;
it takes the value $0$ at the resolvents of the vectors of $W\setminus Y$, which
belong to $\oaideal{J}_{W\setminus Y}$, and the value of $\oastate[\psi_{\xi}]$
at those of $Y$.
The resolvents span a dense subspace (Section \ref{sec:setting}), and the
states are bounded by $1$, which gives the weak-$\ast$ convergence to
$\oastate[\psi_{\xi,\infty}]$.
The proof of Theorem \ref{thm:random-variable-ideals}(3), with $\xi$ on $Y$ in
place of the restriction of a family on $W$, gives the values of
$\oastate[\psi_{\xi,\infty}]$ at the products of resolvents.
The field laws converge against $r_{\lambda}$ and $\cmpconj{r_{\lambda}}$ and
hence against all $h\in\conti(\gtoneptcpt{\fldreal})$, by
Lemma \ref{lem:single-field-algebra}(2).
By Theorem \ref{thm:subspace-complement-quotient}(3), the restriction of
$\oarepn_{\psi_{\xi},\infty}$ to $\oaresolventalgebra(W)$ is the GNS
representation of $\oastate[\psi_{\xi,\infty}]$; it vanishes on
$\oaideal{J}_{W\setminus Y}$ by Theorem \ref{thm:subspace-complement-quotient}(2),
and Theorem \ref{thm:subspace-complement-quotient}(1) gives the equality of the
images in \eqref{eq:escape-limit-algebra-reduction}.
\end{proof}

Theorem \ref{thm:escape-to-infinity} is the abstract form of the removal of an infrared cutoff: the fields of the directions of \(W\setminus Y\), in the models directions with divergent cutoff means, escape to infinity, the limit state vanishes on the ideal of these directions, and the directions of \(W\setminus Y\) belong to the set \(X_{\infty}\) of Proposition \ref{prop:infinite-field-reduction}. By \eqref{eq:escape-limit-algebra-reduction}, the ideal of the escaping directions is contained in the kernel of the GNS representation of the limit state, and the represented algebra of observables is reduced from that of \(\oaresolventalgebra(W)\) to that of the resolvent algebra \(\oaresolventalgebra(Y)\) of the directions whose fields stay finite, which Theorem \ref{thm:subspace-complement-quotient}(2) identifies with the quotient \(\setquot{\oaresolventalgebra(W)}{\oaideal{J}_{W\setminus Y}}\). The ideal of the correlated directions of a direct integral behaves in the opposite way (Proposition \ref{prop:direct-integral-faithful-center}): for \(Y_{\Gamma}\neq X\), it is nonzero and not contained in the kernel, and, along a sequence as in Proposition \ref{prop:direct-integral-faithful-center}(2) whose limit shift is not \(\msrprb_{\Gamma}\)-almost everywhere constant, it appears in the center of the weak closure by Proposition \ref{prop:direct-integral-faithful-center}(2)--(3). Combined with Theorem \ref{thm:shifted-fields-correlated-ideal}, a limit in which the escaping directions \(W\setminus Y\) contain all directions whose fields are correlated with the sub-\(\sigma\)-algebra \(\mblfmlfrak{S}_0\) of the shifts, that is, \(Y\subset Y_0\), gives a limit state that carries no information on \(\mblfmlfrak{S}_0\): the ideal of the correlated directions is contained in the ideal of the escaping directions.

\begin{prop}[Domination of the field laws and the absence of escape]\label{prop:field-law-domination}
Let $W\subset X$ be isotropic.
For $n$ in a directed set $I$, let $\xi_n$ be a family of real random variables
on $W$ on a probability space $\pairbk{\Omega_n,\mblfmlfrak{S}_n,\msrprb_n}$ with
\eqref{eq:random-variable-linearity}, and let $\eta$ be such a family on
$\pairbk{\Omega,\mblfmlfrak{S},\msrprb}$.
Assume that
\begin{equation}\label{eq:field-law-domination}
\begin{aligned}
\sqfun{\prbexp_{\msrprb_n}}{\napiernum^{\imunit u\xi_n(f)}}
&\in
\fldreal,
\quad
\sqfun{\prbexp_{\msrprb_n}}{\napiernum^{\imunit u\xi_n(f)}}
\geq
\sqfun{\prbexp_{\msrprb}}{\napiernum^{\imunit u\eta(f)}}
\geq
0
\end{aligned}
\end{equation}
for all $n\in I$, $f\in W$, and $u\in\fldreal$.
\begin{enumerate}
\item For $n\in I$, $f\in W$, and $\lambda\in\fldreal\setminus\setone{0}$,
\begin{equation}\label{eq:field-law-domination-resolvent}
\begin{aligned}
\abs{\fun{\oastate[\psi_{\xi_n}]}{\fun{\oaresolvent}{\lambda,f}}}
&\geq
\abs{\fun{\oastate[\psi_{\eta}]}{\fun{\oaresolvent}{\lambda,f}}}
>
0.
\end{aligned}
\end{equation}
\item Every weak-$\ast$ limit point $\oastate[\psi]$ of the net
$\rbk{\oastate[\psi_{\xi_n}]}_{n\in I}$ of states of $\oaresolventalgebra(W)$
satisfies $\fun{\oastate[\psi]}{\fun{\oaresolvent}{\lambda,f}}\neq0$ for
$f\in W$ and $\lambda\in\fldreal\setminus\setone{0}$; hence
$\oaideal{J}_{W\setminus Y}\not\subset\Ker\oastate[\psi]$ for every real linear
subspace $Y\neq W$.
\item For $f\in W$, $\lambda\in\fldreal\setminus\setone{0}$, $M>0$, and $n\in I$,
\begin{equation}\label{eq:field-law-domination-tightness}
\begin{aligned}
\fun{\msrprb_n}{\abs{\xi_n(f)}\leq M}
&\geq
\abs{\lambda}
\rbk{\abs{\fun{\oastate[\psi_{\eta}]}{\fun{\oaresolvent}{\lambda,f}}}
-\frac{1}{\sqrt{\lambda^{2}+M^{2}}}};
\end{aligned}
\end{equation}
in particular, hypothesis (1) of Theorem \ref{thm:escape-to-infinity} holds for
no $f\in W$.
\end{enumerate}
\end{prop}
\begin{proof}
(1): If the characteristic function of a real random variable $\zeta$ is real,
then $\zeta$ and $-\zeta$ have the same law, and the expectation of the odd part
of $\fun{r_{\lambda}}{u}=\frac{-\imunit\lambda-u}{\lambda^{2}+u^{2}}$ vanishes.
With
$\frac{1}{\lambda^{2}+u^{2}}=\frac{1}{\abs{\lambda}}\int_{0}^{\infty}\napiernum^{-\abs{\lambda}t}\fun{\cos}{tu}\opdmsr{t}$
and Fubini's theorem, this gives
\begin{equation}\label{eq:symmetric-law-resolvent}
\begin{aligned}
\sqfun{\prbexp}{\fun{r_{\lambda}}{\zeta}}
&=
-\frac{\imunit\lambda}{\abs{\lambda}}
\int_{0}^{\infty}\napiernum^{-\abs{\lambda}t}
\sqfun{\prbexp}{\napiernum^{\imunit t\zeta}}
\opdmsr{t}.
\end{aligned}
\end{equation}
By \eqref{eq:field-law-domination}, \eqref{eq:symmetric-law-resolvent} applies
to $\xi_n(f)$ and to $\eta(f)$, and gives the inequality; the right-hand side is
nonzero by Theorem \ref{thm:random-variable-ideals}(4).

(2): The value $\fun{\oastate[\psi]}{\fun{\oaresolvent}{\lambda,f}}$ is a limit
point of the values $\fun{\oastate[\psi_{\xi_n}]}{\fun{\oaresolvent}{\lambda,f}}$,
whose absolute values are bounded below by (1), and the resolvents of the
vectors of $W\setminus Y$ belong to $\oaideal{J}_{W\setminus Y}$.

(3): The bound
$\abs{\sqfun{\prbexp_{\msrprb_n}}{\fun{r_{\lambda}}{\xi_n(f)}}}\leq\rbk{\lambda^{2}+M^{2}}^{-\frac{1}{2}}+\abs{\lambda}^{-1}\fun{\msrprb_n}{\abs{\xi_n(f)}\leq M}$
of the proof of Theorem \ref{thm:escape-to-infinity} and (1) give
\eqref{eq:field-law-domination-tightness}, whose right-hand side is positive for
large $M$ and does not depend on $n$.
\end{proof}

Proposition \ref{prop:field-law-domination} is the counterpart of Theorem \ref{thm:escape-to-infinity}: fields whose laws are symmetric and whose characteristic functions dominate those of a fixed family do not escape, and no limit state vanishes on an ideal \(\oaideal{J}_{W\setminus Y}\) with \(Y\neq W\). The fields coupled to a source with an imaginary coupling constant in Section \ref{sec:gaussian-source} are of this kind.

\section{KMS states and stochastic processes}\label{sec:stochastic-processes}

In this section, the correspondence of Klein and Landau between stochastically positive KMS systems and stochastic processes \cite[Theorem 6.1]{KleinLandau001} is combined with the ideals. The correspondence requires a faithful state on an algebra generated by the time translates of a commutative subalgebra; for the resolvent algebra, this algebra is in general a proper subalgebra, and the state becomes faithful only on the quotient by the kernel of its representation. The Gaussian \(\sminvtemperature\)-Markov path spaces provide the processes for the free dynamics.

We first recall the notions of \cite{KleinLandau001} that are used. Let \(\oa{B}\) be a \(\oacstar\)-algebra, let \(\fml{\alpha_t}{t\in\fldreal}\) be a one-parameter group of \(\ast\)-automorphisms of \(\oa{B}\), with no continuity of \(t\mapsto\fun{\alpha_t}{A}\) assumed, and let \(\sminvtemperature>0\).

\begin{defn}[KMS condition {\cite[Definition 1.1]{KleinLandau001}}]\label{def:kms-condition}
A state $\oastate[\psi]$ of $\oa{B}$ satisfies the KMS condition at
$\sminvtemperature$ relative to $\alpha$ if, for all $A,B\in\oa{B}$, there is a
continuous function $F_{A,B}$ on the closed strip
$\set{z\in\fldcmp}{0\leq\opimag z\leq\sminvtemperature}$, analytic in the open
strip $\set{z\in\fldcmp}{0<\opimag z<\sminvtemperature}$, with
\begin{equation}\label{eq:kms-condition}
\begin{aligned}
\fun{F_{A,B}}{t}
&=
\fun{\oastate[\psi]}{A\fun{\alpha_t}{B}},
\quad
\fun{F_{A,B}}{t+\imunit\sminvtemperature}
=
\fun{\oastate[\psi]}{\fun{\alpha_t}{B}A}
&&\rbk{t\in\fldreal}.
\end{aligned}
\end{equation}
\end{defn}

Let \(\oastate[\psi]\) satisfy the KMS condition at \(\sminvtemperature\) relative to \(\alpha\). Then \(\oastate[\psi]\circ\alpha_t=\oastate[\psi]\) for every \(t\), and there is a unique self-adjoint operator \(\physliouvilean_{\psi}\) on \(\sphilbfrak{H}_{\psi}\) with \begin{equation}\label{eq:kms-liouvillean}
\begin{aligned}
\fun{\oarepn_{\psi}}{\fun{\alpha_t}{A}}
&=
\napiernum^{\imunit t\physliouvilean_{\psi}}\fun{\oarepn_{\psi}}{A}\napiernum^{-\imunit t\physliouvilean_{\psi}},
\quad
\napiernum^{\imunit t\physliouvilean_{\psi}}\oagnsvector[\Psi]
=
\oagnsvector[\Psi]
&&\rbk{A\in\oa{B},\ t\in\fldreal}
\end{aligned}
\end{equation} \cite[Section 1, after Definition 1.1]{KleinLandau001}. For \(A_1,\ldots,A_n\in\oa{B}\), the correlation function \(\pairbk{t_1,\ldots,t_n}\mapsto\fun{\oastate[\psi]}{\fun{\alpha_{t_1}}{A_1}\cdots\fun{\alpha_{t_n}}{A_n}}\) has an analytic continuation \(\Gamma_{\psi;A_1,\ldots,A_n}\) to the imaginary variables \(t_j=\imunit s_j\) with \(s_1\leq\cdots\leq s_n\) and \(s_n-s_1\leq\sminvtemperature\) \cite[Theorem 2.1 and Section 3]{KleinLandau001}. For \(0\leq k\leq n\) and \(-\frac{\sminvtemperature}{2}\leq s_1\leq\cdots\leq s_k\leq0\leq s_{k+1}\leq\cdots\leq s_n\leq\frac{\sminvtemperature}{2}\), the vectors in \eqref{eq:kms-correlation} lie in the domains of the exponentials of \(\physliouvilean_{\psi}\), and \begin{equation}\label{eq:kms-correlation}
\begin{aligned}
&
\fun{\Gamma_{\psi;A_1,\ldots,A_n}}{\imunit s_1,\ldots,\imunit s_n}
\\ %%%%%%%%%%%%%%%%
&=
\Bigl\langle
\napiernum^{s_k\physliouvilean_{\psi}}\fun{\oarepn_{\psi}}{\faadj{A_k}}
\napiernum^{-\rbk{s_k-s_{k-1}}\physliouvilean_{\psi}}\fun{\oarepn_{\psi}}{\faadj{A_{k-1}}}
\cdots
\napiernum^{-\rbk{s_2-s_1}\physliouvilean_{\psi}}\fun{\oarepn_{\psi}}{\faadj{A_1}}\oagnsvector[\Psi],
\\ %%%%%%%%%%%%%%%%
&\qquad
\napiernum^{-s_{k+1}\physliouvilean_{\psi}}\fun{\oarepn_{\psi}}{A_{k+1}}
\napiernum^{-\rbk{s_{k+2}-s_{k+1}}\physliouvilean_{\psi}}\fun{\oarepn_{\psi}}{A_{k+2}}
\cdots
\napiernum^{-\rbk{s_n-s_{n-1}}\physliouvilean_{\psi}}\fun{\oarepn_{\psi}}{A_n}\oagnsvector[\Psi]
\Bigr\rangle
\end{aligned}
\end{equation} \cite[(3.1)]{KleinLandau001}, where the left vector is \(\oagnsvector[\Psi]\) for \(k=0\) and the right vector is \(\oagnsvector[\Psi]\) for \(k=n\).

\begin{defn}[Stochastically positive KMS systems {\cite[Definitions 3.1 and 3.2]{KleinLandau001}}]\label{def:stochastic-positivity}
Let $\oastate[\psi]$ be a state of $\oa{B}$ that satisfies the KMS condition
at $\sminvtemperature$ relative to $\alpha$, and let $\oa{A}\subset\oa{B}$ be a
commutative $\oacstar$-subalgebra.
\begin{enumerate}
\item The state $\oastate[\psi]$ is stochastically positive with respect to
$\oa{A}$ if the values \eqref{eq:kms-correlation} are nonnegative for every $n$,
all positive $A_1,\ldots,A_n\in\oa{A}$, and all $s_1,\ldots,s_n$ as there.
\item The quadruple $\rbk{\oa{B},\oa{A},\alpha,\oastate[\psi]}$ is a
stochastically positive KMS system at $\sminvtemperature$ if
$\bigcup_{t\in\fldreal}\fun{\alpha_t}{\oa{A}}$ generates $\oa{B}$ as a
$\oacstar$-algebra, $\oastate[\psi]$ is faithful, and $\oastate[\psi]$ is
stochastically positive with respect to $\oa{A}$.
\end{enumerate}
\end{defn}

\subsection{The algebra generated by the time translates of an isotropic field algebra}\label{sec:time-translates}

Let \(\fml{T_t}{t\in\fldreal}\) be a one-parameter group of symplectic linear bijections of \(X\), let \(\alpha_t=\alpha_{T_t}\), let \(Z\subset X\) be isotropic, and let \(\oa{B}_Z\) be the \(\oacstar\)-subalgebra generated by \(\bigcup_{t\in\fldreal}\fun{\alpha_t}{\oaresolventalgebra(Z)}\). Since \(\fun{\alpha_t}{\oaresolventalgebra(Z)}=\oaresolventalgebra(T_tZ)\), the algebra \(\oa{B}_Z\) is the norm closure of the polynomials in the resolvents of the vectors of \(\bigcup_{t\in\fldreal}T_tZ\), it contains \(\idone\), and \(\fun{\alpha_s}{\oa{B}_Z}=\oa{B}_Z\) for every \(s\).

\begin{prop}[Generation by time translates]\label{prop:time-translates-generation}
\begin{enumerate}
\item If $f\in X$ belongs to no $T_tZ$, $t\in\fldreal$, then
$\fun{\oaresolvent}{\lambda,f}\notin\oa{B}_Z$ for every
$\lambda\in\fldreal\setminus\setone{0}$.
\item If every $f\in X$ belongs to some $T_tZ$, then
$\oa{B}_Z=\fun{\oaresolventalgebra}{X,\opform{\sigma}}$.
\item If $\dim X=2n<\infty$ with $n\geq2$ and $t\mapsto T_t$ is continuous, then
$\bigcup_{t\in\fldreal}T_tZ\neq X$.
\item If the linear span of $\bigcup_{t\in\fldreal}T_tZ$ is $X$, then
$\fun{\oarepn}{\fun{\oaresolvent}{\lambda,f}}$ belongs to the von Neumann algebra
generated by $\fun{\oarepn}{\oa{B}_Z}$ for every representation $\oarepn$ that is
regular on $X$, every $\lambda\in\fldreal\setminus\setone{0}$, and every $f\in X$.
\end{enumerate}
\end{prop}

\begin{proof}
(1): The characters $\chi_{f,1},\chi_{f,\infty}\in\Delta_X$ of
Section \ref{sec:characters} take the same value $\fun{r_{\mu}}{\infty}$ at
$\fun{\oaresolvent}{\mu,g}$ for $g\in T_tZ\setminus\setone{0}$, since such a $g$ is
not a multiple of $f$, and the value $\fun{r_{\mu}}{0}$ for $g=0$; hence they
agree on $\oa{B}_Z$, while
$\fun{\chi_{f,1}}{\fun{\oaresolvent}{\lambda,f}}=\fun{r_{\lambda}}{1}\neq0=\fun{\chi_{f,\infty}}{\fun{\oaresolvent}{\lambda,f}}$.

(2): Every generator $\fun{\oaresolvent}{\lambda,f}$ then lies in some
$\oaresolventalgebra(T_tZ)=\fun{\alpha_t}{\oaresolventalgebra(Z)}\subset\oa{B}_Z$,
and the $\oacstar$-subalgebra $\oa{B}_Z$ is the whole algebra.

(3): An isotropic subspace has dimension $m\leq n$ by the nondegeneracy of
$\opform{\sigma}$.
The continuous one-parameter group of matrices is $T_t=\napiernum^{tA}$, and,
for a basis $z_1,\ldots,z_m$ of $Z$, the map
$\pairbk{t,c}\mapsto\napiernum^{tA}\sum_jc_jz_j$ from $\fldreal^{1+m}$ to $X$ is
continuously differentiable with $1+m<2n$; its image
$\bigcup_{t}T_tZ$ is therefore a Lebesgue null set and not $X$.

(4): By \cite[Theorem 4.2(v)]{BuchholzGrundling002}, for $f,h\in X$ and
$\nu\in\fldreal$,
$\fun{\oarepn}{\fun{\oaresolvent}{1,\nu f+h}}$ lies in the double commutant of
$\setone{\fun{\oarepn}{\fun{\oaresolvent}{1,f}},\fun{\oarepn}{\fun{\oaresolvent}{1,h}}}$.
Induction on the number of terms of a linear combination of vectors of
$\bigcup_tT_tZ$ gives the statement for $\lambda=1$, and relation
\eqref{eq:resolvent-relation-scaling} with $\nu=\lambda^{-1}$,
$\fun{\oaresolvent}{\lambda,f}=\lambda^{-1}\fun{\oaresolvent}{1,\lambda^{-1}f}$,
gives it for every $\lambda$.
\end{proof}

By Proposition \ref{prop:time-translates-generation}(3), for a finite-dimensional system with at least two degrees of freedom and a continuous group, the time translates of the algebra of an isotropic subspace never generate the resolvent algebra as a \(\oacstar\)-algebra, while by (4), when the linear span of \(\bigcup_{t\in\fldreal}T_tZ\) is \(X\), they generate it on the level of the von Neumann algebras of regular representations.

\subsection{Reduction of KMS states}\label{sec:kms-reduction}

Let \(\sminvtemperature>0\), and let \(\oastate[\psi]\) be a state of \(\fun{\oaresolventalgebra}{X,\opform{\sigma}}\) that satisfies the KMS condition of Definition \ref{def:kms-condition} at \(\sminvtemperature\) relative to \(\alpha\), which does not require the continuity of \(t\mapsto\fun{\alpha_t}{A}\) in norm. Let \(\oastate[\psi_Z]\) and \(\alpha_{Z,t}\) be the restrictions of \(\oastate[\psi]\) and \(\alpha_t\) to \(\oa{B}_Z\), let \(\oaideal{N}_Z=\Ker\oarepn_{\psi}\cap\oa{B}_Z\), let \(\widetilde{\oa{B}}_Z\) be the quotient \(\oacstar\)-algebra \(\setquot{\oa{B}_Z}{\oaideal{N}_Z}\) with the quotient map \(q_Z\colon\oa{B}_Z\to\widetilde{\oa{B}}_Z\), and let \(\msrprb_{\psi,Z}\) be the Radon probability measure on the spectrum \(\Delta_Z\) of Section \ref{sec:characters} of the commutative algebra \(\oaresolventalgebra(Z)\) that represents the restriction of \(\oastate[\psi]\), with the Gelfand transform \(a\mapsto\hat{a}\).

\begin{prop}[Reduction of a KMS state]\label{prop:kms-reduction}
\begin{enumerate}
\item The state $\oastate[\psi_Z]$ of $\oa{B}_Z$ satisfies the KMS condition of
Definition \ref{def:kms-condition} at $\sminvtemperature$ relative to
$\fml{\alpha_{Z,t}}{t\in\fldreal}$, and
$$\begin{aligned}
\Ker\oarepn_{\psi_Z}
=
\set{b\in\oa{B}_Z}{\fun{\oastate[\psi]}{\faadj{b}b}=0}
=
\oaideal{N}_Z,
\end{aligned}$$
which is invariant under every $\alpha_t$.
\item For $t\in\fldreal$, a $\ast$-automorphism $\widetilde{\alpha}_t$ of
$\widetilde{\oa{B}}_Z$ is well defined by
\begin{equation}\label{eq:kms-reduction-quotient-dynamics}
\begin{aligned}
\fun{\widetilde{\alpha}_t}{\fun{q_Z}{b}}
&=
\fun{q_Z}{\fun{\alpha_{Z,t}}{b}}
&&\rbk{b\in\oa{B}_Z},
\end{aligned}
\end{equation}
and $\widetilde{\alpha}=\fml{\widetilde{\alpha}_t}{t\in\fldreal}$ is a one-parameter
group of $\ast$-automorphisms of $\widetilde{\oa{B}}_Z$.
The image $\fun{q_Z}{\oaresolventalgebra(Z)}$ is a commutative unital
$\oacstar$-subalgebra of $\widetilde{\oa{B}}_Z$, and its time translates under
$\widetilde{\alpha}$ generate $\widetilde{\oa{B}}_Z$.
\item A state $\oastate[\widetilde{\psi}_Z]$ of the quotient
$\widetilde{\oa{B}}_Z$ is well defined by
\begin{equation}\label{eq:kms-reduction-quotient-state}
\begin{aligned}
\fun{\oastate[\widetilde{\psi}_Z]}{\fun{q_Z}{b}}
&=
\fun{\oastate[\psi_Z]}{b}
&&\rbk{b\in\oa{B}_Z},
\end{aligned}
\end{equation}
and it is faithful and satisfies the KMS condition of
Definition \ref{def:kms-condition} at $\sminvtemperature$ relative to the group
$\widetilde{\alpha}$ of \eqref{eq:kms-reduction-quotient-dynamics}.
\item A map $\tau_Z$ from $\fun{q_Z}{\oaresolventalgebra(Z)}$ to
$\fun{\conti}{\supp\msrprb_{\psi,Z}}$ is well defined by
\begin{equation}\label{eq:kms-reduction-gelfand-map}
\begin{aligned}
\fun{\tau_Z}{\fun{q_Z}{a}}
&=
\fnrestr{\hat{a}}{\supp\msrprb_{\psi,Z}}
&&\rbk{a\in\oaresolventalgebra(Z)},
\end{aligned}
\end{equation}
and it is a $\ast$-isomorphism of $\fun{q_Z}{\oaresolventalgebra(Z)}$ onto
$\fun{\conti}{\supp\msrprb_{\psi,Z}}$; in particular,
$\oaresolventalgebra(Z)\cap\Ker\oarepn_{\psi}$ is the set of $a$ with
$\hat{a}=0$ on $\supp\msrprb_{\psi,Z}$.
\end{enumerate}
\end{prop}

\begin{proof}
(1): For $A,B\in\oa{B}_Z$, let $F_{A,B}$ be the function of
\eqref{eq:kms-condition} for $\oastate[\psi]$.
Since $\fun{\alpha_t}{\oa{B}_Z}=\oa{B}_Z$, its boundary values are
$\fun{F_{A,B}}{t}=\fun{\oastate[\psi_Z]}{A\fun{\alpha_{Z,t}}{B}}$ and
$\fun{F_{A,B}}{t+\imunit\sminvtemperature}=\fun{\oastate[\psi_Z]}{\fun{\alpha_{Z,t}}{B}A}$,
which is \eqref{eq:kms-condition} for $\oastate[\psi_Z]$ on $\oa{B}_Z$.
By \cite[Proposition 1.4]{KleinLandau001}, for a state $\oastate[\psi]$ that
satisfies the KMS condition, the vector $\oagnsvector[\Psi]$ of
\eqref{eq:gns-vector-state} is separating for $\fun{\oarepn_{\psi}}{\oa{B}}$.
Hence $\fun{\oarepn_{\psi}}{b}=0$ if and only if
$\fun{\oarepn_{\psi}}{b}\oagnsvector[\Psi]=0$, that is,
$\fun{\oastate[\psi]}{\faadj{b}b}=\norm{\fun{\oarepn_{\psi}}{b}\oagnsvector[\Psi]}^{2}=0$, and
this closed two-sided ideal is invariant under every $\alpha_t$, since
$\oastate[\psi]\circ\alpha_t=\oastate[\psi]$.
Applied to $\oastate[\psi]$ and to $\oastate[\psi_Z]$, this gives (1), since
$\fun{\oastate[\psi_Z]}{\faadj{b}b}=\fun{\oastate[\psi]}{\faadj{b}b}$ for
$b\in\oa{B}_Z$.

(2): Invariance of $\oaideal{N}_Z$ makes $\widetilde{\alpha}_t$ well defined;
$\fun{q_Z}{\oaresolventalgebra(Z)}$ is closed as the image of a $\ast$-homomorphism
of $\oacstar$-algebras; the preimage under $q_Z$ of the $\oacstar$-algebra
generated by its time translates contains the time translates of
$\oaresolventalgebra(Z)$ and is therefore $\oa{B}_Z$.

(3): The state is well defined because $\oastate[\psi_Z]$ vanishes on
$\oaideal{N}_Z$, and it is faithful by (1).
For $A,B\in\oa{B}_Z$, the function $F_{A,B}$ of (1) satisfies
\eqref{eq:kms-condition} for $\oastate[\widetilde{\psi}_Z]$ with
$\fun{q_Z}{A},\fun{q_Z}{B}$ and $\widetilde{\alpha}$ in place of $A,B$ and
$\alpha$, since
$\fun{\oastate[\widetilde{\psi}_Z]}{\fun{q_Z}{A}\fun{\widetilde{\alpha}_t}{\fun{q_Z}{B}}}=\fun{\oastate[\psi_Z]}{A\fun{\alpha_{Z,t}}{B}}$
and
$\fun{\oastate[\widetilde{\psi}_Z]}{\fun{\widetilde{\alpha}_t}{\fun{q_Z}{B}}\fun{q_Z}{A}}=\fun{\oastate[\psi_Z]}{\fun{\alpha_{Z,t}}{B}A}$.

(4): By (1), $a\in\Ker\oarepn_{\psi}$ if and only if
$\sqfun{\prbexp_{\msrprb_{\psi,Z}}}{\abs{\hat{a}}^{2}}=0$, that is, $\hat{a}=0$ on
the support; hence $\tau_Z$ is well defined and injective, and its range is dense
by the Stone--Weierstrass theorem and closed.
\end{proof}

By Proposition \ref{prop:kms-reduction}(2)--(3), the quadruple \(\rbk{\widetilde{\oa{B}}_Z,\fun{q_Z}{\oaresolventalgebra(Z)},\widetilde{\alpha},\oastate[\widetilde{\psi}_Z]}\) satisfies all conditions of Definition \ref{def:stochastic-positivity}(2) except the stochastic positivity, which is a condition on the state. The following corollary gives the stochastic process of the reduced system and the resolvents of \(\oaresolventalgebra(Z)\) that the reduction removes.

\begin{cor}[Stochastic processes of reduced KMS states]\label{cor:kms-reduction-process}
In the setting of Proposition \ref{prop:kms-reduction}, the following hold.
\begin{enumerate}
\item For $f\in Z\setminus\setone{0}$ and $\lambda\in\fldreal\setminus\setone{0}$,
$\fun{\oaresolvent}{\lambda,f}\in\Ker\oarepn_{\psi}$ if and only if
\begin{equation}\label{eq:kms-infinite-field}
\begin{aligned}
\fun{\msrprb_{\psi,Z}}{\set{\chi\in\Delta_Z}{\fun{\chi}{\fun{\oaresolvent}{1,f}}=0}}
&=
1.
\end{aligned}
\end{equation}
\item If $\oastate[\widetilde{\psi}_Z]$ is stochastically positive with respect
to $\fun{q_Z}{\oaresolventalgebra(Z)}$, then
$\rbk{\widetilde{\oa{B}}_Z,\fun{q_Z}{\oaresolventalgebra(Z)},\widetilde{\alpha},\oastate[\widetilde{\psi}_Z]}$
is a stochastically positive KMS system at $\sminvtemperature$, and there are a
probability space $\pairbk{\Omega,\mblfmlfrak{S},\msrprb}$ and random variables
$\xi_t\colon\Omega\to\supp\msrprb_{\psi,Z}$, $t\in\fldreal$, such that the law
of every $\xi_t$ is $\msrprb_{\psi,Z}$ and
\begin{equation}\label{eq:kms-process-correlation}
\begin{aligned}
\sqfun{\prbexp_{\msrprb}}{\fun{\hat{a}_1}{\xi_{t_1}}\cdots\fun{\hat{a}_n}{\xi_{t_n}}}
&=
\fun{\Gamma_{\widetilde{\psi}_Z;\fun{q_Z}{a_1},\ldots,\fun{q_Z}{a_n}}}{\imunit t_1,\ldots,\imunit t_n}
\end{aligned}
\end{equation}
for all $a_1,\ldots,a_n\in\oaresolventalgebra(Z)$ and
$t_1\leq\cdots\leq t_n$ with $t_n-t_1\leq\sminvtemperature$.
\end{enumerate}
\end{cor}

\begin{proof}
(1): Since $f\in Z$, the resolvent $\fun{\oaresolvent}{\lambda,f}$ belongs to
$\oaresolventalgebra(Z)$, and by Proposition \ref{prop:kms-reduction}(4) it
belongs to $\Ker\oarepn_{\psi}$ if and only if its Gelfand transform
$\chi\mapsto\fun{\chi}{\fun{\oaresolvent}{\lambda,f}}$ vanishes on
$\supp\msrprb_{\psi,Z}$.
By Proposition \ref{prop:character-classification}, every $\chi\in\Delta_Z$ is
the restriction of $\chi_{Z',\ell}$ for a subspace $Z'\subset Z$ and a linear
$\ell\colon Z'\to\fldreal$, and by \eqref{eq:character-values} and
\eqref{eq:resolvent-value-function},
$\fun{\chi}{\fun{\oaresolvent}{\lambda,f}}=\fun{r_{\lambda}}{\fun{\ell}{f}}\neq0$
for $f\in Z'$ and $\fun{\chi}{\fun{\oaresolvent}{\lambda,f}}=\fun{r_{\lambda}}{\infty}=0$
for $f\notin Z'$.
Hence the zero set of the Gelfand transform is the set in
\eqref{eq:kms-infinite-field} for every $\lambda$, and it is closed, since
$\chi\mapsto\fun{\chi}{\fun{\oaresolvent}{1,f}}$ is weak-$\ast$ continuous.
Its complement is open, and an open set has measure zero exactly when it does
not meet the support; hence the Gelfand transform vanishes on
$\supp\msrprb_{\psi,Z}$ if and only if \eqref{eq:kms-infinite-field} holds.

(2): By Proposition \ref{prop:kms-reduction}(2), $\fun{q_Z}{\oaresolventalgebra(Z)}$
is a commutative $\oacstar$-subalgebra of $\widetilde{\oa{B}}_Z$ whose time translates under
$\widetilde{\alpha}$ generate $\widetilde{\oa{B}}_Z$, and by
Proposition \ref{prop:kms-reduction}(3), $\oastate[\widetilde{\psi}_Z]$ is a
faithful state that satisfies the KMS condition at $\sminvtemperature$ relative to
$\widetilde{\alpha}$; with the stochastic positivity, the quadruple is a
stochastically positive KMS system at $\sminvtemperature$ by
Definition \ref{def:stochastic-positivity}(2).
By \cite[Theorem 6.1]{KleinLandau001}, there is a stochastic process
$\fml{X_t}{t\in\fldreal}$ on a probability space
$\pairbk{\Omega,\mblfmlfrak{S},\msrprb}$ with values in the spectrum $K$ of
$\fun{q_Z}{\oaresolventalgebra(Z)}$ such that
$\sqfun{\prbexp_{\msrprb}}{\fun{A_1}{X_{t_1}}\cdots\fun{A_n}{X_{t_n}}}=\fun{\Gamma_{\widetilde{\psi}_Z;A_1,\ldots,A_n}}{\imunit t_1,\ldots,\imunit t_n}$
for $A_1,\ldots,A_n\in\fun{q_Z}{\oaresolventalgebra(Z)}$, identified with
their Gelfand transforms on $K$, and for $t_1\leq\cdots\leq t_n$ with
$t_n-t_1\leq\sminvtemperature$.
By Proposition \ref{prop:kms-reduction}(4), $\tau_Z$ is a $\ast$-isomorphism of
$\fun{q_Z}{\oaresolventalgebra(Z)}$ onto $\fun{\conti}{\supp\msrprb_{\psi,Z}}$,
so that there is a homeomorphism $h\colon K\to\supp\msrprb_{\psi,Z}$ with
$\fun{A}{x}=\fun{\fun{\tau_Z}{A}}{\fun{h}{x}}$ for
$A\in\fun{q_Z}{\oaresolventalgebra(Z)}$ and $x\in K$.
The random variables $\xi_t=h\circ X_t$ satisfy
\eqref{eq:kms-process-correlation}, since
$\fun{\tau_Z}{\fun{q_Z}{a}}=\fnrestr{\hat{a}}{\supp\msrprb_{\psi,Z}}$.
For $n=1$, the function
$t\mapsto\fun{\oastate[\widetilde{\psi}_Z]}{\fun{\widetilde{\alpha}_t}{\fun{q_Z}{a}}}=\fun{\oastate[\psi]}{a}$
is constant, so that its analytic continuation is the constant
$\fun{\oastate[\psi]}{a}=\sqfun{\prbexp_{\msrprb_{\psi,Z}}}{\hat{a}}$; hence
$\sqfun{\prbexp_{\msrprb}}{\fun{\hat{a}}{\xi_t}}=\sqfun{\prbexp_{\msrprb_{\psi,Z}}}{\hat{a}}$
for every $a\in\oaresolventalgebra(Z)$, and the law of $\xi_t$ is
$\msrprb_{\psi,Z}$ by the Riesz representation theorem, since the restrictions
$\fnrestr{\hat{a}}{\supp\msrprb_{\psi,Z}}$ exhaust
$\fun{\conti}{\supp\msrprb_{\psi,Z}}$.
\end{proof}

By Corollary \ref{cor:kms-reduction-process}(2), the process sees only the quotient \(\widetilde{\oa{B}}_Z\), and its correlations are the analytic continuations of the functions \(\pairbk{t_1,\ldots,t_n}\mapsto\fun{\oastate[\psi]}{\fun{\alpha_{t_1}}{a_1}\cdots\fun{\alpha_{t_n}}{a_n}}\), which are the values of \(\fun{\oastate[\widetilde{\psi}_Z]}{\fun{\widetilde{\alpha}_{t_1}}{\fun{q_Z}{a_1}}\cdots\fun{\widetilde{\alpha}_{t_n}}{\fun{q_Z}{a_n}}}\) at real times. By Proposition \ref{prop:character-classification}, a character \(\chi\in\Delta_Z\) assigns a real value to the fields of a subspace of \(Z\) and the value \(\infty\) to the others, and the process takes values in these assignments. By Corollary \ref{cor:kms-reduction-process}(1), the resolvents of \(\oaresolventalgebra(Z)\) removed by the reduction are those of the directions \(f\) whose field is infinite at almost every configuration of the one-time law.

\subsection{\texorpdfstring{Gaussian \(\sminvtemperature\)-Markov path spaces}{Gaussian \textbackslash sminvtemperature-Markov path spaces}}\label{sec:beta-markov-path-spaces}

Let \(\sminvtemperature>0\), let \(\sphilbfrak{h}\neq\setone{0}\) be a separable complex Hilbert space, and let \(\physham[h]_{\txtbsn}\) be an injective nonnegative self-adjoint operator on \(\sphilbfrak{h}\), a one-particle Hamiltonian possibly without a spectral gap \cite[Definition 2.1]{YoshitsuguSekine018}. The singular Gaussian \(\sminvtemperature\)-Markov path space of \cite[Definition 4.4 and Theorem 4.6]{YoshitsuguSekine018}, which extends \cite[Section 21.4]{DerezinskiGerard001}, realizes the left Araki--Woods representation \cite[Theorem 5.4]{YoshitsuguSekine018}. Let \(\pairbk{\prbqspace,\mblfmlfrak{S},\msrprb_{\sminvtemperature,0}}\) be this path space, with the time circle \(S_{\sminvtemperature}=\fldreal/\sminvtemperature\ringratint\) and the Gaussian field \(\opfocksegal\) of \cite[Sections 2.2 and 4.1]{YoshitsuguSekine018}. Let \(\sphilbfrak{h}_{\fldreal}\) be \(\sphilbfrak{h}\) regarded as a real Hilbert space with the inner product \(\opreal\bkt{f}{g}\), let \(\smnumberdensity_{\sminvtemperature}=\rbk{\napiernum^{\sminvtemperature\physham[h]_{\txtbsn}}-1}^{-1}\) be the Planck density, and let \(\sphilbfrak{h}_{\fldreal,\smnumberdensity}=\sphilbfrak{h}_{\fldreal}\cap\dom\smnumberdensity_{\sminvtemperature}^{\frac{1}{2}}\) \cite[(26)]{YoshitsuguSekine018}. For the isometry \(\iota_{\physham[h]_{\txtbsn}}\) of \cite[Lemma 5.2]{YoshitsuguSekine018} and the sharp-time vectors \(j_{\sminvtemperature,t}\), written \(j_t\) in \cite[Definition 3.3]{YoshitsuguSekine018}, the sharp-time field of \(f\in\sphilbfrak{h}_{\fldreal,\smnumberdensity}\) at \(t\in S_{\sminvtemperature}\) is the Segal field \(\fun{\opfocksegal}{j_{\sminvtemperature,t}\fun{\iota_{\physham[h]_{\txtbsn}}}{f}}\) of \cite[Corollary 5.7]{YoshitsuguSekine018}, and let \(\oastate[\psi_{\txtarakiwoods}]\) be the state of the resolvent algebra given by the Araki--Woods vacuum vector in the left representation of \cite[Theorem 5.4]{YoshitsuguSekine018}.

\begin{prop}[Sharp-time fields and the Araki--Woods state]\label{prop:path-space-random-variables}
Let $\sphilbfrak{w}\subset\sphilbfrak{h}_{\fldreal,\smnumberdensity}$ be an
isotropic subspace,
and for $t\in S_{\sminvtemperature}$ let
$\xi_t(f)=\fun{\opfocksegal}{j_{\sminvtemperature,t}\fun{\iota_{\physham[h]_{\txtbsn}}}{f}}$,
$f\in\sphilbfrak{w}$.
\begin{enumerate}
\item The family $\xi_t$ satisfies \eqref{eq:random-variable-linearity}, and the
state $\oastate[\psi_{\xi_t}]$ of Theorem \ref{thm:random-variable-ideals}(1) is the
restriction of $\oastate[\psi_{\txtarakiwoods}]$ to
$\oaresolventalgebra(\sphilbfrak{w})$ for
every $t$.
\item Consequently, Theorems \ref{thm:random-variable-ideals},
\ref{thm:random-variable-gns}, and \ref{thm:ideal-inclusion-sigma-independence}
hold for the sharp-time fields at any fixed time, with $W=\sphilbfrak{w}$: the
resolvent subalgebra of a
subspace $\sphilbfrak{y}\subset\sphilbfrak{w}$ corresponds to the
$\sigma$-algebra of the fields of $\sphilbfrak{y}$ at
the time $t$, and the quotient by
$\oaideal{J}_{\sphilbfrak{w}\setminus\sphilbfrak{y}}$ to the restriction to
it with the value $\infty$ for the other fields.
\item For $0\leq t_1\leq\cdots\leq t_n\leq\frac{\sminvtemperature}{2}$, $f_j\in\sphilbfrak{w}$,
and $\lambda_j\in\fldreal\setminus\setone{0}$, the expectations of the products
$\prod_j\fun{r_{\lambda_j}}{\xi_{t_j}(f_j)}$ are the imaginary-time correlations
of the resolvents in the Araki--Woods representation
\cite[Proposition 5.6]{YoshitsuguSekine018}.
\end{enumerate}
\end{prop}

\begin{proof}
(1): The fields are real linear in $f$ almost everywhere as values of a linear
map into $\lp^{2}$ \cite[Corollary 5.7]{YoshitsuguSekine018}.
The case of equal times $t_1=\cdots=t_n=0$ of
\cite[Proposition 5.6]{YoshitsuguSekine018} gives the statement for $t=0$, and
the invariance of $\msrprb_{\sminvtemperature,0}$ under the time translations,
which map the fields at $0$ to those at $t$, gives it for every $t$.

(2): By (1), the theorems apply to the family $\xi_t$ with $W=\sphilbfrak{w}$.

(3): This is \cite[Proposition 5.6]{YoshitsuguSekine018}.
\end{proof}

By Proposition \ref{prop:path-space-random-variables}, the general dictionary of Sections \ref{sec:random-variables}--\ref{sec:escape-to-infinity} holds on the path space of a free Bose field at positive temperature, and, by (3), the path space carries the correlations of all imaginary times, as the process of Klein and Landau does for the algebra generated by the time translates of \(\oaresolventalgebra(\sphilbfrak{w})\).

\subsection{Gaussian fields coupled to a source}\label{sec:gaussian-source}

In this section, a Gaussian field is coupled to a source depending on the paths of a second system, with a real or with an imaginary coupling constant; these are the functional integrals of the linear couplings and of the minimal coupling in the models of Section \ref{sec:applications}. Let \(\sphilbfrak{k}\) be a real vector space with a nonnegative symmetric bilinear form \(\opform{q}\), written \(\fun{\opform{q}}{F}=\fun{\opform{q}}{F,F}\), and let \(\pairbk{\prbqspace,\mblfmlfrak{S}_{2},\msrprb_{2}}\) be a probability space with a family \(\opfocksegal=\fml{\opfocksegal(F)}{F\in\sphilbfrak{k}}\) of jointly Gaussian centered real random variables, real linear in \(F\) almost everywhere, with \begin{equation}\label{eq:gaussian-source-covariance}
\begin{aligned}
\sqfun{\prbexp_{\msrprb_{2}}}{\opfocksegal(F)\opfocksegal(G)}
&=
\onehalf\fun{\opform{q}}{F,G}
\quad
\rbk{F,G\in\sphilbfrak{k}}.
\end{aligned}
\end{equation} Let \(\pairbk{\Omega_{1},\mblfmlfrak{S}_{1},\msrprb_{1}}\) be a probability space, the paths of the second system, and \(\mathsf{J}\colon\Omega_{1}\to\sphilbfrak{k}\) a source such that \(\fun{\opform{q}}{F,\mathsf{J}}\) for \(F\in\sphilbfrak{k}\) and \(\fun{\opform{q}}{\mathsf{J}}\) are \(\mblfmlfrak{S}_{1}\)-measurable, and assume that a measurable function \(\opfocksegal_{\mathsf{J}}\) on \(\Omega_{1}\times\prbqspace\) satisfies \(\opfocksegal_{\mathsf{J}}(\omega,\cdot)=\opfocksegal(\mathsf{J}(\omega))\) \(\msrprb_{2}\)-almost everywhere for \(\msrprb_{1}\)-almost every \(\omega\). In the models, \(\sphilbfrak{k}\) is a space of Euclidean test functions of a Gaussian path space as in Section \ref{sec:beta-markov-path-spaces}, \(\mathsf{J}\) is the source along the path of the particle or of the spin, and \(\opfocksegal_{\mathsf{J}}\) is the field integrated along the path. Let \(W\subset X\) be isotropic and \(T\colon W\to\sphilbfrak{k}\) real linear, the embedding of the fields at a fixed time, and regard random variables on \(\Omega_{1}\) and on \(\prbqspace\) as random variables on \(\Omega_{1}\times\prbqspace\).

\begin{prop}[Gaussian fields coupled to a source]\label{prop:gaussian-source}
In the setting of Section \ref{sec:gaussian-source}, let
$\theta(f)=\fun{\opform{q}}{Tf,\mathsf{J}}$ for $f\in W$.
\begin{enumerate}
\item Let $c\in\fldreal$ with
$Z_{c}=\sqfun{\prbexp_{\msrprb_{1}}}{\napiernum^{\frac{c^{2}}{4}\fun{\opform{q}}{\mathsf{J}}}}<\infty$.
Then
$\opdmsr{\msrprb_{c}}=\frac{1}{Z_{c}}\napiernum^{c\opfocksegal_{\mathsf{J}}}\opdmsr{\rbk{\msrprb_{1}\otimes\msrprb_{2}}}$
is a probability measure, and under $\msrprb_{c}$ the fields
$\xi(f)=\opfocksegal(Tf)$ are $\xi(f)=\eta(f)+\frac{c}{2}\theta(f)$, where
$\eta(f)=\xi(f)-\frac{c}{2}\theta(f)$ has the joint laws of
$\opfocksegal(Tf)$ under $\msrprb_{2}$ and $\mblfmlfrak{S}_{\eta}(W)$ is
independent of $\mblfmlfrak{S}_{1}$.
Theorem \ref{thm:shifted-fields-correlated-ideal} applies with
$\mblfmlfrak{S}_{0}=\mblfmlfrak{S}_{1}$ and $\ell=\frac{c}{2}\theta$.
\item Let $c\in\fldreal\setminus\setone{0}$, let
$\opdmsr{\widetilde{\msrprb}_{1}}=\frac{1}{Z}\napiernum^{-\frac{c^{2}}{4}\fun{\opform{q}}{\mathsf{J}}}\opdmsr{\msrprb_{1}}$
with
$Z=\sqfun{\prbexp_{\msrprb_{1}}}{\napiernum^{-\frac{c^{2}}{4}\fun{\opform{q}}{\mathsf{J}}}}$,
and let $\xi$ be a family of real random variables on $W$ with
\eqref{eq:random-variable-linearity} and
\begin{equation}\label{eq:phase-coupled-characteristic}
\begin{aligned}
\sqfun{\prbexp}{\napiernum^{\imunit u\xi(f)}}
&=
\frac{1}{Z}
\sqfun{\prbexp_{\msrprb_{1}\otimes\msrprb_{2}}}
{\napiernum^{\imunit u\opfocksegal(Tf)-\imunit c\opfocksegal_{\mathsf{J}}}}
\quad
\rbk{u\in\fldreal,\ f\in W}.
\end{aligned}
\end{equation}
Then the following hold for $f\in W$ and $u\in\fldreal$.
\begin{enumerate}
\item
It holds that
$$\sqfun{\prbexp}{\napiernum^{\imunit u\xi(f)}}
=\napiernum^{-\frac{u^{2}}{4}\fun{\opform{q}}{Tf}}
\sqfun{\prbexp_{\widetilde{\msrprb}_{1}}}{\napiernum^{\frac{uc}{2}\theta(f)}},
\quad
\sqfun{\prbexp_{\widetilde{\msrprb}_{1}}}{\napiernum^{\frac{uc}{2}\theta(f)}}\leq\napiernum^{\frac{u^{2}}{4}\fun{\opform{q}}{Tf}},$$
and $\theta(f)$ and $-\theta(f)$ have the same law under
$\widetilde{\msrprb}_{1}$.
\item The characteristic functions of $\xi(f)$ are real and satisfy
\eqref{eq:field-law-domination} with $\eta(f)=\opfocksegal(Tf)$ under
$\msrprb_{2}$; hence Proposition \ref{prop:field-law-domination} applies to every
net of such families with the same $\opform{q}$ and $T$, for arbitrary sources
and coupling constants.
\item It follows that $$\sqfun{\prbexp}{\xi(f)}=0,
\quad
\sqfun{\prbexp}{\xi(f)^{2}}=\onehalf\fun{\opform{q}}{Tf}-\frac{c^{2}}{4}\sqfun{\prbexp_{\widetilde{\msrprb}_{1}}}{\theta(f)^{2}}.$$
\item If $\fun{\widetilde{\msrprb}_{1}}{\theta(f)\neq0}>0$, there are no
independent real random variables $\eta'$ and $\ell'$ such that $\eta'$ has the
law of $\opfocksegal(Tf)$ and $\eta'+\ell'$ the law of $\xi(f)$.
\end{enumerate}
\item Let $\opform{q}=\opform{q}_{0}+\opform{q}_{1}$ with nonnegative symmetric
bilinear forms such that $\fun{\opform{q}_{0}}{\mathsf{J}}=0$
$\msrprb_{1}$-almost everywhere.
Then the characteristic functions of $\xi(f)$ in (1) and in (2) are
$\napiernum^{-\frac{u^{2}}{4}\fun{\opform{q}_{0}}{Tf}}$ times the functions
$\chi_{1,f}(u)$ obtained from them with $\opform{q}_{1}$ in place of
$\opform{q}$.
Let $\xi_{1}$ be a family of real random variables on $W$ with
\eqref{eq:random-variable-linearity} and the characteristic functions
$\chi_{1,f}$; in the case of (2), its existence is a hypothesis, since removing a
Gaussian factor need not leave a characteristic function.
Let $\zeta$ be a family of
jointly Gaussian centered real random variables on $W$, independent of $\xi_{1}$,
with
$\sqfun{\prbexp}{\zeta(f)\zeta(g)}=\onehalf\fun{\opform{q}_{0}}{Tf,Tg}$.
Then $\xi_{1}+\zeta$ has the joint laws of the family $\xi$ of (1) or (2),
Theorem \ref{thm:shifted-fields-correlated-ideal} applies with the
$\sigma$-algebra $\mblfmlfrak{S}_{\zeta}(W)$ of the order coordinate as
$\mblfmlfrak{S}_{0}$, $\eta=\xi_{1}$, and $\ell=\zeta$, and
\begin{equation}\label{eq:gaussian-source-condensate-directions}
\begin{aligned}
Y_{0}
&=
\set{f\in W}{\fun{\opform{q}_{0}}{Tf}=0};
\end{aligned}
\end{equation}
the ideal $\oaideal{J}_{W\setminus Y_{0}}$ of the correlated directions does not
depend on $c$ and $\mathsf{J}$.
\end{enumerate}
\end{prop}
\begin{proof}
For $\msrprb_{1}$-almost every $\omega$ and $a,u\in\fldreal$, the Gaussian law of
$a\opfocksegal(\mathsf{J}(\omega))+\imunit u\opfocksegal(F)$ and
\eqref{eq:gaussian-source-covariance} give
\begin{equation}\label{eq:gaussian-source-tilt}
\begin{aligned}
\sqfun{\prbexp_{\msrprb_{2}}}
{\napiernum^{a\opfocksegal(\mathsf{J}(\omega))+\imunit u\opfocksegal(F)}}
&=
\fnexp{\frac{a^{2}}{4}\fun{\opform{q}}{\mathsf{J}(\omega)}
+\frac{\imunit au}{2}\fun{\opform{q}}{F,\mathsf{J}(\omega)}
-\frac{u^{2}}{4}\fun{\opform{q}}{F}},
\end{aligned}
\end{equation}
and both sides extend to entire functions of $a\in\fldcmp$.

(1): The case $u=0$, $a=c$ of \eqref{eq:gaussian-source-tilt} and Fubini's
theorem give the normalization of $\msrprb_{c}$.
For $B\in\mblfmlfrak{S}_{1}$ and $f\in W$, \eqref{eq:gaussian-source-tilt} with
$a=c$ and $F=Tf$ gives
$$\sqfun{\prbexp_{\msrprb_{c}}}{\fndef{B}\napiernum^{\imunit u\eta(f)}}
=\frac{1}{Z_{c}}\sqfun{\prbexp_{\msrprb_{1}}}{\fndef{B}\napiernum^{\frac{c^{2}}{4}\fun{\opform{q}}{\mathsf{J}}}}\napiernum^{-\frac{u^{2}}{4}\fun{\opform{q}}{Tf}}.$$
By \eqref{eq:random-variable-linearity}, the joint laws of finitely many
$\eta(f_k)$ are determined by these characteristic functions of the linear
combinations, which gives the law of $\eta$ and the independence.
The random variables $\frac{c}{2}\theta(f)$ are
$\mblfmlfrak{S}_{1}$-measurable.

(2-a): The equality is \eqref{eq:gaussian-source-tilt} with $a=-\imunit c$ and
$F=Tf$, integrated over $\omega$.
The left-hand side has absolute value at most $1$, which gives the bound.
The right-hand side is real, so the characteristic function of $\xi(f)$ is real
and takes complex conjugate, hence equal, values at $u$ and $-u$; the moment
generating functions of $\frac{c}{2}\theta(f)$ and $-\frac{c}{2}\theta(f)$ under
$\widetilde{\msrprb}_{1}$ are finite on $\fldreal$ by the bound and coincide, so
that the two random variables have the same law, and $c\neq0$ gives the
symmetry of $\theta(f)$.

(2-b): By (2-a), $\theta(f)$ has mean $0$ under $\widetilde{\msrprb}_{1}$, and
Jensen's inequality gives
$\sqfun{\prbexp_{\widetilde{\msrprb}_{1}}}{\napiernum^{\frac{uc}{2}\theta(f)}}\geq1$.

(2-c): The bound of (2-a) allows the differentiation of the equality of (a) twice at
$u=0$ under the expectation.

(2-d): For independent $\eta'$ and $\ell'$ with $\eta'$ of the law of
$\opfocksegal(Tf)$, the characteristic function of $\eta'+\ell'$ at $u=1$ has
absolute value at most $\napiernum^{-\frac{1}{4}\fun{\opform{q}}{Tf}}$.
By (2-a), the characteristic function of $\xi(f)$ at $1$ is
$\napiernum^{-\frac{1}{4}\fun{\opform{q}}{Tf}}\sqfun{\prbexp_{\widetilde{\msrprb}_{1}}}{\napiernum^{\frac{c}{2}\theta(f)}}$,
which exceeds $\napiernum^{-\frac{1}{4}\fun{\opform{q}}{Tf}}$ by the strict
Jensen inequality for the nonconstant random variable $\theta(f)$ of mean $0$.

(3): By the Cauchy--Schwarz inequality for $\opform{q}_{0}$, the assumption gives
$\fun{\opform{q}_{0}}{F,\mathsf{J}}=0$ for all $F\in\sphilbfrak{k}$ almost
everywhere, hence
$\fun{\opform{q}}{F,\mathsf{J}}=\fun{\opform{q}_{1}}{F,\mathsf{J}}$ and
$\fun{\opform{q}}{\mathsf{J}}=\fun{\opform{q}_{1}}{\mathsf{J}}$ almost
everywhere, and \eqref{eq:gaussian-source-tilt} together with
$\fun{\opform{q}}{Tf}=\fun{\opform{q}_{0}}{Tf}+\fun{\opform{q}_{1}}{Tf}$ gives the
factorization of the characteristic functions.
The characteristic functions of $\xi_{1}(f)+\zeta(f)$ are the same products, and
by \eqref{eq:random-variable-linearity} the joint laws coincide.
The family $\xi_{1}$ is independent of $\mblfmlfrak{S}_{\zeta}(W)$, $\zeta(f)$ is
$\mblfmlfrak{S}_{\zeta}(W)$-measurable, and the centered Gaussian random
variable $\zeta(f)$ is almost everywhere constant exactly when its variance
$\onehalf\fun{\opform{q}_{0}}{Tf}$ vanishes, which gives
\eqref{eq:gaussian-source-condensate-directions}.
\end{proof}

With a real coupling constant, the source shifts the fields by random variables of the paths, the form of Section \ref{sec:shifted-fields}; with an imaginary one, it does not shift them: the field laws are symmetric with mean \(0\), the variances are reduced by the coupling, and the characteristic functions dominate those of the uncoupled fields, so that by Proposition \ref{prop:field-law-domination} no direction escapes along any net of sources. In both cases, a source orthogonal to the condensate form \(\opform{q}_{0}\) leaves the condensate directions of the uncoupled field unchanged, provided that, with an imaginary coupling constant, the fields of the reduced form \(\opform{q}_{1}\) exist.

\section{Applications to the models}\label{sec:applications}

The results are used in the models as follows; the model-specific statements are proved in the papers on the models.

\begin{itemize}
\item Free Bose gas \cite{YoshitsuguSekine004}.
The BEC state is a mixture of component states with fields shifted by the
condensate parts, which are functions of the order coordinate.
On the real-valued test functions, the sharp-time fields of the BEC measure are of
the form of Section \ref{sec:shifted-fields}, and the ideal of the condensate
directions is the ideal $\oaideal{J}_{W\setminus Y_0}$ of
Theorem \ref{thm:shifted-fields-correlated-ideal}: its quotient is the resolvent
algebra of the test functions on which the zero-mode form vanishes, and its
generators are the resolvents of the directions whose fields are correlated with
the order coordinate.
The GNS representation of the BEC state is faithful, and the order coordinate
appears in the center of its weak closure through strong limits of resolvents of
condensate directions, as described at the end of
Section \ref{sec:shifted-fields}.
\item Van Hove model \cite{YoshitsuguSekine006,YoshitsuguSekine007}.
The linear coupling shifts the fields by a mean functional, defined on a subspace
of test functions, and the removal of the infrared cutoff makes the fields of the
directions with divergent real cutoff means escape to infinity, as in
Theorem \ref{thm:escape-to-infinity} and
Corollary \ref{cor:shifted-vacuum-reduction}; these directions lie outside the
domain of the mean functional, but not every direction outside it escapes.
The ideal of the infrared-singular directions is the ideal generated by the
resolvents of the directions outside this domain.
The quotient by it reduces the algebra of observables to the resolvent algebra
of the domain (Theorem \ref{thm:subspace-complement-quotient}(2)), and the
limit states annihilate the smaller ideal of the escaping directions and agree
on the resolvent algebra of the domain with the ground state of the model.
The no-go statement is the inclusion of the ideal of the condensate
directions in the ideal of the infrared-singular directions, equivalently, by
Theorem \ref{thm:shifted-fields-correlated-ideal}(3), the independence of the
fields of the admissible test functions from the order coordinate.
\item Spin--boson and Nelson models
\cite{YoshitsuguSekine008,YoshitsuguSekine009}.
The Bose field is coupled linearly to the spin or to the particle.
With the cutoffs, the functional integral is of the form of
Proposition \ref{prop:gaussian-source}(1), with the $\sigma$-algebra of the paths
of the spin or of the particle as $\mblfmlfrak{S}_{1}$: conditioned on the paths,
the field is Gaussian with the covariance of the free field and a mean that
depends on the paths, and Theorem \ref{thm:shifted-fields-correlated-ideal}
gives the ideal of the directions correlated with the paths.
The cutoff sources have no zero-momentum component, so that
Proposition \ref{prop:gaussian-source}(3) separates the zero mode, and the
condensate directions are those of the free field.
In the spin--boson model, the conditions on the source
\cite[Section 2.1]{YoshitsuguSekine008} bound the covariance form of the
sources uniformly in the infrared cutoff, so that the shifts
$\abs{\frac{c}{2}\theta(f)}\leq\frac{\abs{c}}{2}\fun{\opform{q}}{Tf}^{\frac{1}{2}}\fun{\opform{q}}{\mathsf{J}}^{\frac{1}{2}}$
are bounded uniformly and no direction escapes.
In the Nelson model, the shifts converge on the physical test functions
\cite{YoshitsuguSekine009}; for the other directions, the hypotheses of
Theorem \ref{thm:escape-to-infinity} are not established there.
\item Pauli--Fierz models \cite{YoshitsuguSekine009}.
The minimal coupling is not linear in the field.
In the model without spin, the vector potential enters the functional integral
through the phase of its stochastic integral along the path of the particle, the
form of Proposition \ref{prop:gaussian-source}(2) with the coupling constant
given by the charge, and the weight of the paths is positive after the
integration over the field.
The fields at a fixed time therefore have symmetric laws with mean $0$, their
variances are reduced by the coupling, and, in the directions with a nonzero
pairing with the source, they are not independent shifts of the free fields.
By Proposition \ref{prop:gaussian-source}(2-b) and
Proposition \ref{prop:field-law-domination}, no direction escapes when the
cutoffs are removed: every limit point of the cutoff states is nonzero at the
resolvents of all real-valued test functions of the free field, which form an
isotropic subspace.
The real-valued directions removed by the restriction to the physical test
functions in \cite{YoshitsuguSekine009}, if any, therefore do not escape; on
them, the restriction is not a reduction of the algebra of observables by an
annihilated ideal, but restricts the algebra to the directions in which the
removal of the cutoffs is proved.
The current of a closed path of the particle has no zero-momentum component,
and Proposition \ref{prop:gaussian-source}(3), with the state of
\cite{YoshitsuguSekine009} for vanishing condensate density as $\xi_{1}$,
separates the zero mode; the condensate directions are those of the free
radiation field, independently of the coupling.
In the model with spin, the jumps of the spin path insert values of the field,
and the weight is of neither form of Proposition \ref{prop:gaussian-source}.
The current and the sources of the Pauli term, which carry a factor of the
momentum, have no zero-momentum component, and the factorization in the proof of
Proposition \ref{prop:gaussian-source}(3) applies to the Gaussian integrals of
\cite[Definition 5.7]{YoshitsuguSekine009}; only this separation of the
zero mode is used for this model.
\end{itemize}

For the free fields, the probabilistic statements are made on the singular Gaussian \(\sminvtemperature\)-Markov path spaces through Proposition \ref{prop:path-space-random-variables}; for the interacting models, through Proposition \ref{prop:gaussian-source} on the path spaces of the papers on these models, for the Pauli--Fierz model with spin only through the separation of the zero mode; their operator-algebraic counterparts follow from the dictionary of Sections \ref{sec:resolvent-ideals-probabilistic}--\ref{sec:escape-to-infinity}.

\section{AI statement}\label{ai-statement}

Large language models were used to draft and proofread the English text, to check cross-references and bibliographic citations, and to check the details of the proofs for errors and inconsistencies. The author reviewed and verified all suggested changes and remains fully responsible for the mathematical content and the text of this paper.

\bibliography{myref.bib}

\end{document}

%% file: mycommands.tex
\makeatletter
\providecommand*{\dashv}{\mathrel{\mathpalette\@Dashv\vDash}}
\newcommand*{\@dashv}[2]{\reflectbox{$\m@th#1#2$}}
\makeatother
\newcommand{\abs}[1]{\left| #1 \right|} % 絶対値
\NewDocumentCommand{\weaknorm}{O{\dbk} m}{#1{#2}} % 弱(準)ノルム
\newcommand{\norm}[1]{\left\Vert #1 \right\Vert} % ノルム
\newcommand{\bkt}[2]{\left\langle #1,\,#2 \right\rangle} % 内積
\newcommand{\rbkt}[2]{\left( #1,\,#2 \right)} % 内積に利用
\DeclareMathOperator*{\slim}{\mathrm{s} \hyphen \lim} % 強極限
\newcommand{\isomto}{\mathrel{\rightarrowtail\kern-1.9ex\twoheadrightarrow}} % 全単射, 同型射, ホモロジー代数・圏論でよく使う
\newcommand{\cbk}[1]{\left\{ #1 \right\}} % 中括弧
\newcommand{\dbk}[1]{\left\langle #1 \right\rangle} % <f>形式の括弧, Dirac bracketの意味でdbk
\newcommand{\pairbk}[1]{\rbk{#1}} % 適当な対のための括弧
\newcommand{\rbk}[1]{\left( #1 \right)} % 丸括弧, 共通
\newcommand{\sqbk}[1]{\left[ #1 \right]} % いわゆる大括弧
\newcommand{\funrbk}[2]{\fun{\rbk{#1}}{#2}} % 関数の部分に丸括弧を入れてまとめる
\newcommand{\fun}[2]{#1 \rbk{#2}} % 関数テンプレート
\newcommand{\sqfuncond}[3]{\sqfun{#1}{#2 \middle| #3}} % 角括弧を使う条件つき期待値などに利用する関数
\newcommand{\sqfun}[2]{#1 \sqbk{#2}} % 角括弧を使う関数, 確率論の期待値や完全な熱力学関数に利用
\newcommand{\commutator}[2]{\sqbk{#1,\,#2}} % 交換子
\NewDocumentCommand{\imunit}{O{\mathsf{i}}}{#1} % 虚数単位, 共通
\NewDocumentCommand{\placeholder}{O{\bullet}}{#1} % 関手やホモロジーなどでハイフンや点で表す要素
\NewDocumentCommand{\trace}{O{\operatorname{Tr}}}{#1} % トレース
\newcommand{\Ker}{\operatorname{Ker}} % 核の集合
\newcommand{\bigmiddleslash}[2]{\left. #1 \middle/ #2 \right.} % 主に商集合\setquotに利用
\newcommand{\cmpconj}[1]{\overline{#1}} % 複素共役
\newcommand{\dom}{\operatorname{dom}} % 定義域
\newcommand{\eqcsq}[1]{\sqbk{#1}} % 同値類の記号
\newcommand{\fml}[2]{\cbk{#1}_{#2}} % 族の基本系
\newcommand{\hyphen}{\hbox{-}} % 記号定義の補助
\newcommand{\idone}{1} % 1で表した恒等写像
\newcommand{\inv}[1]{#1^{-1}} % 逆像・逆写像, \invもよく使うため独立させる
\newcommand{\napiernum}{\mathsf{e}} % ネイピア数, 自然対数の底, 素電荷など他にもeが出てきて紛らわしい場合に利用
\newcommand{\net}[2]{\rbk{#1}_{#2}} % ネット
\newcommand{\onehalf}{\frac{1}{2}} % 1/2
\newcommand{\opchern}[1]{\operatorname{ch}} % チャーン指標
\newcommand{\opimag}{\operatorname{Im}} % 複素数の虚部
\newcommand{\opreal}{\operatorname{Re}} % 複素数の実部
\newcommand{\setSymbolDownLeft}[2]{{\vphantom{#2}}_{#1}{#2}} % 左下に添字をつけるvphantomのエイリアス
\newcommand{\setSymbolUpLeft}[2]{{\vphantom{#2}}^{#1}{#2}} % 左上に添字をつけるvphantomのエイリアス
\newcommand{\supp}{\operatorname{supp}} % 台, 共通
\NewDocumentCommand{\agvariety}{O{\mathcal}}{#1} % 代数多様体, 特にイデアルが生成する代数多様体
\NewDocumentCommand{\cmdrel}{O{\omega}}{#1} % 分散関係
\NewDocumentCommand{\dfsp}{O{A}}{#1} % 微分形式の空間
\NewDocumentCommand{\eqcpointed}{O{\eqcsq} m}{#1{#2}_{\ast}} % 基点つき同値類
\NewDocumentCommand{\fnheaviside}{O{H}}{#1} % ヘビサイド関数
\NewDocumentCommand{\fthol}{O{\mathcal{O}}}{#1} % 正則関数の空間用
\NewDocumentCommand{\ftmero}{O{\mathcal{M}}}{#1} % 有理型関数の空間用
\NewDocumentCommand{\grcentralizer}{O{Z}}{#1} % 中心化群, expedition0005395
\NewDocumentCommand{\grmetform}{O{2} m m}{\grmet[#1] \! \rbkt{#2}{#3}} % (一般)相対性理論, 計量の二次形式
\NewDocumentCommand{\grmet}{O{2}}{\setSymbolDownLeft{#1}{g}} % (一般)相対性理論, 計量の文字
\NewDocumentCommand{\grnormalizer}{O{N}}{#1} % 正規化群, expedition0005395
\NewDocumentCommand{\gropasym}{O{A}}{#1} % 反対称化作用素, expedition0001365
\NewDocumentCommand{\gropsym}{O{S}}{#1} % 対称化作用素, expedition0001365
\NewDocumentCommand{\grpermorderedpair}{O{\mathcal{P}}}{#1} % 偶数個の要素の異なる順序対への分解の全体, expedition0009453
\NewDocumentCommand{\grsym}{O{\mathfrak{S}} m}{#1_{#2}} % 対称群, expedition0001319
\NewDocumentCommand{\gtbase}{O{\mathcal}}{#1} % 位相の基底, 開集合の基底, expedition0000259
\NewDocumentCommand{\gtfilter}{O{\mathcal}}{#1} % フィルター, expedition0000349
\NewDocumentCommand{\gtfmlclosed}{O{\mathcal}}{#1} % 閉集合系だけではなく, ある集合上の一般の閉集合族を表す, expedition0000227
\NewDocumentCommand{\gtfmlopen}{O{\mathcal}}{#1} % 「位相」の意味での開集合系だけではなく, ある集合上の一般の開集合族を表す, expedition0000227
\NewDocumentCommand{\gtopenball}{O{B}}{#1} % 開球
\NewDocumentCommand{\gtopencover}{O{\mathcal}}{#1} % 開被覆
\NewDocumentCommand{\gtopennbh}{O{\mathcal}}{#1} % 開近傍系
\NewDocumentCommand{\gtpreopencover}{O{\mathcal}}{#1} % 内部を取ると開被覆になる集合の族, expedition0009107
\NewDocumentCommand{\gtsubbase}{O{\mathcal}}{#1} % 位相の準基, expedition0005835
\NewDocumentCommand{\gtvicinity}{O{\mathcal}}{#1} % 一様構造での近縁
\NewDocumentCommand{\lasp}{O{\mathcal}}{#1} % 線型空間
\NewDocumentCommand{\latprightrbk}{O{\top} m}{\rbk{#2}^{#1}} % 転置, expedition0009877
\NewDocumentCommand{\latpright}{O{\top} m}{#2^{#1}} % 転置, expedition0009877
\NewDocumentCommand{\latp}{O{t} m}{\setSymbolUpLeft{#1}{#2}} % 転置, expedition0009877
\NewDocumentCommand{\lpdistribution}{O{\mu} m}{#2_{\ast,#1}} % 分布関数, expedition0010310
\NewDocumentCommand{\lpmollifier}{O{\rho}}{#1} % フリードリクスの軟化子
\NewDocumentCommand{\lpofpositive}{O{\chi}}{#1} % 正型関数
\NewDocumentCommand{\manliederiv}{O{L}}{#1} % リー微分
\NewDocumentCommand{\mansmoothnbh}{O{\mathcal}}{#1} % 多様体上の滑らかな開近傍の族
\NewDocumentCommand{\mblfmldsysgenerated}{O{d} m}{\fun{#1}{#2}} % 生成された$d$-系
\NewDocumentCommand{\mblfmlgenerated}{O{\sigma} m}{\fun{#1}{#2}} % 生成された加法族
\NewDocumentCommand{\oacorrfn}{O{\Gamma}}{#1} % 汎関数積分表示とも関わる相関関数, expedition0011645
\NewDocumentCommand{\oagnsvector}{O{\Omega}}{#1} % GNS表現のベクトル状態
\NewDocumentCommand{\oaideal}{O{\mathcal}}{#1} % 作用素環のイデアル：環や代数のイデアルとは分けておく
\NewDocumentCommand{\oanumberoperator}{O{A}}{#1} % レゾルベント環での個数作用素
\NewDocumentCommand{\oaposcone}{O{\mathcal{P}}}{#1} % 冨田-竹崎理論での正錐
\NewDocumentCommand{\oapressure}{O{P}}{#1} % トレースによる通常の圧力は大文字でP: expedition0009842, トレース状態による修正圧力は小文字でp: expedition0009797
\NewDocumentCommand{\oarepn}{O{\pi}}{#1} % 作用素環の表現
\NewDocumentCommand{\oaspkmsstate}{O{\txtkms}}{#1} % KMS状態の空間
\NewDocumentCommand{\oaspnormalstate}{O{N}}{#1} % 作用素環上の正規状態の空間
\NewDocumentCommand{\oasppurestate}{O{P}}{#1} % 作用素環上の純粋状態の空間
\NewDocumentCommand{\oaspstate}{O{E}}{#1} % 作用素環の状態がなす空間
\NewDocumentCommand{\oastatevector}{O{\Omega}}{#1} % GNS表現に付随する状態ベクトル
\NewDocumentCommand{\oastate}{O{\omega}}{#1} % 作用素環の状態
\NewDocumentCommand{\opdilation}{O{\delta}}{#1} % スケーリング(伸張)の作用素, expedition0010830
\NewDocumentCommand{\opdmat}{O{\rho}}{#1} % 密度行列(密度作用素)
\NewDocumentCommand{\opfockan}{O{a}}{#1} % 場の量子論での消滅作用素
\NewDocumentCommand{\opfockcran}{O{a}}{#1^{\#}} % 場の量子論での生成・消滅作用素をまとめて表す記法
\NewDocumentCommand{\opfockcrdagger}{O{a}}{#1^{\dagger}} % 場の量子論での生成作用素のバージョン
\NewDocumentCommand{\opfockcr}{O{a}}{#1^{\ast}} % 場の量子論での生成作用素
\NewDocumentCommand{\opfocknumber}{O{N}}{#1} % 個数作用素
\NewDocumentCommand{\opfocksegalconj}{O{\pi}}{#1} % 場の量子論でのシーガルの場の作用素, expedition0009962
\NewDocumentCommand{\opfocksegal}{O{\phi}}{#1} % 場の量子論でのシーガルの場の作用素
\NewDocumentCommand{\opfockvac}{O{\Omega}}{#1} % フォック真空
\NewDocumentCommand{\opspecmeas}{O{E}}{#1} % スペクトル測度の標準的な記号
\NewDocumentCommand{\opspec}{O{} m}{\fun{\sigma_{#1}}{#2}} % 作用素のスペクトル, expedition0001551
\NewDocumentCommand{\optransl}{O{\tau}}{#1} % 平行移動の作用素, expedition0010829
\NewDocumentCommand{\opvarspec}{O{} m}{\operatorname{spec}_{#1} #2} % 作用素のスペクトル, スピン系の議論など \sigma ではまぎらわしい場合に利用, expedition0001551
\NewDocumentCommand{\physaction}{O{\mathcal{A}}}{#1} % 物理の作用
\NewDocumentCommand{\physcharge}{O{e}}{\mathrm{#1}} % 電荷, 素電荷などの物理量を表す. e以外にもqに利用する
\NewDocumentCommand{\physcplconst}{O{\mathsf{g}}}{#1} % 結合定数, coupling constant
\NewDocumentCommand{\physelectrostaticcapasity}{O{\mathrm{Cap}}}{#1} % 静電容量, expedition0010917
\NewDocumentCommand{\physenergy}{O{E}}{#1} % エネルギー
\NewDocumentCommand{\physgse}{O{E}}{#1_{0}} % 基底エネルギー, ground state energy
\NewDocumentCommand{\physham}{O{H}}{#1} % ハミルトニアンの雛形, たいていは下つき添字で水素原子, 自由電磁場などを表す
\NewDocumentCommand{\physlagdensity}{O{\mathcal{L}}}{#1} % ラグランジアン密度の雛形
\NewDocumentCommand{\physlag}{O{L}}{#1} % ラグランジアンの雛形
\NewDocumentCommand{\physliouvilean}{O{L}}{#1} % リウビリアンの雛形
\NewDocumentCommand{\physmass}{O{m}}{#1} % 物質の質量
\NewDocumentCommand{\prbbrownmv}{O{B}}{#1} % ブラウン運動
\NewDocumentCommand{\prbcharfun}{O{\chi}}{#1} % 特性関数
\NewDocumentCommand{\prbdist}{O{\mathcal{P}}}{#1} % 確率分布
\NewDocumentCommand{\prbgaussianmeasure}{O{\msrcal{N}}}{#1} % ガウス測度
\NewDocumentCommand{\prbnormaldist}{O{N}}{#1} % 正規分布
\NewDocumentCommand{\prbpoissonprocess}{O{N}}{#1} % ポアソン過程
\NewDocumentCommand{\prbprocess}{O{X}}{#1} % 確率過程
\NewDocumentCommand{\prbqspace}{O{\mathcal{Q}}}{#1} % $Q$-表示の土台の集合を表す
\NewDocumentCommand{\prbspsample}{O{\Omega}}{#1} % 確率空間に対する標本空間, expedition0010114
\NewDocumentCommand{\prbspstate}{O{\mathsf{S}}}{#1} % 確率論での状態空間
\NewDocumentCommand{\psh}{O{\mathfrak}}{#1} % 前層
\NewDocumentCommand{\qtquantumchannel}{O{\mathcal{L}}}{#1} % 量子通信路, expedition0005561, expedition0006844
\NewDocumentCommand{\repn}{O{\pi}}{#1} % 表現
\NewDocumentCommand{\schattencls}{O{\mathbb{K}}}{#1} % コンパクト作用素またはシャッテンクラス
\NewDocumentCommand{\setfmlcylinder}{O{\mathcal{C}}}{#1} % 筒集合族, expedition0000313, expedition0004973
\NewDocumentCommand{\setfml}{O{\mathcal}}{#1} % 集合族
\NewDocumentCommand{\setindex}{O{\mathcal} m}{#1{#2}} % 添字集合
\NewDocumentCommand{\setlattice}{O{\Gamma}}{#1} % 格子, expedition0010278
\NewDocumentCommand{\setspecial}{O{\mathcal} m}{#1{#2}} % 特別な集合を表すための記号, 一点集合・二点集合などを太字で表すといった用途
\NewDocumentCommand{\shdiffform}{O{\sheaf{A}}}{#1} % 微分形式がなす層
\NewDocumentCommand{\sheaf}{O{\mathfrak}}{#1} % 層の記号
\NewDocumentCommand{\smchemicalpotential}{O{\mu}}{#1} % 化学ポテンシャル
\NewDocumentCommand{\smenergydensity}{O{\varrho}}{#1} % エネルギー密度, expedition0011356
\NewDocumentCommand{\smfluctuationwithdmat}{O{\beta} m}{\smuncertaintywithdmat[#1]{#2}^2} % 密度行列に付随する正規状態でのゆらぎ, expedition0011059
\NewDocumentCommand{\sminvtemperature}{O{\beta}}{#1} % 逆温度
\NewDocumentCommand{\smlocaldensityoperator}{O{\rho}}{#1} % 大正準状態に対する局所密度作用素, expedition0011294
\NewDocumentCommand{\smmicrocanonicalstate}{O{\beta} m}{\physmean{#2}_{#1}} % 密度行列に付随する正規状態, expedition0011059
\NewDocumentCommand{\smnumberdensity}{O{\rho}}{#1} % 粒子数密度, expedition0011351
\NewDocumentCommand{\smooth}{O{\mathcal{E}}}{#1} % 滑らかな関数の空間, 特に微分形式関連でも利用, expedition0010649
\NewDocumentCommand{\smparticlenumber}{O{N}}{#1} % 平均粒子数, expedition0011351
\NewDocumentCommand{\smpressure}{O{p}}{#1} % 統計力学での圧力, expedition0011341
\NewDocumentCommand{\smspecificfreeenergy}{O{\bar{f}}}{#1} % (ヘルムホルツの)比自由エネルギー
\NewDocumentCommand{\smthermalvac}{O{\beta}}{\Omega_{#1}} % 熱的真空, expedition0011060
\NewDocumentCommand{\smuncertaintywithdmat}{O{\beta} m}{\rbk{\triangle #2}_{#1}} % 密度行列に付随する正規状態での不確かさ, expedition0011059
\NewDocumentCommand{\sphilbfrak}{O{\mathfrak}}{#1} % ヒルベルト空間
\NewDocumentCommand{\sphilb}{O{\mathcal}}{#1} % ヒルベルト空間
\NewDocumentCommand{\splowerhalf}{O{\mathbb{H}}}{#1_{\txtneg}} % 開下半空間
\NewDocumentCommand{\spupperhalf}{O{\mathbb{H}}}{#1_{\txtnonneg}} % 開上半空間
\NewDocumentCommand{\topmetric}{O{d}}{#1} % 位相空間上の距離関数
\NewDocumentCommand{\vaoutnormal}{O{\widehat}}{#1} % 外向き外法線
\newcommand{\category}[1]{\mathop{\mathsf{#1}}} % 圏を生成する
\newcommand{\catpresheaf}[1]{\category{PSh}} % 前層がなす圏
\newcommand{\conti}{C} % 連続または滑らかな関数の空間
\newcommand{\faadj}[1]{#1^{\ast}} % 行列・作用素に対する共役・随伴. 線型代数でも内積の必要性からこれに統合. 双対空間にはdualをあてる
\newcommand{\fldcmp}{\fld{C}} % 複素数体
\newcommand{\fldreal}{\fld{R}} % 実数体
\newcommand{\fld}[1]{\mathbb{#1}} % 実数体・複素数体・四元数体, またはこれらをまとめて表す体用のコマンド
\newcommand{\fndef}[1]{\boldsymbol{1}_{#1}} % 定義関数
\newcommand{\fnexp}[1]{\fun{\exp}{#1}} % 指数関数
\newcommand{\fnrestr}[2]{\left. #1 \right|_{#2}} % 写像の制限
\newcommand{\gtclos}[1]{\overline{#1}} % 一般位相の閉包
\newcommand{\gtoneptcpt}[1]{\dot{#1}} % 一点コンパクト化
\newcommand{\lp}{L} % ルベーグ空間, L_{\txtloc}^p(X), L^p(X)
\newcommand{\mblfmlfrak}[1]{\mathfrak{#1}} % 加法族の mathfrak 表記
\newcommand{\monnat}{\mathbb{N}} % 0を含めた自然数全体がなすモノイド
\newcommand{\msras}[1]{#1 \hyphen \txtprobas} % ある測度に対するほとんど確実に
\newcommand{\msrcal}[1]{\mathcal{#1}} % mathcalで表す測度
\newcommand{\msrdirac}{\delta} % ディラックのデルタ測度
\newcommand{\msrprb}{\mathrm{Pr}} % たまに役立つであろう, 立体のPrで表す確率測度
\newcommand{\msr}[1]{#1} % 測度の識別用
\newcommand{\oacenter}{Z} % 作用素環の中心
\newcommand{\oacommutant}[1]{#1^{\prime}} % 作用素環での可換子環
\newcommand{\oacstar}{C^{\ast}} % $C^{\ast}$-環
\newcommand{\oadoublecommutant}[1]{#1^{\prime \prime}} % 二重可換子環
\newcommand{\oaresolventalgebra}{\oa{R}} % レゾルベント環
\newcommand{\oaresolvent}{R} % レゾルベント環の要素としてのレゾルベント
\newcommand{\oa}[1]{\mathcal{#1}} % 作用素環を表す記号用
\newcommand{\opdmsr}[1]{\mathop{d #1}} % 積分で現れる測度用のコマンド, dxのような単純な対象も含む
\newcommand{\opform}[1]{\mathsf{#1}} % 二次形式
\newcommand{\opspbddlin}[1]{\fun{\mathbb{B}}{#1}} % 有界線型作用素がなす空間
\newcommand{\opspecint}[1]{\mathcal{E}} % スペクトル積分, expedition0001621
\newcommand{\physmean}[1]{\dbk{#1}} % 物理, 特に統計力学によく出てくる平均の記号
\newcommand{\prbexp}{\mathbb{E}} % 期待値, 条件つき期待値にも応用
\newcommand{\ringratint}{\mathbb{Z}} % 有理整数環
\newcommand{\semigrposint}{\monnat_1} % 正の整数の全体または1以上の自然数全体

\newcommand{\seq}[2]{\if\relax\detokenize{#1}\relax \rbk{#1} \else \rbk{#1}_{#2} \fi} % 点列の基本系
\newcommand{\setisomorphism}[1]{\operatorname{Iso}} % 同型写像の集合
\newcommand{\setone}[1]{\cbk{#1}}
\newcommand{\setquot}[2]{\bigmiddleslash{#1}{#2}} % 商集合
\newcommand{\set}[2]{\left\{#1 \, \middle| \, #2\right\}}
\newcommand{\splinspan}{\operatorname{span}} % 線型包
\newcommand{\txtarakiwoods}{\mathrm{AW},\mathrm{b}} % 荒木-ウッズ
\newcommand{\txtbsn}{\mathrm{b}} % ボソンのb
\newcommand{\txtfock}{\mathrm{F}} % フォック
\newcommand{\txtkms}{\mathrm{KMS}} % KMS状態の集合を表す
\newcommand{\txtneg}{\mathrm{-}} % 負の部分
\newcommand{\txtnonneg}{\mathrm{+}} % 正または非負の部分
\newcommand{\txtprobas}{\mathrm{a.s.}} % ほとんど確実に
\newcommand{\txtregular}{\mathrm{reg}} % 正則な測度などの正則
\newcommand{\txtsingular}{\mathrm{s}} % 特異
\newcommand{\txtvac}{\mathrm{vac}} % 真空, vacuum
\newcommand{\vainnprod}{\cdot} % ベクトル解析枠, 高校以来の内積の記号